\documentclass[final]{IEEEtran}
\usepackage{amsthm,amssymb,graphicx,multirow,amsmath,color,amsfonts}
\usepackage[caption=false,font=footnotesize]{subfig}
\usepackage[update,prepend]{epstopdf}
\usepackage[noadjust]{cite}
\usepackage[latin1]{inputenc}
\usepackage{tikz}
\usepackage{bbm} 
\usepackage{pdfpages}
\usepackage{multirow}
\usepackage{tabulary}
\usepackage{comment}
\usepackage{algorithm}
\usepackage{url}
\usepackage{algpseudocode}
\def\BSTATE{\STATE\hskip-\ALG@thistlm}
\makeatother

\def\nbv{{\mathbf{v}}}

\def\nbx{{\mathbf{x}}}

\def\nb0{{\mathbf{0}}}
\def\nb1{{\mathbf{1}}}

\def\nbA{{\mathbf{A}}}

\def\ncalL{{\mathcal{L}}}

\def\ncalQ{{\mathcal{Q}}}

\def\nbbB{{\mathbb{B}}}

\def\nbbE{{\mathbb{E}}}

\def\nbbR{{\mathbb{R}}}

\newtheorem{ndef}{Definition}

\newtheorem{theorem}{Theorem}
\newtheorem{prop}{Proposition}
\newtheorem{cor}{Corollary}

\newtheorem{remark}{Remark}

\def\a{\overset{({\rm a})}{=}}
\def\b{\overset{({\rm b})}{=}}
\def\c{\overset{({\rm c})}{=}}
\def\d{\overset{({\rm d})}{=}}

\begin{document}
\graphicspath{{./Figures/}}
\title{
Closed-form Characterization of the MGF of AoI in Energy Harvesting Status Update Systems 
}
\author{
Mohamed A. Abd-Elmagid, \textit{Member}, \textit{IEEE}, and Harpreet S. Dhillon, \textit{Senior Member}, \textit{IEEE}
\thanks{M. A. Abd-Elmagid and H. S. Dhillon are with Wireless@VT, Department of ECE, Virginia Tech, Blacksburg, VA. Email: \{maelaziz,\ hdhillon\}@vt.edu. The support of the U.S. NSF (Grants CPS-1739642 and CNS-1814477) is gratefully acknowledged. This work was presented in part at the IEEE/IFIP WiOpt, 2021 \cite{abd2021distributional}. 

Copyright (c) 2022 IEEE. Personal use of this material is permitted.  However, permission to use this material for any other purposes must be obtained from the IEEE by sending a request to pubs-permissions@ieee.org.
}
}

\maketitle

\begin{abstract}
This paper considers a real-time status update system in which an energy harvesting (EH)-powered transmitter node observes some physical process, and sends its sensed measurements in the form of \textit{status updates} to a destination node. The status update and harvested energy packets are assumed to arrive at the transmitter according to independent Poisson processes, and the service time of each status update is assumed to be exponentially distributed. We quantify the \textit{freshness} of status updates when they reach the destination using the concept of \textit{Age of Information (AoI)}. Unlike most of the existing analyses of AoI focusing on the evaluation of its average value when the transmitter is not subject to energy constraints, our analysis is focused on understanding the \textit{distributional properties} of AoI through the characterization of its moment generating function (MGF). In particular, we use the stochastic hybrid systems (SHS) framework to derive closed-form expressions of the MGF of AoI under several queueing disciplines at the transmitter, including non-preemptive and preemptive in service/waiting strategies. 
Using these MGF results, we further obtain closed-form expressions for the first and second moments of AoI in each queueing discipline. 
We demonstrate the generality of this analysis by recovering several existing results for the corresponding system with no energy constraints as special cases of the new results. Our numerical results verify the analytical findings, and 
demonstrate the necessity of incorporating the higher moments of AoI in the implementation/optimization of real-time status update systems rather than just relying on its average value.
\end{abstract}
\section{Introduction} \label{sec:intro}
Timely delivery of real-time status updates is necessary for many critical and emerging applications, such as predicting and controlling forest fires, factory automation, and intelligent transportation systems, to name a few. A typical real-time status update system consists of a {\it transmitter node} that generates status updates about some physical process of interest, and then sends them through a communication system to a {\it destination node}. Examples of the transmitter node include Internet of Things (IoT) devices, sensors and aggregators, while of the destination node include cellular base stations \cite{abd2018role}. Clearly, the performance of real-time status update applications 
is restricted by the freshness of status updates when they reach the destination node. The performance bottleneck in many such cases is the energy-constrained nature of these low-power transmitters, which may result in the loss or out-of-order reception of status updates at the destination node. Thus, in order to accurately characterize the performance of a real-time status update system/application, one needs a performance metric that is able to quantify the freshness of information status at the destination node while incorporating the energy constraints at the transmitter node.
 
 Since conventional performance metrics of communication networks, specifically throughput and delay, do not explicitly account for the generation time of status updates, they are not suitable for characterizing the performance of real-time status update systems. Motivated by this, the authors of \cite{kaul2012real} introduced the AoI performance metric which tracks the generation time of each status update at the transmitter node, and thus being able to quantify the freshness of status updates upon their reception at the destination node. 
 In particular, for a queuing-theoretic model in which randomly generated status updates arrive at the transmitter node according to a Poisson process, AoI was defined in \cite{kaul2012real} as the time elapsed since the latest successfully received status update at the destination node was generated at the transmitter node. As will be discussed next in detail, the queueing-theoretic analysis of AoI has mostly been limited to the characterization of its average value when the transmitter does not suffer from energy limitations. To accurately quantify the performance of real-time status update systems, this paper presents a novel analysis for deriving distributional properties of AoI through the characterization of the MGF for several queueing disciplines. In our analysis, we consider that the transmitter node is powered by EH such that the harvested energy is stored in a battery of finite capacity.
\subsection{Related Work} 
For systems in which the transmitter node has a single source (that generates status updates about some physical process) and a single server (that sends the generated updates to the destination node), referred to as single-source single-server systems, the authors of \cite{kaul2012real} first derived a closed-form expression of the average AoI under first-come-first-served (FCFS) queueing discipline (for the case in which the transmitter does not have energy limitations). The average value of AoI or peak AoI (a related metric based on the maximum values of AoI over time) is then characterized under several queueing disciplines in a series of subsequent prior works \cite{kaul2012status,soysal,costa2016age,chen2016age,kam2018age,kavitha2018controlling,zou2019waiting,tripathi2019age}. In particular, the authors of \cite{kaul2012status} studied the M/M/1 last-come-first-served with preemption in service (LCFS-PS) queue, where a new arriving status update preempts the packet in service (if any) upon its arrival. The G/G/1 system was studied in \cite{soysal} under the LCFS-PS and LCFS with no preemption (LCFS-NP) queueing disciplines, where in the LCFS-NP queueing discipline, a new arriving status update will be discarded if the server is busy. The concept of peak AoI was introduced in \cite{costa2016age}, where closed-form expressions for the average values of both AoI and peak AoI were derived under the LCFS-NP and LCFS with preemption in waiting (LCFS-PW) queueing disciplines. Note that under the LCFS-PW queuing discipline, there is a data buffer of size one which stores the latest arriving status update while the server is busy. In \cite{chen2016age}, the average peak AoI was also analyzed for several queueing disciplines when a status update delivery error may occur probabilistically. The authors of \cite{kam2018age,kavitha2018controlling,zou2019waiting} demonstrated that the achievable average values of AoI and peak AoI could be improved by: i) introducing deadlines for status updates waiting in the queue for service \cite{kam2018age}, ii) controlling status update drops (when arriving status updates find the server busy) \cite{kavitha2018controlling}, and iii) introducing a waiting duration before service (for the M/G/1 queue under LCFS-NP and LCFS-PW) \cite{zou2019waiting}. The average AoI and average peak AoI were also analyzed  for various discrete time queueing systems in \cite{tripathi2019age} under FCFS and LCFS disciplines. Further, without considering energy constraints at the transmitter, the average AoI/peak AoI was analyzed for the single-source multi-server systems \cite{kam2014effect,kam2015effect,Boji_21}, and for the multi-source single-server systems in which the transmitter node has multiple sources of information (generating status updates about different physical processes) and a single server (delivering the status updates generated by the sources to the destination node) \cite{yates2012real,huang2015optimizing,pappas2015age,yates2017status,najm2018status,kosta2019age,Moltafet_multisource,Xu_21}. 

While the average of AoI or peak AoI was derived for a variety of queueing systems \cite{kaul2012real,kaul2012status,soysal,costa2016age,chen2016age,kam2018age,kavitha2018controlling,zou2019waiting,tripathi2019age,kam2014effect,kam2015effect,Boji_21,yates2012real,huang2015optimizing,pappas2015age,yates2017status,najm2018status,kosta2019age,Moltafet_multisource,Xu_21}, only a handful of recent works aimed to characterize their distributions (or some distributional properties) \cite{Inoue19,kosta2020non,Champati19,Chiariotti_dist,ayan2020probability,olga20,Akar21,jiang2020joint}. In particular, the distribution/distributional properties of AoI/peak AoI were investigated for single-source systems in \cite{Inoue19,kosta2020non,Champati19,Chiariotti_dist,ayan2020probability,olga20}, and for multi-source systems in \cite{Akar21,jiang2020joint}. The authors of \cite{Inoue19} (\cite{kosta2020non}) demonstrated that the stationary distribution of AoI for a continuous (discrete) time queue can be expressed in terms of the stationary distributions of the system delay and the peak AoI. Using this result, the stationary distributions of AoI and peak AoI were analyzed in \cite{Inoue19,kosta2020non} for various queueing disciplines such as FCFC, LCFC-WP, LCFC-PS, and LCFC-PW. The distribution of AoI for the G/G/1 queue (under LCFS-NP and LCFS-PW disciplines) was studied in \cite{Champati19}. The authors of \cite{Chiariotti_dist} characterized the distribution of peak AoI for a system having two M/M/1 FCFC queues in tandem. The distribution of AoI/peak AoI was also analyzed for multi-hop networks with/without priorities in \cite{ayan2020probability,olga20}. For multi-source single-server systems, the distributions of AoI and PAoI were characterized for various discrete time queues in \cite{Akar21}, and the joint Laplace transform of the AoI values associated with different sources was derived in \cite{jiang2020joint}. Note that a common assumption in \cite{Inoue19,kosta2020non,Champati19,Chiariotti_dist,ayan2020probability,olga20,Akar21,jiang2020joint} for the analysis of the distribution/distributional properties of AoI/peak AoI is that the transmitter node does not have energy limitations. Different from \cite{Inoue19,kosta2020non,Champati19,Chiariotti_dist,ayan2020probability,olga20,Akar21,jiang2020joint}, our focus in this paper is on the characterization of distributional properties of AoI in the case where the transmitter node is powered by EH.

Due to the common ergodicity assumption of the AoI process in the above works, their analyses were mainly based on identifying the properties of the AoI sample functions and applying geometric arguments to evaluate the average value, distribution, or distributional properties of AoI/peak AoI. These approaches require a careful choice of random variables for representing the AoI sample function, and often need involved calculations of the joint moments. This has motivated the authors of \cite{yates2018age} to build on the SHS framework in \cite{hespanha2006modelling}, and derive promising results allowing the use of the SHS approach for the characterization of average AoI under an arbitrary queueing discipline. The results of \cite{yates2018age} were then generalized in \cite{yates2020age} to the characterization of higher moments of AoI and the MGF. The SHS approach requires us to track the discrete state of the system (for example, represented by the number of status updates in the system) using a continuous-time Markov chain, as well as the continuous state of the system that is modeled by a vector containing the relevant age processes\footnote{A more detailed description of the SHS will be provided in Subsection \ref{Sub:SHSs}.}. Following \cite{yates2018age,yates2020age}, the SHS approach was then used to evaluate the average AoI for a variety of queueing disciplines in \cite{SHS_1,SHS_2,SHS_3,SHS_4,SHS_5,SHS_6,SHS_7}, and the MGF of AoI for a two-source single-server system with status update management in \cite{SHS_8}. Compared to the analyses of \cite{SHS_1,SHS_2,SHS_3,SHS_4,SHS_5,SHS_6,SHS_7,SHS_8}, where there are no energy constraints at the transmitter node, the analysis of AoI using the SHS approach becomes much more challenging when the transmitter suffers from energy limitations. The difficulty in the analysis of this case emerges from the fact that the decisions of discarding, serving or storing (if possible) a new arriving status update are not only a function of the system occupancy with status updates (as it is the case in \cite{SHS_1,SHS_2,SHS_3,SHS_4,SHS_5,SHS_6,SHS_7,SHS_8}), but also a function of the evolution of the battery state over time at the transmitter node. Thus, the joint evolution of the energy battery state and the system occupancy with status updates has to be incorporated in the process of decision-making. In order to achieve that, one needs to analyze a two-dimensional continuous-time Markov chain modeling the system discrete state with new transitions associated with the events of harvested energy packet arrivals/departures, compared to the conventional one-dimensional Markov chain used in \cite{SHS_1,SHS_2,SHS_3,SHS_4,SHS_5,SHS_6,SHS_7,SHS_8} to track the number of status updates in the system when the transmitter is not subject to energy constraints.

\begin{table*}
\centering
{\caption{A Summary of the Queueing Theory-based Analyses of AoI in the Existing Literature.} 
\label{table:summary}
\scalebox{.9}
{ \begin{tabular}{ |c |c|c|}
\hline
    & No energy constraints at the transmitter  & The transmitter is powered by EH\\ \hline
Average of AoI/peak AoI& \cite{kaul2012real,kaul2012status,soysal,costa2016age,chen2016age,kam2018age,kavitha2018controlling,zou2019waiting,tripathi2019age,kam2014effect,kam2015effect,Boji_21,yates2012real,huang2015optimizing,pappas2015age,yates2017status,najm2018status,kosta2019age,Moltafet_multisource,Xu_21,yates2018age,SHS_1,SHS_2,SHS_3,SHS_4,SHS_5,SHS_6,SHS_7} & \cite{Yates_EH,zheng2019closed,farazi2018average,farazi2018bverage} \\ \hline
Distribution/distributional properties of AoI/peak AoI& \cite{Inoue19,kosta2020non,Champati19,Chiariotti_dist,ayan2020probability,olga20,Akar21,jiang2020joint,yates2020age,SHS_8} &  This paper\\ \hline
\end{tabular}}} 
\end{table*} 
For the scenario where the transmitter node is powered by EH, there are a handful of prior works \cite{Yates_EH,zheng2019closed,farazi2018average,farazi2018bverage} analyzing AoI in the single-source single-server systems by applying geometric arguments \cite{Yates_EH,zheng2019closed}, and by using the SHS approach \cite{farazi2018average,farazi2018bverage}. However, the analyses of \cite{Yates_EH,zheng2019closed,farazi2018average,farazi2018bverage} have been limited to the evaluation of the average AoI. Different from these, this paper makes the first attempt at deriving the distributional properties of AoI through the characterization of the MGF for a variety of queueing disciplines. Table \ref{table:summary} further highlights the gap in the literature that we aim to fill in this paper. Before going into more details about our contributions, it is instructive to note that besides the above queueing theory-based analyses of AoI, there have been efforts to evaluate and optimize AoI or some other AoI-related metrics in different communication systems that deal with time-sensitive information (see \cite{roy_survey} for a comprehensive survey). For instance, AoI has been studied in the context of age-optimal transmission scheduling policies \cite{sun2017update,bedewy2016optimizing,Qing_he,lu2018age,jiang2019timely,huang2020age,han2020fairness}, multi-hop networks \cite{talak2017,bedewy2017age,Buyukates_ulu}, broadcast networks \cite{kadota2016,hsu2019scheduling}, ultra-reliable low-latency vehicular networks \cite{abdel2018ultra}, unmanned aerial vehicle (UAV)-assisted communication systems \cite{abd2018average,AbdElmagid2019Globecom_b,ferdowsi2021neural}, EH systems \cite{Baran_EH,Jing_EH,arafa2019age,AbdElmagid2019Globecom_a,hatami2020age,abd2019tcom,AbdElmagid_joint,nouri2020age,khorsandmanesh2020average}, large-scale analysis of IoT networks \cite{emara2019spatiotemporal,mankar2020stochastic_GC2,Praful_GC1}, information-theoretic analysis \cite{Sun_IT,bastopcu2020partial,wang2020value,shisher2021age}, timely source coding \cite{zhong2016timeliness,feng2019age}, cache updating systems \cite{tang2020age,ma2020age,bastopcu2020information}, economic systems \cite{zhang2019price}, timely communication in federated learning \cite{yang2020age,buyukates2020timely}, and remote estimation \cite{ornee2019sampling}.
\subsection{Contributions}\label{sub:con}
This paper studies a queueing model for real-time status update systems consisting of an EH-powered transmitter node equipped with an energy queue of finite capacity, and a destination node. In particular, the transmitter is responsible for keeping the status of its observed physical process fresh at the destination, by frequently generating status update packets over time and sending them to the destination. The update/energy packets are assumed to arrive at the transmitter according to a Poisson process, and the service time of each status update is assumed to be exponentially distributed. It is worth noting that this system setup is of practical interest since the transmitter may represent an aggregator (that is subject to energy constraints) in an IoT network which is supposed to send sensed measurements about some physical process to a remote destination node. For this setup, our main contributions are listed next.

{\it A novel analysis for deriving the MGF of the AoI at the destination.} We use the SHS framework to characterize distributional properties of AoI at the destination node. In particular, we derive closed-form expressions of the MGFs of AoI at the destination under the LCFS-NP, LCFS-PS, and LCFS-PW queueing disciplines. For each queueing discipline, we study the following two different cases: i) the transmitter can harvest energy only if the system is empty, and ii) the transmitter can simultaneously harvest energy and serve status updates. Since the practical scenario where the transmitter is subject to energy constraints is considered here, we have to model the discrete state of the system as a two-dimensional continuous-time Markov chain to track both the numbers of update and energy packets in the system. 

{\it Analytical characterization of the first and second moments of AoI at the destination.} For each queueing discipline, we use the derived MGF of AoI at the destination to obtain closed-form expressions for both the first and second moments of AoI. These results allow us to gain useful insights about the achievable AoI performance by each of the considered queueing disciplines. For instance, we analytically demonstrate the superiority of the LCFS-PS queueing discipline over the LCFS-NP and LCFS-PW queueing disciplines. We further provide asymptotic results showing the behavior of the difference (or gap) between the achievable AoI performance by the LCFS-PS queueing discipline and that of the LCFS-NP/LCFS-PW queueing discipline as a function of the system parameters. Our results also demonstrate that the ability of the transmitter node to simultaneously harvest energy and serve update packets reduces the gap between the achievable AoI performances by the LCFS-PS and LCFS-NP queueing disciplines (with respect to the case where the transmitter can harvest energy only if the system is empty). Another interesting insight obtained from our results is that the second moment of AoI associated with the LCFS-NP queueing discipline is invariant to exchanging the arrival rates of status update and harvested energy packets.

{\it Asymptotic results demonstrating the generality of the derived expressions.} We demonstrate that as the harvested energy arrival rate at the transmitter node becomes large, the first and second moments of AoI for each queueing discipline converge to their counterparts in the case where the transmitter does not have energy limitations. In particular, the AoI performance of the LCFS-NP queueing discipline converges to that of the M/M/1/1 case in \cite{costa2016age}, the AoI performance of the LCFS-PS queueing discipline converges to its counterpart in \cite{yates2018age} for the single-source case, and the AoI performance of the LCFS-PW queueing discipline (when the transmitter is able to simultaneously harvest energy and serve status updates) converges to its counterpart in \cite{yates2018age} for the single-source case. 

{\it System design insights.} Our numerical results provide several useful system design insights. For instance, they quantify the improvement in the achievable AoI performance by each queueing discipline associated with the increase in either the battery capacity or the harvested energy arrival rate at the transmitter node. They also show the superiority of the LCFS-PS queueing discipline over the LCFS-NP and LCFS-PW queueing disciplines in terms of the achievable AoI performance. Further, the results reveal the necessity of incorporating the higher moments of AoI in the implementation/optimization of real-time status updates systems rather than just relying on the average value (as has been mostly done in the existing literature on AoI).
\subsection{Organization}
The rest of the paper is organized as follows. Section \ref{sec:Model} presents our system model. The problem statement and the solution approach are then presented in Section \ref{sec:problem_statement}. Afterwards, we present our analysis used to characterize the MGFs of AoI at the destination associated with the LCFS-NP, LCFS-PS and LCFS-PW queueing disciplines in Sections \ref{sec:WP}, \ref{sec:PS} and \ref{sec:PW}, respectively. Section \ref{sec:numerical} presents our numerical results, which verify the analytical findings from Sections \ref{sec:WP}-\ref{sec:PW} as well as show the impact of system parameters on the achievable AoI performance by each of the considered queueing disciplines. Finally, Section \ref{sec:con} concludes the paper.
\section{System Model}\label{sec:Model}
\subsection{Network Model}
\begin{figure}[t!]
\centering
\includegraphics[width=0.9\columnwidth]{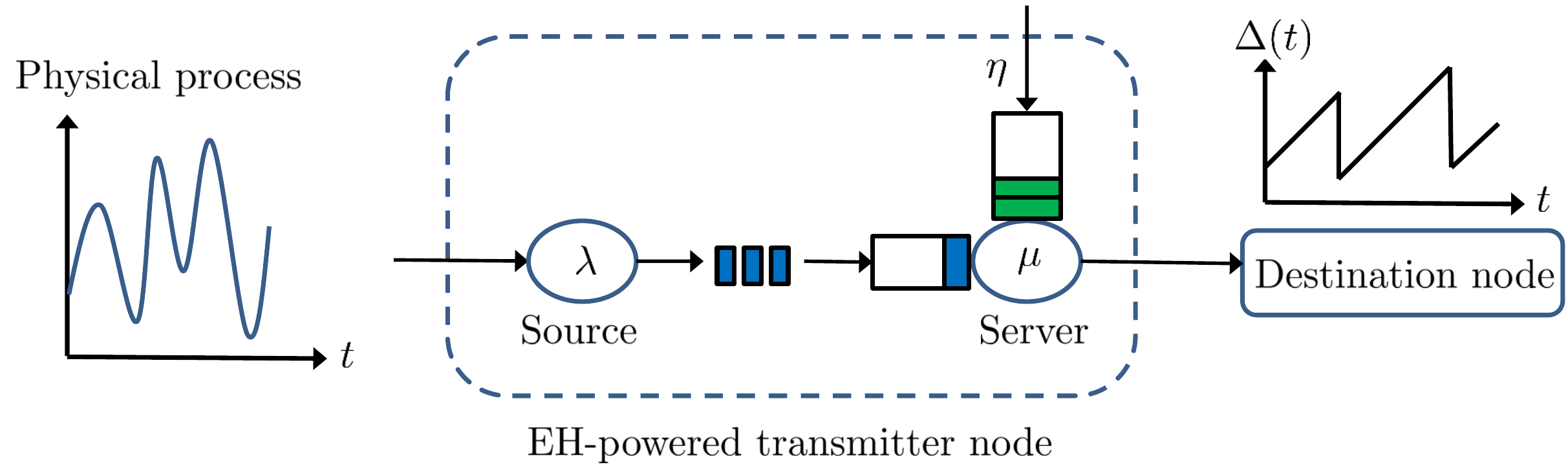}
\caption{An illustration of the system setup.}
\label{f:sys_setup}
\end{figure}

We consider a real-time status update system in which an EH-powered transmitter node monitors some physical process, and sends its measurements to a destination node in the form of status update packets. As shown in Fig. \ref{f:sys_setup}, the transmitter node contains a single source generating the status update packets and a single server delivering the generated packets to the destination. Each status update packet carries some information about the value of the physical process and a time stamp indicating the time at which that information was measured. 
As already pointed out in Subsection \ref{sub:con}, this system setup can be mapped to many scenarios of practical interest, such as an IoT network in which an aggregator (represents the transmitter node in our model) delivers sensed measurements 
to a destination node. 

The status update packets are assumed to be generated by the source at the transmitter node according to a Poisson process with rate $\lambda$. Further, the transmitter harvests energy in the form of energy packets such that each energy packet contains the energy required for sending one status update packet to the destination node \cite{Yates_EH,zheng2019closed,farazi2018average,farazi2018bverage}. In particular, the harvested energy packets are assumed to arrive at the transmitter according to a Poisson process with rate $\eta$, and are stored in a battery queue of length $B$ packets at the server (for serving the generated update packets). Given that the transmitter node has at least one energy packet in its battery queue, the time needed by its server to send a status update packet is assumed to be distributed as an exponential random variable with rate $\mu$ \cite{kaul2012real,kaul2012status,costa2016age}. Let $\rho = \frac{\lambda}{\mu}$ and $\beta = \frac{\eta}{\mu}$ denote the server utilization and energy utilization factors, respectively. 
We quantify the freshness of information status about the physical process at the destination node (as a consequence of receiving status update packets from the transmitter node) using the concept of AoI. Formally, AoI is defined as the following \cite{kaul2012real}:
\begin{ndef}
Let $t_i$ and $t'_i$ denote the arrival and reception time instants of the $i$-th update packet at the transmitter and destination, respectively. Further, define $L(t)$ to be the index of the latest update packet received at the destination by time $t$, i.e., $L(t) = {\rm max}\{i|t'_i \leq t\}$. Then, AoI is defined as the following random process
\begin{align}
\Delta(t) = t - t_{L(t)}.
\end{align}
\end{ndef}
\subsection{Queueing Disciplines Considered in this Paper}\label{sub:disciplines}
For the above system setup, we analyze the AoI performance at the destination under three different queueing disciplines for managing update packet arrivals at the transmitter node. In the following, we describe each of these queueing disciplines:
\begin{itemize}
    \item {\it LCFS-NP queueing discipline}: Under this queueing discipline, a new arriving update packet enters the service upon its arrival if the server is idle (i.e., the system is empty) and the battery contains at least one energy packet; otherwise, the new arriving update packet is discarded. 
    \item {\it LCFS-PS queueing discipline}: When the server is idle, the management of a new arriving update packet under this queueing discipline is similar to the LCFS-NP one. However, when the server is busy, a new arriving update packet replaces the current packet being served and the old packet in service is discarded.   
    \item {\it LCFS-PW queueing discipline}: Under this queueing discipline, the management of a new arriving update packet when the server is idle is similar to the previous two queuing disciplines. However, when the server is busy, the transmitter node stores the latest arriving update packet in a queue of length one packet at the server. In particular, a new arriving update packet when the server is busy replaces the current one waiting in the queue at the server, and starts its service time (in case it is not replaced by another new update packet that arrives before the delivery of the current packet in service to the destination) once the current packet in service is delivered to the destination. Note that in order to have two status updates in the system (one in service and the other in waiting), there must be at least two energy packets in the battery queue (required for serving the two  update packets). This means that if the server is busy and the battery queue contains one energy packet, a new arriving update packet will be discarded.
\end{itemize}

\begin{figure}[t!]
\centering
\subfloat[]{\includegraphics[width=0.7\columnwidth]{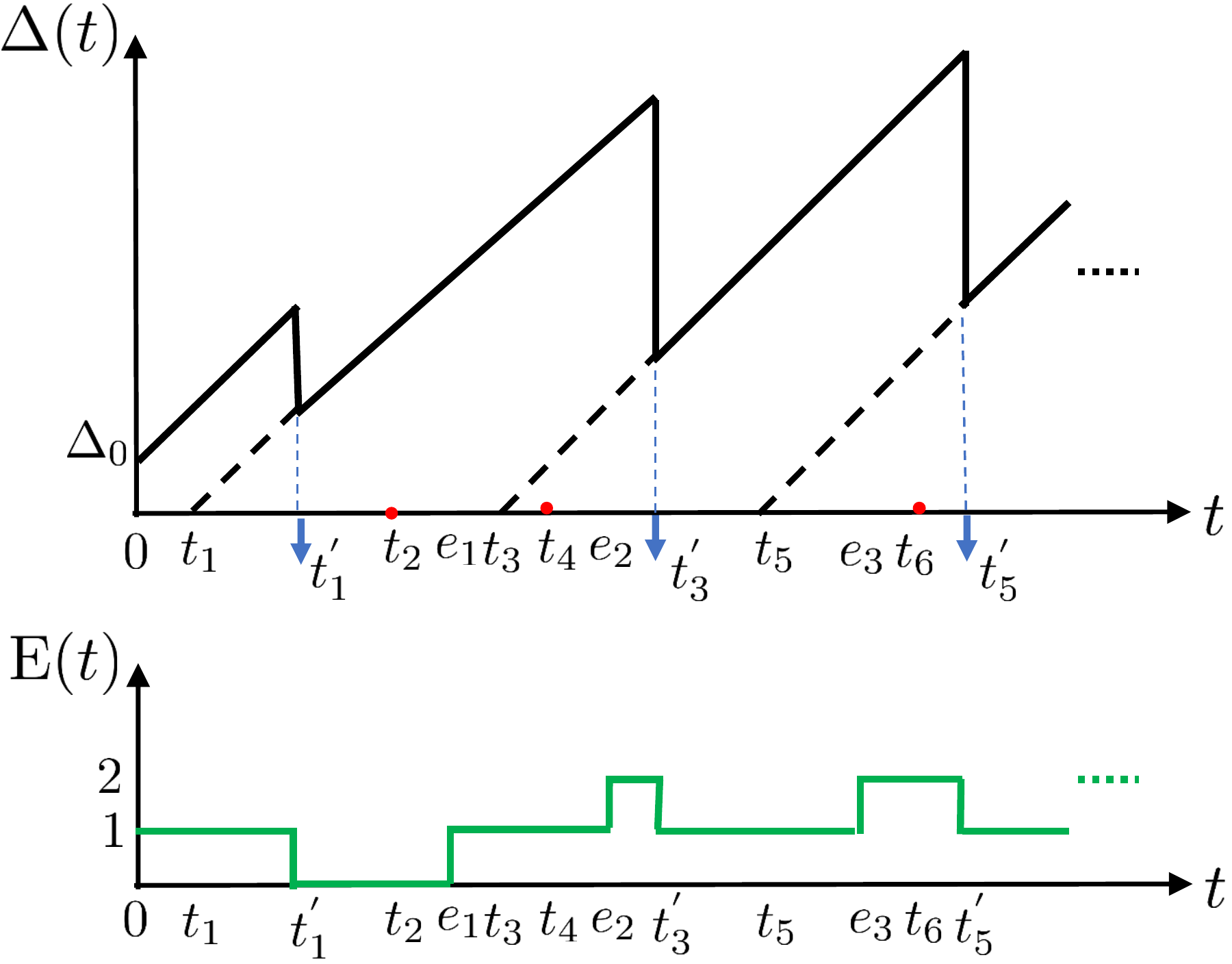}%
\label{f:LCFS_WP_des}} \vfil
\subfloat[]{\includegraphics[width=0.7\columnwidth]{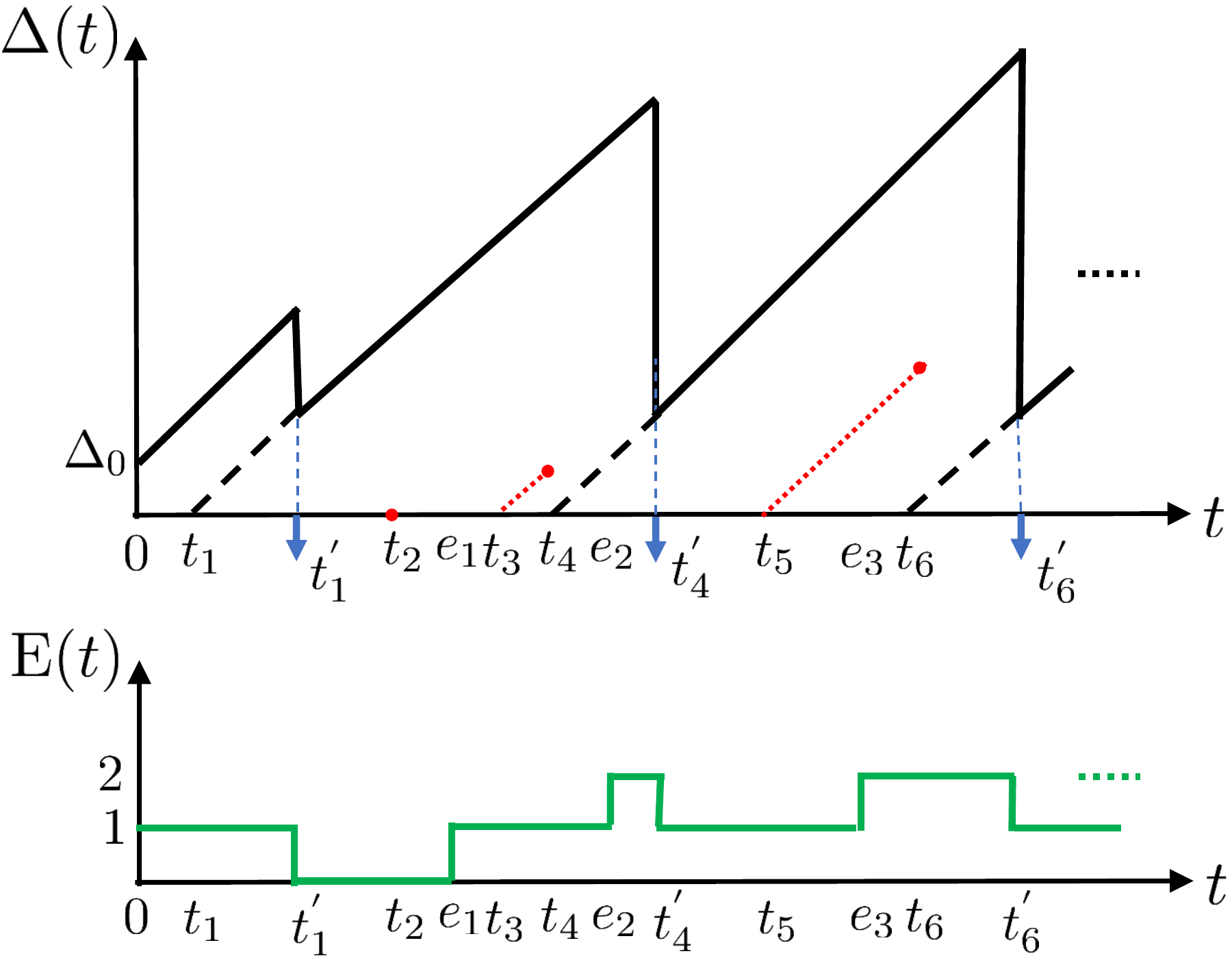}%
\label{f:LCFS_PS_des}} \vfil
\subfloat[]{\includegraphics[width=0.7\columnwidth]{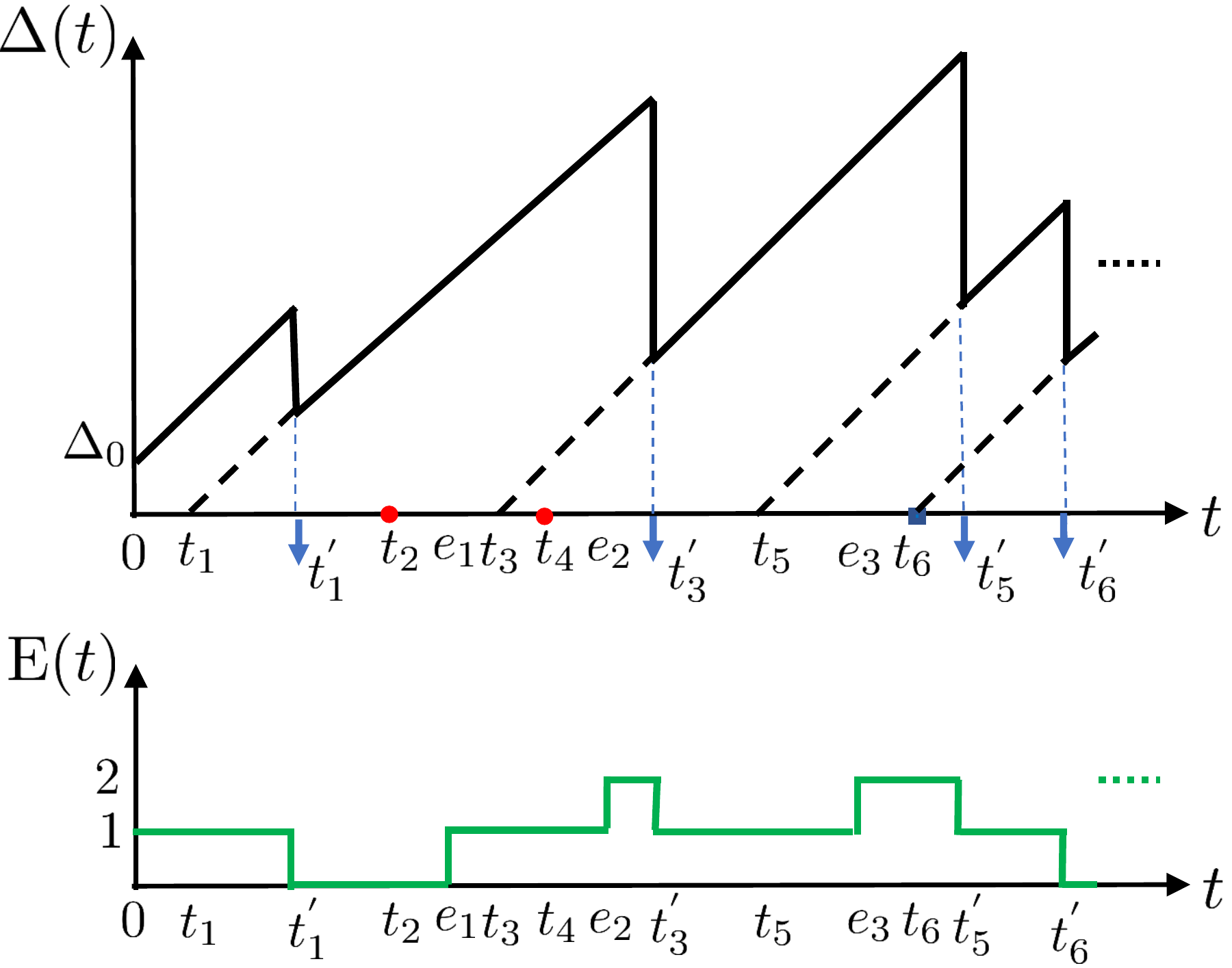}%
\label{f:LCFS_PW_des}}
\caption{An illustration of the evolution of the AoI $\Delta(t)$ and the battery level ${\rm E}(t)$ versus time: (a) the LCFS-NP queueing discipline, (b)  the LCFS-PS queueing discipline, and (c) the LCFS-PW queueing discipline. We denote the arrival time instant of the $i$-th harvested energy packet by $e_i$, and assume that $\Delta(0) = \Delta_0, {\rm E}(0) = 1$ and $B = 2$. Further, we mark a status update that is discarded upon its arrival using a red dot, a status update that gets preempted while being served in the LCFS-PS queueing discipline using  a dotted red line, and a status updates that is stored in the queue upon its arrival (waiting for service) in the LCFS-PW queueing discipline by a blue square.}  \label{f:Des_QDs}
\end{figure}

Fig. \ref{f:Des_QDs} further provides an illustration of the joint evolution of AoI at the destination and the battery level at the transmitter under each of the considered queueing disciplines. As already conveyed, we consider that an energy packet contains the amount of energy required for transmitting one status update packet to the destination node. Therefore, we assume that the length of the energy battery queue reduces by one whenever a status update packet is successfully transmitted to the destination node. Further, with regards to the EH process at the transmitter node, we study two different cases for each of the above queueing disciplines. In the first case, we consider that the transmitter can harvest energy only if its server is idle (i.e., there are no status update packets in the system). This case corresponds to the scenario where the transmitter can harvest energy only from radio frequency (RF) signals, and is equipped with a single RF chain. Therefore, it can either transmit a status update packet or harvest energy at a certain time instant. On the other hand, the transmitter is assumed to be able to harvest energy anytime (i.e., even its server is busy) in the second case. This case corresponds to: i) a scenario of having two separate RF chains for EH and data communications at the transmitter when it is powered solely by harvesting energy from RF signals, or ii) a scenario where the energy is harvested from some other sources, such as solar, if the transmitter has a single RF chain. Clearly, in both scenarios, the transmitter can simultaneously harvest energy and serve status update packets. Our study of the above two cases will also allow us to gain useful insights about the impact of the structure of the EH process at the transmitter on the achievable AoI performance at the destination, as will be clear in Remark \ref{rem:asyminsight}. Before presenting our problem statement and solution approach in the next section, we summarize the different cases that we study in Table \ref{table:sum_cases}. 
\begin{table*}
\centering
{\caption{Different Cases Studied in This Paper.} 
\label{table:sum_cases}
\scalebox{.9}
{ \begin{tabular}{ |c |c|c|}
\hline
    & EH is only allowed when the system is empty  & EH is allowed anytime\\ \hline
LCFS-NP queueing discipline & Subsection \ref{sub:WP_a}&  Subsection \ref{sub:WP_b}\\ \hline 
LCFS-PS queueing discipline & Subsection \ref{sub:PS_a}&  Subsection \ref{sub:PS_b}\\ \hline
LCFS-PW queueing discipline & Subsection \ref{sub:PW_a}& Subsection  \ref{sub:PW_b}\\ \hline
\end{tabular}}} 
\end{table*} 
\section{Problem Statement and Solution Approach}\label{sec:problem_statement}
\subsection{Problem Statement}
Our goal is to analytically characterize the AoI performance at the destination node as a function of: i) the rate of status update packet arrivals $\lambda$, ii) the rate of harvesting energy packets $\eta$, iii) the rate of serving status update packets $\mu$, and iv) the finite capacity of the energy battery queue $B$, at the transmitter node. Unlike most of the analyses of AoI in the existing literature which were focused on deriving the time-average of AoI, our analysis is focused on deriving distributional properties of AoI through the characterization of the MGF. A key benefit of such analysis lies in the fact that it allows one to judge the accuracy/reliability of solely relying on the time-average value of AoI in the design/optimization of real-time status update systems. As will be demonstrated in Section \ref{sec:numerical}, the implementation of real-time status update systems based on just the average value of AoI does not ensure reliability, and it is crucial to incorporate the higher moments of AoI in the design of such systems. This, in turn, highlights the significance of the analytical distributional properties of AoI derived in this paper. 
\subsection{Stochastic Hybrid Systems: A Brief Introduction}\label{Sub:SHSs}
To derive the MGF of AoI for the considered queueing disciplines at the transmitter node (presented in Subsection \ref{sub:disciplines}), we resort to the SHS framework in \cite{hespanha2006modelling}, which was first tailored for the analysis of AoI by \cite{yates2018age} and \cite{yates2020age}. In the following, we provide a very brief\footnote{Interested readers are advised to refer to \cite{yates2018age} and \cite{yates2020age} for a detailed discussion about the use of the SHS approach in the analysis of AoI.} description of the SHS framework, which will be useful in understanding our AoI MGF analysis in the subsequent sections. The SHS technique is used to analyze hybrid queueing systems that can be modeled by a combination of discrete and continuous state parameters. In particular, the SHS technique models the discrete state of the system $q(t) \in \ncalQ = \{1,\cdots,m\}$ by a continuous-time finite-state Markov chain, where $\ncalQ$ is the discrete state space. This continuous-time Markov chain governs the dynamics of the system discrete state that usually describes the occupancy of the system, e.g., $q(t)$ represents the numbers of status update and energy packets in the system for our problem. On the other hand, the evolution of the continuous state of the system is described by a continuous process ${\bf x}(t) = [x_0(t), \cdots,x_n(t)] \in \nbbR^{1 \times \left(n+1\right)}$, e.g., $x(t)$ models the evolution of the age-related processes in our system setting. 

A transition $l \in \ncalL$ from state $q_l$ to state $q'_l$ (in the Markov chain modeling $q(t)$) occurs due to the arrival of a status update/energy packet or the delivery of a status update to the destination (i.e., the departure of a status update from the system), where $\ncalL$ denotes the set of all transitions. Therefore, the transition $l$ takes place with the exponential rate $\lambda^{(l)} \delta_{q_l,q(t)}$ due to the fact the time elapsed between departures and arrivals is exponentially distributed, where the Kronecker delta function $\delta_{q_l,q(t)}$ ensures that the transition $l$
occurs only when the discrete state $q(t)$ is equal to $q_l$.
As a consequence of the occurrence of transition $l$, the discrete state of the system moves from state $q_l$  to state $q'_
l$, and the continuous state $\nbx$ is reset to $\nbx'$ according to a binary
reset map matrix $\nbA_l \in \nbbB^{(n+1)\times(n+1)}$ as $\nbx'=\nbx \nbA_l$. Further, $\overset{\cdot}{\nbx}(t) \triangleq \dfrac{\partial\nbx(t)}{\partial t} = {\bf 1}$ holds
as long as the state $q(t)$ is unchanged, where ${\bf 1}$ is the row vector $[1, \cdots,1] \in \nbbR^{1\times(n+1)}$. Different from ordinary continuous-time Markov chains, an inherent feature of SHSs is the possibility of having self-transitions in the Markov chain modeling the system discrete state. In particular, although a self-transition keeps $q(t)$ unchanged, it causes a change in the continuous process $x(t)$.

Now, we define some useful quantities for the characterization of the MGF of AoI at the destination node using the SHS technique. Denote by $\pi_q(t)$ the probability of being in state $q$ of the continuous-time Markov chain at time $t$. Further, let $\nbv_q(t) = [v_{q0}(t), \cdots,v_{qn}(t)] \in \nbbR^{1\times(n+1)}$ denote the correlation vector between $q(t)$ and $x(t)$, and $\nbv^s
_q(t) = [v^s_{q0}(t), \cdots, v^s_{qn}(t)] \in
\nbbR^{1\times(n+1)}$ denote the correlation vector between $q(t)$ and the exponential function $e^{s\nbx(t)}$, where $s \in \nbbR$. Thus, we can respectively express $\pi_q(t)$, $\nbv_q(t)$ and $\nbv^s_q(t)$ as
\begin{align}
\pi_q(t) = {\rm Pr}\left(q(t) = q\right) = \nbbE[\delta_{q,q(t)}],\; \forall q \in \ncalQ,
\end{align}
\begin{align}
\nbv_q(t) = [v_{q0}(t),\cdots,v_{qn}(t)] = \nbbE[\nbx(t)\delta_{q,q(t)}],\; \forall q \in \ncalQ,
\end{align}
\begin{align}
\nbv^s_q(t) = [v^s_{q0}(t),\cdots,v^s_{qn}(t)] = \nbbE[e^{s\nbx(t)}\delta_{q,q(t)}],\; \forall q \in \ncalQ.
\end{align}

According to the ergodicity assumption of the continuous-time Markov chain modeling $q(t)$ in the AoI analysis \cite{yates2018age,yates2020age}, the state probability vector $\pi(t) =
[\pi_0(t),\cdots,\pi_m(t)]$ converges uniquely to the stationary vector $\bar{\pi} = [\bar{\pi}_0,\cdots,\bar{\pi}_m]$ satisfying
\begin{align}\label{gen_steady}
\bar{\pi}_q\sum_{l\in\ncalL_q}{\lambda^{(l)}} = \sum_{l\in\ncalL'_q}{\lambda^{(l)}\bar{\pi}_{q_l}},\; q \in \ncalQ,\;\; \sum_{q\in\ncalQ}{\bar{\pi}_q} = 1,
\end{align}
where $\ncalL'_q = \{l\in\ncalL: q'_l = q\}$ and $\ncalL_q = \{l\in\ncalL: q_l = q\}$ denote the sets of incoming and outgoing transitions for state $q, \forall q \in \ncalQ$. 

Using the above notations, it has been shown in \cite[Theorem~1]{yates2020age} that under the
ergodicity assumption of the Markov chain modeling $q(t)$, if we can find a non-negative limit $\bar{\nbv}_q = [\bar{v}_{q0},\cdots,\bar{v}_{qn}],\;\forall q \in \ncalQ$, for the correlation vector $\nbv_q(t)$ satisfying
\begin{align}\label{gen_vavg}
\bar{\nbv}_q \sum_{l \in \ncalL_q}{\lambda^{(l)}} = \bar{\pi}_q {\bf 1} + \sum_{l\in\ncalL'_q}{\lambda^{(l)}} \bar{\nbv}_{q_l}\nbA_l,\; q \in \ncalQ,
\end{align} 
then:
\begin{itemize}
    \item The expectation of $x(t)$, $\nbbE[x(t)]$, converges to the following stationary vector:
    \begin{align}\label{gen_conv_Mom1}
\nbbE[x] = \sum_{q \in \ncalQ}{\bar{\nbv}_q}.
\end{align}
    \item There exists $s_0 > 0$ such that for all $s < s_0$, $\nbv_q^s(t)$ converges to $\bar{\nbv}^s_q$ that satisfies
\begin{align}\label{gen_vMGF}
\bar{\nbv}_q^s \sum_{l \in \ncalL_q}{\lambda^{(l)}} = s\bar{\nbv}_q^s + \sum_{l\in\ncalL'_q}{\lambda^{(l)}} [\bar{\nbv}_{q_l}^s\nbA_l + \bar{\pi}_{q_l} {\bf 1}\hat{\nbA}_l],\; q \in \ncalQ,
\end{align}
where $\hat{\nbA}_l \in \nbbB^{(n+1)\times(n+1)}$ is a binary matrix whose elements are constructed as: $\hat{\nbA}_l(k,j) = 1$ if $k=j$ and the $j$-th column of $\nbA_l$ is a zero vector; otherwise, $\hat{\nbA}_l(k,j) = 0$. Further, the MGF of the state $\nbx(t)$, which can be obtained as $\nbbE[e^{s\nbx(t)}]$, converges to the following stationary vector:
\begin{align}\label{gen_conv_MGF}
\nbbE[e^{s\nbx}] = \sum_{q \in \ncalQ}{\bar{\nbv}_q^s}.
\end{align}
\end{itemize}

From (\ref{gen_conv_Mom1}) and (\ref{gen_conv_MGF}), when the first element of the continuous state $\nbx(t)$
represents the AoI at the destination node, the expectation and the MGF of AoI at the destination node respectively converges to:
\begin{align}\label{gen_Mom1}
\overset{(1)}{\Delta} = \sum_{q\in\ncalQ}{\bar{v}_{q0}},
\end{align}
\begin{align}\label{gen_MGF}
M(s) = \sum_{q\in\ncalQ}{\bar{v}^{s}_{q0}}.
\end{align}

\section{The MGF Analysis of LCFS-NP Queueing Discipline}\label{sec:WP}
In this section, we present the analysis of the MGF of AoI for the LCFS-NP queueing discipline. It is clear from \cite[Theorem~1]{yates2020age} that in order to use (\ref{gen_vMGF}) to derive the MGF of AoI at the destination, one needs to find a non-negative limit $\bar{\nbv}_q, \forall q \in \ncalQ$, satisfying (\ref{gen_vavg}). It can be shown that the sets of equations in (\ref{gen_vavg}) and (\ref{gen_vMGF}) can be solved using the same approach. Therefore, for the sake of brevity, we next focus on evaluating $\bar{v}^s_{q0}, \forall q \in \ncalQ$, satisfying the set of equations in (\ref{gen_vMGF}), using which the MGF of AoI at the destination can be calculated as in (\ref{gen_MGF}). Even though we do not show this explicitly, it is easy to check (along the same lines as the solution of (\ref{gen_vMGF}) given in the paper) that the solution of (\ref{gen_vavg}), $\bar{\nbv}_q, \forall q \in \ncalQ$, is non-negative. 

Using the notations of the SHS approach (presented in Subsection \ref{Sub:SHSs}), the discrete state space $\ncalQ$ in this queueing discipline is given by $\ncalQ = \{1,2,\cdots,2B+1\}$. Each state in $\ncalQ$ represents a potential combination of the numbers of the update packets in the system and the energy packets in the battery queue at the server. For instance, a state $j=(e_j,u_j)$ indicates that the system has $u_j$ status update packets and the energy battery queue at the server contains $e_j$ energy packets. Note that since the transmitter node does not store arriving update packets when the server is busy in this queueing discipline, the system can have at most one status update packet at any time instant, and thus we have $u_j \in \{0,1\}$. In particular, $u_j=0$ indicates that the system is empty and hence the server is idle, and $u_j=1$ indicates that the server is serving the existing update packet in the system. Further, since the battery queue at the server has a capacity of $B$ packets, we have $e_j \in \{0,1,\cdots,B\}$. On the other hand, the continuous process $\nbx(t)$ in this queueing discipline is given by $\nbx(t)=[x_0(t),x_1(t)]$, where $x_0(t)$ represents the value of AoI at the destination node at time instant $t$, and $x_1(t)$ indicates the value that the AoI at the destination node will become if the existing update packet in the system completes its service at time instant $t$ (i.e., the packet is delivered to the destination at $t$). Recall from Subsection \ref{Sub:SHSs} that as long as there is no change (due to the arrival/departure of an update/energy packet) in the discrete state $q(t)$, we have $\dfrac{\partial \nbx(t)}{\partial t} = 1$, i.e., the elements of the age vector $\nbx(t)$ increase linearly with time. In the following, we derive the MGF of AoI at the destination node for the following two scenarios: i) the transmitter node can harvest energy only if the system is empty, and ii) the transmitter node is able to harvest energy anytime.
\begin{figure}[t!]
\centering
\subfloat[]{\includegraphics[width=0.8\columnwidth]{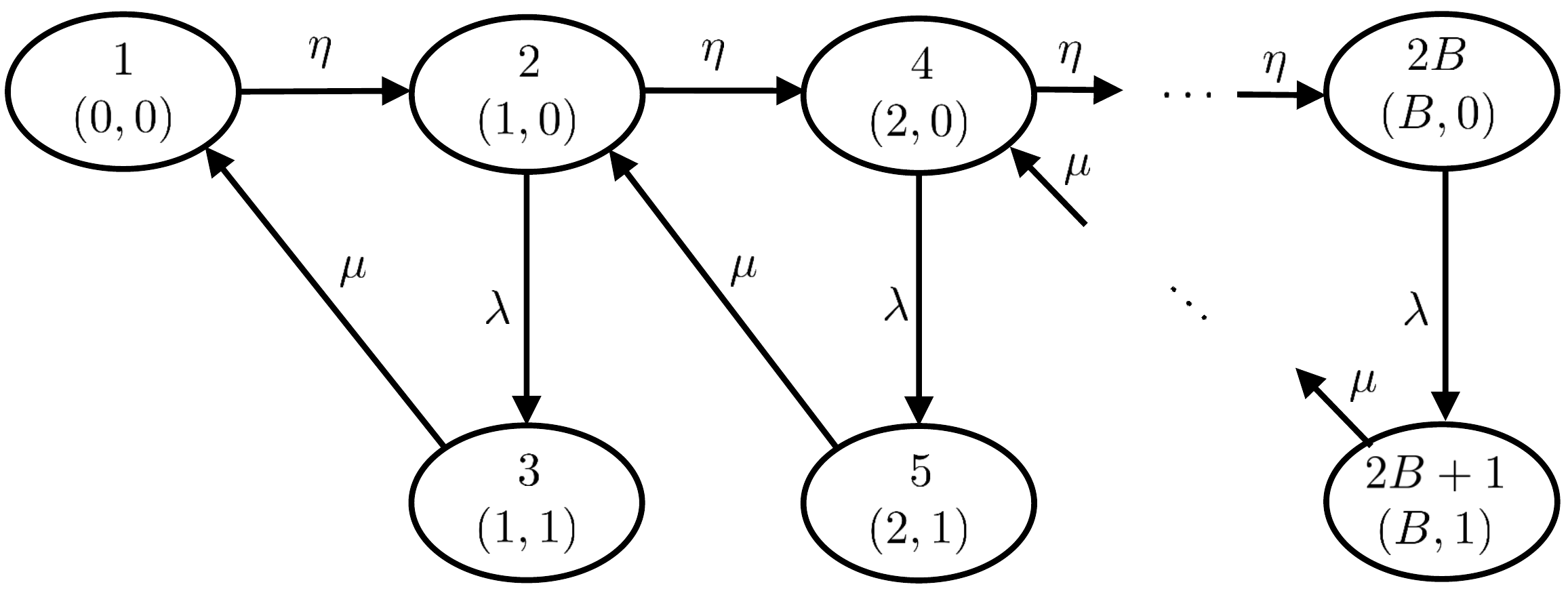}%
\label{f:WP_MC}} \vfil 
\subfloat[]{\includegraphics[width=0.8\columnwidth]{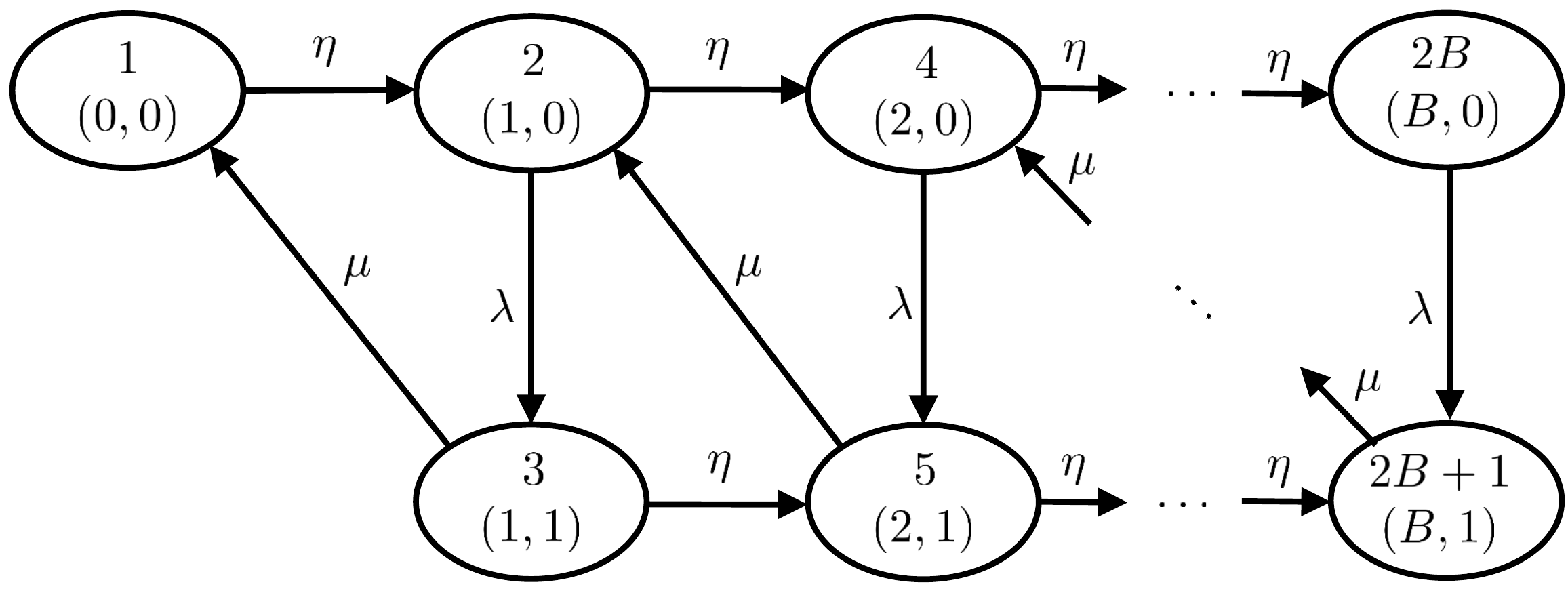}%
\label{f:WP_harvall_MC}}
\caption{Markov chains modeling the discrete state in the LCFS-NP: (a) the transmitter node can harvest only if the system is empty, and (b) the transmitter node can simultaneously harvest energy and serve status updates.}\label{f:WP}
\end{figure}  
\subsection{EH is Only Allowed When the System is Empty} \label{sub:WP_a}
When the transmitter node is only allowed to harvest energy if the system is empty, the continuous-time Markov chain modeling the discrete state of the system $q(t)$ is depicted in Fig. \ref{f:WP_MC}. We denote the set of states in the $i-$th row of the Markov chain by ${\rm r}_i$. Further, Table \ref{table:WP} presents the set of different transitions $\ncalL$ and their impact on the values of both $q(t)$ and $\nbx(t)$. Before proceeding to the characterization of the MGF of AoI at the destination in this case using (\ref{gen_steady})-(\ref{gen_MGF}), we first describe the set of transitions as follows
\begin{itemize}
    \item $l = 3k -2$: This subset of transitions takes place between the states of the Markov chain in ${\rm r}_1$, corresponding to the time when the system is empty. In particular, a transition from this set of transitions occurs when a new energy packet is harvested by the transmitter. Clearly, since harvesting a new energy packet does not impact the value of AoI at the destination node, we observe that the first element in the updated value of the age vector $\nbx \nbA_l$ (as a consequence of this transition) is $x_0$, i.e., this transition does not induce any change in the value of AoI at the destination. Further, since the server is idle in the states of ${\rm r}_1$, the second component of $\nbx(t)$ (quantifying the age of the packet in service) becomes irrelevant for such set of states. Note that whenever a component of $\nbx(t)$ is/becomes irrelevant after the occurrence of some transition $l$, its value in the updated age vector $\nbx \nbA_l$ can be set arbitrarily. Following the convention used in \cite{yates2018age}, we set the value corresponding to such irrelevant components in the updated age value to 0, and thus we observe that the second component of $\nbx \nbA_{3k-2}$ is 0.
    \item $l = 3k -1$: A transition from this subset of transitions occurs when there is a new arriving update packet at the transmitter node. Since the age of this new arriving update packet at the transmitter is 0 and it does not have any impact on the AoI value at the destination, we note that the updated age vector $\nbx \nbA_l$ is set to be $[x_0,0]$.
    \item $l = 3k$: This subset of transitions occurs when a status update packet is delivered to the destination node. Clearly, as a consequence of this transition, the AoI value at the destination is reset to the age of the new received update packet, and thus the first component of $\nbx \nbA_l$ is $x_1$. In addition, since the system becomes empty after the occurrence of this transition, the second component of the age vector $\nbx(t)$ becomes irrelevant, and thus its corresponding value in the updated age vector $\nbx \nbA_l$ is 0.
\end{itemize}
\begin{table*}
\centering
{\caption{Transitions of the LCFS-NP queueing discipline in Fig. \ref{f:WP_MC} $(2 \leq k \leq B)$.} 
\label{table:WP}
\scalebox{.9}
{ \begin{tabular}{ |c |c|c|c|c|c|c|c|}
\hline
 $l$   & $q_l\rightarrow q'_l$  & $\lambda^{(l)}$ & $\nbx \nbA_l$ & $\nbA_l$ & $\hat{\nbA}_l$ & $\bar{\nbv}^s_{q_l} \nbA_l$ & $\bar{\pi}_{q_l} {\bf 1} \hat{\nbA}_l$\\ \hline
1& 1 $\rightarrow$ 2& $\eta$&$[x_0,0]$&
$\begin{bmatrix}
1 & 0\\ 0 & 0\\
\end{bmatrix}$
& $\begin{bmatrix}
0 & 0\\ 0 & 1\\
\end{bmatrix}$ &$[\bar{v}^s_{10},0]$ &$[0,\bar{\pi}_1]$ \\ \hline
2& 2 $\rightarrow$ 3&$\lambda$&$[x_0,0]$& $\begin{bmatrix}
1 & 0\\ 0 & 0\\
\end{bmatrix}$& $\begin{bmatrix}
0 & 0\\ 0 & 1\\
\end{bmatrix}$&$[\bar{v}^s_{20},0]$ &$[0,\bar{\pi}_2]$ \\ \hline
3& 3 $\rightarrow$ 1&$\mu$&$[x_1,0]$ & $\begin{bmatrix}
0 & 0\\ 1 & 0\\
\end{bmatrix}$&$\begin{bmatrix}
0 & 0\\ 0 & 1\\
\end{bmatrix}$ &$[\bar{v}^s_{31},0]$ &$[0,\bar{\pi}_3]$ \\ \hline
$3k-2$& $2k-2 \rightarrow 2k$& $\eta$&$[x_0,0]$& $\begin{bmatrix}
1 & 0\\ 0 & 0\\
\end{bmatrix}$& $\begin{bmatrix}
0 & 0\\ 0 & 1\\
\end{bmatrix}$&$[\bar{v}^s_{2k-2,0},0]$ &$[0,\bar{\pi}_{2k-2}]$ \\ \hline
$3k-1$& $2k \rightarrow 2k+1$&$\lambda$& $[x_0,0]$&$\begin{bmatrix}
1 & 0\\ 0 & 0\\
\end{bmatrix}$ &$\begin{bmatrix}
0 & 0\\ 0 & 1\\
\end{bmatrix}$ &$[\bar{v}^s_{2k,0},0]$ &$[0,\bar{\pi}_{2k}]$ \\ \hline
$3k$& $2k+1 \rightarrow 2k-2$&$\mu$&$[x_1,0]$ &$\begin{bmatrix}
0 & 0\\ 1 & 0\\
\end{bmatrix}$ &$\begin{bmatrix}
0 & 0\\ 0 & 1\\
\end{bmatrix}$ & $[\bar{v}^s_{2k+1,1},0]$&$[0,\bar{\pi}_{2k+1}]$ \\ \hline
\end{tabular}}} 
\end{table*} 

Now, in order to use (\ref{gen_vMGF}) to derive the MGF of AoI at the destination, we observe that the steady state  probabilities $\bar{\pi}_q, q \in \ncalQ$, and the two vectors $\bar{\nbv}^s_{q_l} \nbA_l$ and $\bar{\pi}_{q_l}{\bf 1}\hat{\nbA}_l$ (associated with each transition $l$ in $\ncalL$) need to be computed. The calculations of  $\bar{\nbv}^s_{q_l} \nbA_l$ and $\bar{\pi}_{q_l}{\bf 1}\hat{\nbA}_l$, $l \in \ncalL$, are listed in Table \ref{table:WP}, and the steady state probabilities $\bar{\pi}_q, q \in \ncalQ$ are given by the following proposition.
\begin{prop}\label{prop1}
The steady state probabilities $\{\bar{\pi}_i\}$ can be expressed as
\begin{align}\label{prop1_1}
    \bar{\pi}_{2i}= \left(\frac{\beta}{\rho}\right)^{i} \bar{\pi}_1,
\end{align}
\begin{align}\label{prop1_2}
    \bar{\pi}_{2i+1}= \rho \left(\frac{\beta}{\rho}\right)^{i} \bar{\pi}_1,
\end{align}
where $1 \leq i \leq B$ and $\bar{\pi}_1$ is given by
\begin{align}\label{prop1_3}
    \bar{\pi}_{1}= \begin{cases}
    \dfrac{1}{1 + B (1 + \rho)}, & \;{\rm if}\; \rho = \beta,\\
    \dfrac{\rho^{B}\left(\beta - \rho\right)}{\rho^{B}\left(\beta - \rho\right) + \beta \left(1 + \rho\right) \left(\beta^{B} - \rho^{B}\right)}, & \;{\rm otherwise}.
    \end{cases}
\end{align}
\end{prop}
\begin{IEEEproof}
See Appendix \ref{app:prop1}.
\end{IEEEproof}

Having the steady state probabilities $\{\bar{\pi}_q\}$ in Proposition \ref{prop1} and the set of transitions $\ncalL$ in Table \ref{table:WP}, we are now ready to use (\ref{gen_vMGF}) for deriving the MGF of AoI at the destination in the following theorem.
\begin{theorem}\label{MGF_WP}
When the transmitter node can only harvest energy if the system is empty, the MGF of AoI for the LCFS-NP queueing discipline is given by
\begin{align}\label{theorem_MGF_WP_1}
\overset{\rm NP}{M}\left(\bar{s}\right) = \dfrac{\rho \bar{\pi}_1 \Big[\bar{s}^{2} \theta - \bar{s} \theta \left(1 + \rho + \beta\right) + \beta \left(1+ \theta + \theta \rho\right)\Big]}{\left(1 - \bar{s}\right)^{2} \left(\rho - \bar{s}\right) \left(\beta - \bar{s}\right)},
\end{align}
where $\bar{s} = \frac{s}{\mu}$ and $\theta$ can be expressed as
\begin{align}
\theta = 
\begin{cases}
B,\;\; &{\rm if}\; \rho = \beta,\\
\dfrac{\beta\left(\beta^{B} - \rho^{B}\right)}{\rho^{B}\left(\beta - \rho\right)},\;\; &{\rm otherwise}.
\end{cases}
\end{align}
\end{theorem}
\begin{IEEEproof}
See Appendix \ref{app:MGF_WP}.
\end{IEEEproof}
\begin{cor}\label{cor:WP}
Using $\overset{\rm NP}{M}(\bar{s})$ derived in Theorem \ref{MGF_WP}, the first moment of the LCFS-NP queueing discipline (when the transmitter node can only harvest energy if the system is empty) can be expressed as
\begin{align}\label{mom1_WP}
\overset{(1)}{\Delta}_{\rm NP} = 
\begin{cases}
\dfrac{2B\rho^2 + 2\left(1 + B\right)\rho + B + 2}{\mu\big[B\rho^2 + (1 + B)\rho\big]},\;\;  {\rm if}\;\; \rho = \beta,\\
\dfrac{\beta^{B+2}\left(2\rho^2+2\rho+1\right) - \rho^{B+2} \left(2\beta^2 + 2\beta + 1\right)}{\mu\big[\beta^{B+2}\left(\rho^2+\rho\right) - \rho^{B+2} \left(\beta^2+\beta\right)\big]},
\end{cases}
\end{align}
where the second case in (\ref{mom1_WP}) holds when $\rho \neq \beta$. Further, the second moment of AoI $\overset{(2)}{\Delta}_{\rm NP}$ is given by (\ref{mom2_WP}) [at the top of the next page].
\end{cor}
\begin{figure*}
\begin{align}\label{mom2_WP}
\overset{(2)}{\Delta}_{\rm NP} = 
\begin{cases}
\dfrac{2\big[3B\rho^3+\left(3B+3\right)\rho^2+\left(2B+4\right)\rho+B+3\big]}{\mu^2\rho^2\left(1+B+B\rho\right)},\;\; & {\rm if}\; \rho = \beta,\\
\dfrac{2\big[\beta^{B+3}\left(3\rho^3+3\rho^2+2\rho+1\right) - \rho^{B+3}\left(3\beta^3+3\beta^2+2\beta+1\right)\big]}{\mu^2\big[\beta^{B+3}\rho^2\left(1+\rho\right) - \rho^{B+3} \beta^2 \left(\beta + 1\right)\big]},\;\; & {\rm otherwise}.
\end{cases}
\end{align}
\end{figure*}
\begin{IEEEproof}
The expressions in (\ref{mom1_WP}) and (\ref{mom2_WP}) follow from the fact that the expression of $\overset{\rm NP}{M}(\bar{s})$ (derived in Theorem \ref{MGF_WP}) can be used to compute the $k$-th moment of AoI (denoted by $\overset{(k)}{\Delta}_{\rm NP}$) as follows
\begin{align}\label{MGF_derav}
\overset{(k)}{\Delta}_{\rm NP} = \frac{1}{\mu^{k}}\times\dfrac{{\rm d}^k\big[\overset{\rm NP}{M}(\bar{s})\big]}{{\rm d}\bar{s}^k} \Big|_{\bar{s}=0},
\end{align}
where $\frac{{\rm d}^k}{{\rm d}\bar{s}^k}$ denotes the $k$-th derivative with respect to $\bar{s}$.
\end{IEEEproof}
\begin{remark}\label{rem:1}
Note that the expression of $\overset{(1)}{\Delta}_{\rm NP}$ in (\ref{mom1_WP}) is identical to the average AoI expression derived in \cite[Theorem~1]{farazi2018average}. Further, when $\beta \rightarrow \infty$, $\overset{(1)}{\Delta}_{\rm NP}$ and $\overset{(2)}{\Delta}_{\rm NP}$ in the case of $\rho \neq \beta$ reduce to
\begin{align}
\nonumber &\underset{\beta \rightarrow \infty}{\rm lim} \overset{(1)}{\Delta}_{\rm NP} = \dfrac{2\rho^2+2\rho+1}{\mu\left(\rho^2+\rho\right)},\\
&\underset{\beta \rightarrow \infty}{\rm lim} \overset{(2)}{\Delta}_{\rm NP} = \dfrac{2\left(3\rho^3+3\rho^2+2\rho+1\right)}{\mu^2\rho^2\left(1+\rho\right)}.
\end{align}

Clearly, $\overset{(1)}{\Delta}_{\rm NP}$ reduced to the average AoI expression derived in \cite{costa2016age} for the M/M/1/1 case where the transmitter node does not have energy limitations, and $\overset{(2)}{\Delta}_{\rm NP}$ reduced to the second moment of AoI for the same case.
\end{remark}
\begin{remark}\label{rem:WP_symmetry}
As was the case for $\overset{(1)}{\Delta}_{\rm NP}$ (given by (\ref{mom1_WP})) in \cite{farazi2018average}, we observe from (\ref{mom2_WP}) that $\overset{(2)}{\Delta}_{\rm NP}$ is invariant to exchanging $\rho$ and $\beta$ when $\rho \neq \beta$. This is a counterintuitive insight since the energy and update packets are managed by the transmitter node in a totally different manner, i.e., the capacity of the battery queue at the server is $B$ energy packets whereas under the LCFS-NP queueing discipline, there can be at most one update packet in the system (in service) at any time instant.
\end{remark}
\begin{table*}[t!]
\centering
{\caption{Transitions of the LCFS-NP queueing discipline in Fig. \ref{f:WP_harvall_MC} $(2 \leq k \leq B)$.} 
\label{table:WP_harvall}
\scalebox{.9}
{ \begin{tabular}{ |c |c|c|c|c|c|c|c|}
\hline
 $l$   & $q_l\rightarrow q'_l$  & $\lambda^{(l)}$ & $\nbx \nbA_l$ & $\nbA_l$ & $\hat{\nbA}_l$ & $\bar{\nbv}^s_{q_l} \nbA_l$ & $\bar{\pi}_{q_l} {\bf 1} \hat{\nbA}_l$\\ \hline
 $3B+k-1$&$2k-1 \rightarrow 2k+1$ &$\eta$ & $[x_0,x_1]$&$\begin{bmatrix}
1 & 0\\ 0 & 1\\
\end{bmatrix}$ & $\begin{bmatrix}
0 & 0\\ 0 & 0\\
\end{bmatrix}$&$[\bar{v}^s_{2k-1,0},\bar{v}^s_{2k-1,1}]$ &$[0,0]$ \\ \hline
\end{tabular}}} 
\end{table*} 
\subsection{EH is Allowed Anytime}\label{sub:WP_b}
The discrete state of the system when the transmitter node can harvest energy anytime is modeled by the Markov chain in Fig. \ref{f:WP_harvall_MC}. Clearly, the set of transitions in this case contains the ones presented by Table \ref{table:WP} along with a new subset of transitions given by Table \ref{table:WP_harvall}. In particular, the subset of transitions in Table \ref{table:WP_harvall} (labeled by $l = 3B+k-1$) takes place between the states in ${\rm r}_2$, corresponding to the time when the server is busy. A transition from this subset of transitions occurs when there is a new arriving energy packet at the transmitter while its server is serving an update packet. As already conveyed, the value of AoI at the destination is not influenced by the transitions associated with the energy packet arrivals, and thus the age vector $\nbx(t)$ remains the same after any of these transitions, as indicated by the value of $\nbx \nbA_l$ in Table \ref{table:WP_harvall}. Further, when the transmitter node can simultaneously harvest energy and serve status updates, note that it is not tractable to derive closed-form expressions of the steady state probabilities $\{\pi_i\}$ (as in Proposition \ref{prop1}) for any of the queueing disciplines considered in this paper. That said, in this case of the EH process at the transmitter node, we first derive an expression of the MGF of AoI at the destination for each queueing discipline as a function of an arbitrary value of the battery capacity $B$. Afterwards, we provide the first and second moments of AoI expressions for $B = 1$ and $B = 2$. We also utilize the SHS technique to numerically obtain the first and second moments for larger values of $B$ in Section \ref{sec:numerical}. In the following theorem, we derive the MGF of AoI at the destination node.
\begin{theorem}\label{MGF_WP_harvall}
When the transmitter node can harvest energy anytime, the MGF of AoI for the LCFS-NP queueing discipline is given by (\ref{theorem_MGF_WP_harvall_1}) [at the top of the next page], where $\underset{k \in \;{\rm r}_1 / \{1\}}{\sum}{\bar{\pi}_k}= \gamma \bar{\pi}_1$. 
\end{theorem}
\begin{figure*}[]
\begin{align}\label{theorem_MGF_WP_harvall_1}
\overset{\rm NP}{M}\left(\bar{s}\right) = \dfrac{\rho \bar{\pi}_1 \Big[-\bar{s}^{3} \gamma + \bar{s}^{2} \gamma \left(2\beta + \rho + 2\right) - \bar{s} \gamma \left(\beta^2+\beta\left(2\rho+3\right)+1+\rho\right) + \left(\beta^2+\beta\right)\left(1+ \gamma + \gamma \rho\right)\Big]}{\left(1 - \bar{s}\right)^{2} \left(\rho - \bar{s}\right) \left(\beta - \bar{s}\right) \left(1 + \beta - \bar{s}\right)},
\end{align}
\end{figure*}
\begin{IEEEproof}
See Appendix \ref{app:MGF_WP_harvall}.
\end{IEEEproof}
\begin{cor}\label{cor:WP_harvall}
Given that $B=1$ and the transmitter can harvest energy anytime, the first and second moments of AoI for the LCFS-NP queueing discipline can be respectively expressed as 
\begin{align}\label{mom1_WP_harvall_B1}
\overset{(1)}{\Delta}_{\rm NP} = \dfrac{\beta^2\left(2\rho^2+2\rho+1\right) + \beta \rho\left(1+2\rho\right)+\rho^2}{\mu\rho\beta\Big[\beta\left(1+\rho\right)+\rho\Big]},
\end{align}
\begin{align}\label{mom2_WP_harvall_B1}
\overset{(2)}{\Delta}_{\rm NP} = \dfrac{\sum_{n=0}^3{\beta^n \alpha_n}}{\mu^2\rho^2\beta^2\Big[\beta\left(1+\rho\right)+\rho\Big]},
\end{align}
\begin{align*}
&\alpha_3 = 6\rho^3+6\rho^2+4\rho+2, \;\alpha_2 = 6\rho^3+4\rho^2+2\rho, \\&\alpha_1 = 4\rho^3+2\rho^2, \;\alpha_0 = 2\rho^3.
\end{align*}

Further, the first and second moments of AoI for $B=2$ can be respectively expressed as
\begin{align}\label{mom1_WP_harvall_B2}
\overset{(1)}{\Delta}_{\rm NP} = \dfrac{\bar{\pi}_2 \sum_{n=0}^{5}{\beta^n\bar{\alpha}_n }}{\mu\rho^2\beta^2\left(1+\beta\right)^2},
\end{align}
\begin{align*}
&\bar{\alpha}_5 = 2\rho^2+2\rho+1, \;\bar{\alpha}_4 = 2\rho^3+6\rho^2+5\rho+2, \; \bar{\alpha}_3 = 4\rho^3+\\ &6\rho^2+4\rho+1,\;\bar{\alpha}_2= 3\rho^3+3\rho^2+\rho, \bar{\alpha}_1 = 3\rho^3+\rho^2, \bar{\alpha}_0 = \rho^3,
\end{align*}
\begin{align}\label{mom2_WP_harvall_B2}
\overset{(2)}{\Delta}_{\rm NP} = \dfrac{2\bar{\pi}_2 \sum_{n=0}^7{\beta^n\alpha'_n }}{\mu^2\rho^3\beta^3\left(1+\beta\right)^3},
\end{align}
\begin{align*}
&\alpha'_7 = 3\rho^3+3\rho^2+2\rho+1, \;\alpha'_6 = 3\rho^4+12\rho^3+11\rho^2+7\rho+3, \\& \alpha'_5 = 9\rho^4+18\rho^3+15\rho^2+9\rho+3,\; \alpha'_4= 10\rho^4+13\rho^3+10\rho^2\\&+5\rho+1,\ \alpha'_3 = 7\rho^4+7\rho^3+4\rho^2+\rho, \;\alpha'_2 = 7\rho^4+4\rho^3+\rho^2,\;\\
&\alpha'_1=4\rho^4+\rho^3,\; \alpha'_0 = \rho^4,
\end{align*}
where $\bar{\pi}_2$ is given by (\ref{pi_2_B2}) [at the top of the next page].
\end{cor}
\begin{figure*}
\begin{align}\label{pi_2_B2}
\bar{\pi}_2= \dfrac{\beta^3\rho+\beta^2\rho\left(1+\rho\right)+\beta\rho^2}{\beta^4\left(1+\rho\right)+\beta^3\left(2\rho^2+3\rho+1\right)+\beta^2\left(\rho^3+3\rho^2+2\rho\right)+\beta\left(\rho^3+2\rho^2\right)+\rho^3}.
\end{align}
\hrulefill
\end{figure*}
\begin{IEEEproof}
Using (\ref{MGF_derav}), the result follows from the derivatives of $\overset{\rm NP}{M}(\bar{s})$ in Theorem \ref{MGF_WP_harvall}. Note that $\gamma = \frac{\beta}{\rho}$ for $B = 1$, and $\gamma = \frac{\beta}{\rho^2}\big[\rho + \beta \left(1 + \rho + \beta\right)\big]$ for $B = 2$.
\end{IEEEproof}
\begin{remark}\label{rem:3a}
Similar to Remark \ref{rem:1}, note that as $\beta \rightarrow \infty$, $\overset{(1)}{\Delta}_{\rm NP}$ and $\overset{(2)}{\Delta}_{\rm NP}$ expressions derived for $B = 1$ and $B = 2$ (in (\ref{mom1_WP_harvall_B1})- (\ref{mom2_WP_harvall_B2})) reduce to the first and second moments of AoI for the M/M/1/1 case in \cite{costa2016age}.
\end{remark}
\begin{remark}\label{rem:BigO}
By inspecting (\ref{mom1_WP_harvall_B1}), we note that the polynomials in the numerator and denominator have the same degree with respect to $\rho$ or $\beta$. As a result, the expression in (\ref{mom1_WP_harvall_B1}) is essentially $\Theta(1)$ in terms of $\rho$ or $\beta$, meaning that the growth rate of the first moment of AoI approaches 0 as $\rho$ or $\beta$ approaches $\infty$. This is the case for all the first/second moment of AoI expressions derived in this paper under each queueing discipline.
\end{remark}
\section{The MGF Analysis of LCFS-PS Queueing Discipline}\label{sec:PS}
This section is dedicated to the analysis of the MGF of AoI for the LCFS-PS queueing discipline. The discrete state space $\ncalQ$ and the continuous process $\nbx(t)$ in this queueing discipline are similar to the ones associated with the LCFS-NP queueing discipline (presented in Section \ref{sec:WP}). We start by deriving the MGF of AoI at the destination node when the transmitter node can harvest energy only if the system is empty, and then extend the analysis to characterize the MGF of AoI in the scenario where the transmitter node can simultaneously harvest energy and serve status update packets.
\subsection{EH is Only Allowed When the System is Empty}\label{sub:PS_a}
\begin{figure}[t!]
\centering
\subfloat[]{\includegraphics[width=0.8\columnwidth]{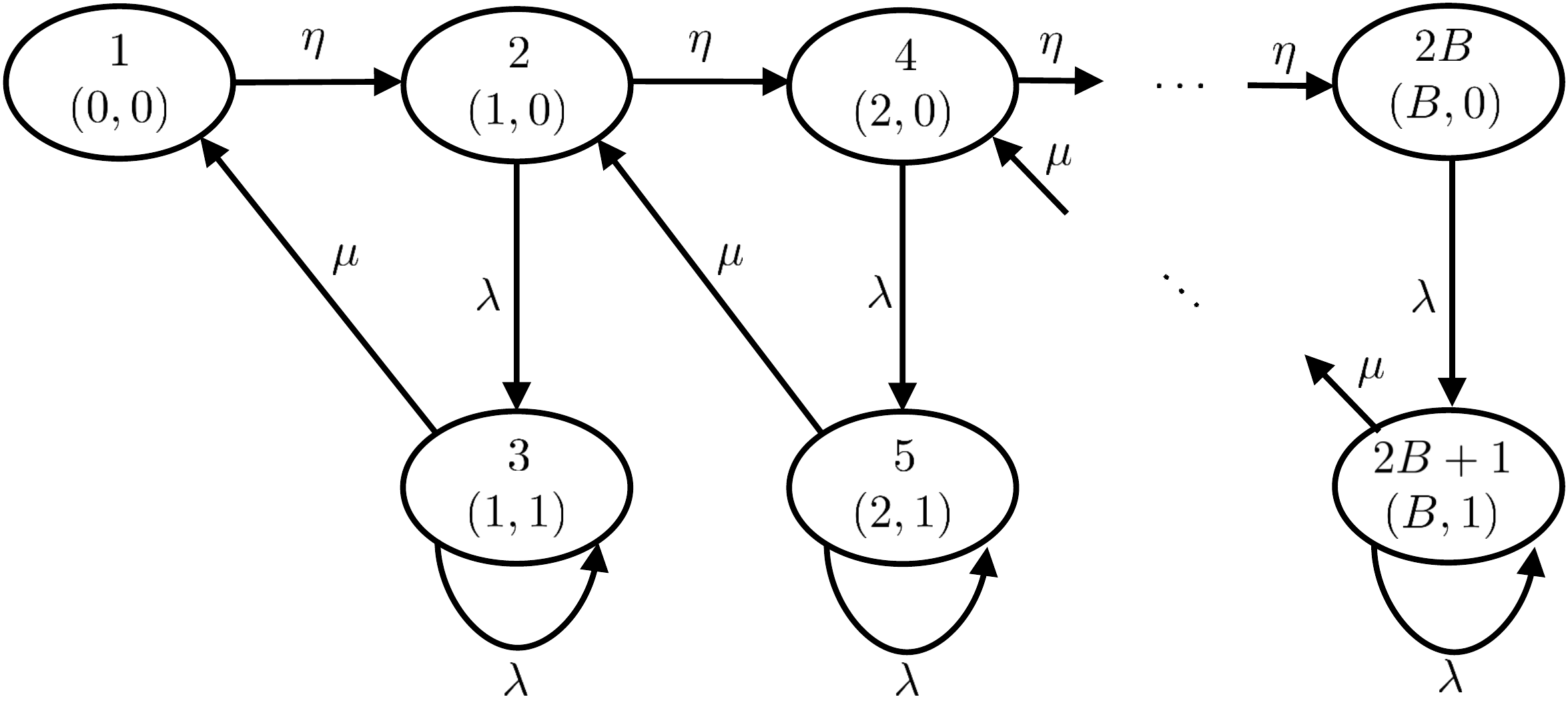}%
\label{f:PS_MC}} \vfil 
\subfloat[]{\includegraphics[width=0.8\columnwidth]{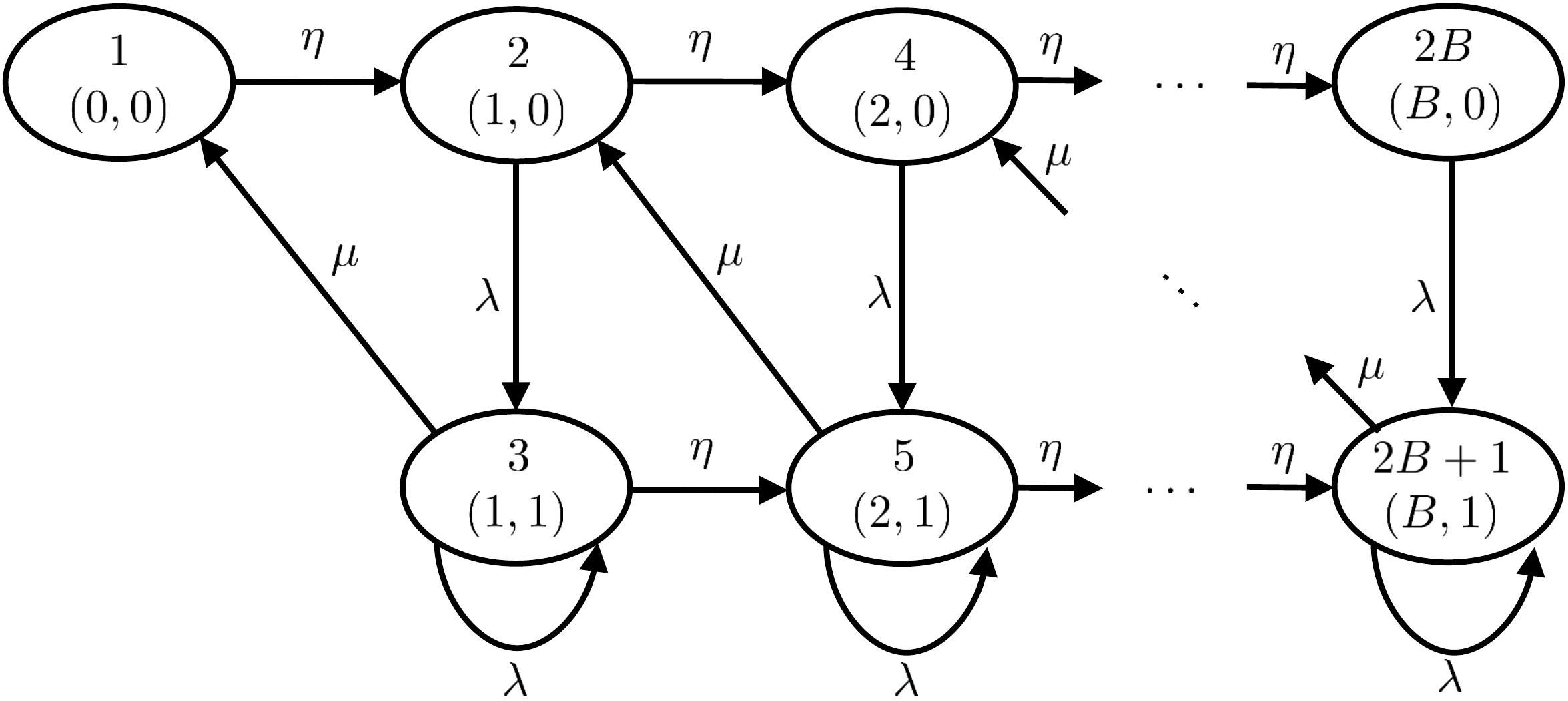}%
\label{f:PS_harvall_MC}}
\caption{Markov chains modeling the discrete state in the LCFS-PS: (a) the transmitter node can harvest only if the system is empty, and (b) the transmitter node can simultaneously harvest energy and serve status updates.}\label{f:PS}
\end{figure}

Fig. \ref{f:PS_MC} depicts the Markov chain representing discrete state of the system in the LCFS-PS queueing discipline when the transmitter node is able to harvest energy only if the system is empty. The set of transitions in this case can be constructed using Tables \ref{table:WP} and \ref{table:PS}. The subset of transitions $l=3B+k-1, 2 \leq k \leq B$, in Table \ref{table:PS} refers to the event of having a new arriving update packet at the transmitter node while its server is serving another update packet. According to the mechanism of the LCFS-PS queueing discipline, the status update that is currently being served will be discarded, and the new arriving one will enter the service upon its arrival. Since the new arriving update packet does not influence the AoI value at the destination and its age is 0, the updated age vector is given by $[x_0,0]$. Furthermore, from (\ref{gen_steady}), we note that the self-transitions do not impact the values of the steady state probabilities $\{\pi_i\}$, and hence $\{\pi_i\}$ in this case can be obtained using Proposition \ref{prop1}. That said, the MGF of AoI at the destination is provided in the next theorem.
\begin{table*}
\centering
{\caption{Transitions of the LCFS-PS queueing discipline in Fig. \ref{f:PS_MC} $(2 \leq k \leq B).$} 
\label{table:PS}
\scalebox{.9}
{ \begin{tabular}{ |c |c|c|c|c|c|c|c|}
\hline
 $l$   & $q_l\rightarrow q'_l$  & $\lambda^{(l)}$ & $\nbx \nbA_l$ & $\nbA_l$ & $\hat{\nbA}_l$ & $\bar{\nbv}^s_{q_l} \nbA_l$ & $\bar{\pi}_{q_l} {\bf 1} \hat{\nbA}_l$\\ \hline
 $3B+k-1$&$2k-1 \rightarrow 2k-1$ &$\lambda$ &$[x_0,0]$ &$\begin{bmatrix}1 &0 \\0 & 0\\ \end{bmatrix}$ & $\begin{bmatrix}0 &0 \\0 & 1\\ \end{bmatrix}$&$[\bar{v}^s_{2k-1,0},0]$ &$[0,\bar{\pi}_{2k-1}]$ \\ \hline
\end{tabular}}} 
\end{table*} 
\begin{theorem} \label{MGF_PS}
When the transmitter node can only harvest energy if the system is empty, the MGF of AoI for the LCFS-PS queueing discipline is given by
\begin{align}\label{theorem_MGF_PS_1}
\overset{\rm PS}{M}\left(\bar{s}\right) = \frac{\rho \left(1 + \rho\right)\bar{\pi}_1 \Big[\bar{s}^{2} \theta - \bar{s} \theta \left(1 + \rho + \beta\right) + \beta \left(1+ \theta + \theta \rho\right)\Big]}{\left(1 - \bar{s}\right) \left(\rho - \bar{s}\right) \left(1 + \rho - \bar{s}\right)\left(\beta - \bar{s}\right)}.
\end{align}
\end{theorem}
\begin{IEEEproof}
See Appendix \ref{app:MGF_PS}.
\end{IEEEproof}
\begin{cor}\label{cor:PS}
Using $\overset{\rm PS}{M}(\bar{s})$ derived in Theorem \ref{MGF_PS}, the first moment of the LCFS-PS queueing discipline (when the transmitter node can only harvest energy if the system is empty) can be expressed as: $\overset{(1)}{\Delta}_{\rm PS} =$ 
\begin{align}\label{mom1_PS}
\begin{cases}
\dfrac{B\rho^3 + \left(3B + 1\right)\rho^2 + \left(3B + 4\right) \rho+ B + 2}{\mu \rho \left(1 + \rho\right) \left(\rho B + B + 1\right)},\;\; {\rm if}\; \rho = \beta,\\
\dfrac{\beta^{B+2}\left(1 + \rho\right)^{3} - \rho^{B+2} \big[\left(\beta^2 + \beta\right) \left(\rho + 2\right) + 1 + \rho \big]}{\mu \left(1 + \rho\right)\big[\beta^{B+2}\left(\rho^2+\rho\right) - \rho^{B+2} \left(\beta^2+\beta\right)\big]},
\end{cases}
\end{align}
where the second case in (\ref{mom1_PS}) holds when $\rho \neq \beta$. Further the second moment of AoI $\overset{(2)}{\Delta}_{\rm PS}$ is given by (\ref{mom2_PS}) [at the top of the next page].
\end{cor}
\begin{figure*}
\begin{align}\label{mom2_PS}
\overset{(2)}{\Delta}_{\rm PS} = 
\begin{cases}
\dfrac{2\big[B\rho^5+\left(4B+1\right)\rho^4+\left(7B+5\right)\rho^3+\left(7B+12\right)\rho^2+\left(4B+10\right)\rho+B+3\big]}{\mu^2\rho^2\left(1 + \rho\right)^2\left(1+B+B\rho\right)}, \;\;&{\rm if}\; \rho = \beta,\\
\dfrac{2\beta^{B+3}\left(1 + \rho\right)^3 \left(1 + \rho + \rho^2\right) - 2\rho^{B+3}\big[\left(\beta^3 + \beta^2 + \beta\right) \left(\rho^2+3\rho +2\right) + \beta^3 +\beta^2 + \left(1 + \rho\right)^2\big]}{\mu^2 \left(1 + \rho\right)^2\big[\beta^{B+3}\rho^2\left(1+\rho\right) - \rho^{B+3} \beta^2 \left(1 + \beta\right)\big]},\;\; & {\rm otherwise}.
\end{cases}
\end{align}
\end{figure*}
\begin{remark}\label{rem:3}
When $\beta \rightarrow \infty$, $\overset{(1)}{\Delta}_{\rm PS}$ and $\overset{(2)}{\Delta}_{\rm PS}$ (in (\ref{mom1_PS}) and (\ref{mom2_PS}) for the case $\rho \neq \beta$) reduce to
\begin{align}
\underset{\beta \rightarrow \infty}{\rm lim} \overset{(1)}{\Delta}_{\rm PS} = \dfrac{1}{\lambda} + \dfrac{1}{\mu},\; \underset{\beta \rightarrow \infty}{\rm lim} \overset{(2)}{\Delta}_{\rm PS} = 2\Big[\dfrac{1}{\lambda^2}+\dfrac{1}{\mu\lambda}+\dfrac{1}{\mu^2}\Big],
\end{align}
which match the first and second moments of AoI for the LCFC-PS queueing discipline in \cite[Theorem~2(a)]{yates2018age}, where the transmitter node does not have energy limitations. 
\end{remark}
\begin{remark}\label{rem:comp_WPandPS}
Note that from Corollaries \ref{cor:WP} and \ref{cor:PS}, we have
\begin{align}\label{WPandPS_diff_mom1}
 \overset{(1)}{\Delta}_{\rm NP} - \overset{(1)}{\Delta}_{\rm PS} = \begin{cases}
 \dfrac{B \rho^2 + \rho \left(B + 1\right) + B}{\mu\big[B \rho^2 + \rho \left(2B + 1\right) + B + 1\big]}, & {\rm if}\; \rho = \beta,\\
 \dfrac{\rho}{\mu\left(1 + \rho \right)},& {\rm otherwise},
 \end{cases}
\end{align}
\begin{figure*}[t!]
\begin{align} \label{WPandPS_diff_mom2}
 \overset{(2)}{\Delta}_{\rm NP} - \overset{(2)}{\Delta}_{\rm PS} = \begin{cases}
 \dfrac{2\big[2B \rho^3 + \rho^2 \left(5B + 2\right) + \rho \left(4B + 5\right) + B + 2 \big]}{\mu^2\big[B \rho^3 + \rho^2 \left(3B + 1\right) + \rho \left(3B + 2\right) + B + 1\big]},\;\; &{\rm if}\; \rho = \beta,\\
 \dfrac{2\beta^{B+3}\left(1 + \rho\right)^2 \left(2\rho^3 + \rho^2\right) - 2\rho^{B+3}\big[\left(\beta^3 + \beta^2 \right) \left(2\rho^2+3\rho \right) + \beta \left(\rho^2 + \rho \right)\big]}{\mu^2 \left(1 + \rho\right)^2\big[\beta^{B+3}\rho^2\left(1+\rho\right) - \rho^{B+3} \beta^2 \left(1 + \beta \right)\big]},&\;\; {\rm otherwise}.
 \end{cases}
\end{align}
\hrulefill
\end{figure*}
and $\overset{(2)}{\Delta}_{\rm NP} - \overset{(2)}{\Delta}_{\rm PS}$ is given by (\ref{WPandPS_diff_mom2}) [at the top of the next page]. We observe from (\ref{WPandPS_diff_mom1}) and (\ref{WPandPS_diff_mom2}) that $ \overset{(1)}{\Delta}_{\rm NP} - \overset{(1)}{\Delta}_{\rm PS} \geq 0$ and $\overset{(2)}{\Delta}_{\rm NP} - \overset{(2)}{\Delta}_{\rm PS} \geq 0$ for any choice of values of the system parameters. This indicates the superiority of the LCFS-PS queueing discipline over the LCFS-NP one, in terms of the achievable AoI performance at the destination node when the transmitter node can harvest energy only if the system is empty. We further observe that when $\rho \neq \beta$, $\overset{(1)}{\Delta}_{\rm NP} - \overset{(1)}{\Delta}_{\rm PS}$ does not depend on the parameters related to the EH process (i.e., $\beta$ and $B$), and the difference monotonically increases as a function of $\rho$ from $\underset{\rho \rightarrow 0}{\rm lim} \overset{(1)}{\Delta}_{\rm NP} - \overset{(1)}{\Delta}_{\rm PS} = 0$ until it approaches $\underset{\rho \rightarrow \infty}{\rm lim} \overset{(1)}{\Delta}_{\rm NP} - \overset{(1)}{\Delta}_{\rm PS} = \frac{1}{\mu}$. On the other hand, $\overset{(2)}{\Delta}_{\rm NP} - \overset{(2)}{\Delta}_{\rm PS}$ monotonically increases as a function of $\rho$ from $\underset{\rho \rightarrow 0}{\rm lim} \overset{(2)}{\Delta}_{\rm NP} - \overset{(2)}{\Delta}_{\rm PS} = \frac{2}{\mu^2}$ until it approaches $\underset{\rho \rightarrow \infty}{\rm lim} \overset{(2)}{\Delta}_{\rm NP} - \overset{(2)}{\Delta}_{\rm PS} = \dfrac{4\beta^2+4\beta+2}{\mu^2\beta(1 + \beta)}$.
\end{remark}
\begin{table*}
\centering
{\caption{Transitions of the LCFS-PS queueing discipline in Fig. \ref{f:PS_harvall_MC} $(2 \leq k \leq B).$} 
\label{table:PS_harvall}
\scalebox{.9}
{ \begin{tabular}{ |c |c|c|c|c|c|c|c|}
\hline
 $l$   & $q_l\rightarrow q'_l$  & $\lambda^{(l)}$ & $\nbx \nbA_l$ & $\nbA_l$ & $\hat{\nbA}_l$ & $\bar{\nbv}^s_{q_l} \nbA_l$ & $\bar{\pi}_{q_l} {\bf 1} \hat{\nbA}_l$\\ \hline
 $3B+2k-3$&$2k-1 \rightarrow 2k-1$ &$\lambda$ &$[x_0,0]$ &$\begin{bmatrix}1 &0 \\0 & 0\\ \end{bmatrix}$ & $\begin{bmatrix}0 &0 \\0 & 1\\ \end{bmatrix}$&$[\bar{v}^s_{2k-1,0},0]$ &$[0,\bar{\pi}_{2k-1}]$ \\ \hline
 $3B+2k-2$&$2k-1 \rightarrow 2k+1$ &$\eta$ & $[x_0,x_1]$&$\begin{bmatrix}
1 & 0\\ 0 & 1\\
\end{bmatrix}$ & $\begin{bmatrix}
0 & 0\\ 0 & 0\\
\end{bmatrix}$&$[\bar{v}^s_{2k-1,0},\bar{v}^s_{2k-1,1}]$ &$[0,0]$ \\ \hline
\end{tabular}}} 
\end{table*} 
\subsection{EH is Allowed Anytime}\label{sub:PS_b}
The discrete state of the system in this case is modeled by the Markov chain in Fig. \ref{f:PS_harvall_MC}. Clearly, the set $\ncalL$ in this queueing discipline is defined by the transitions in Tables \ref{table:WP} and \ref{table:PS_harvall}. For this case, the MGF of AoI at the destination node is given by the following theorem.
\begin{theorem}\label{MGF_PS_harvall}
When the transmitter node can harvest energy anytime, the MGF of AoI at the destination node for the LCFS-PS queueing discipline is given by: $\overset{\rm PS}{M}\left(\bar{s}\right) =$
\begin{align}
 \dfrac{\rho \bar{\pi}_1 \Big[\bar{s}^{2} \gamma' - \bar{s} \gamma' \left(1 + \rho + 2\beta \right)  + \left(1+\gamma'\right)\beta\left(1+\rho+\beta\right)\Big]}{\left(1 - \bar{s}\right) \left(\rho - \bar{s}\right) \left(\beta - \bar{s}\right) \left(1 + \rho + \beta - \bar{s}\right)},
\end{align}
where $\underset{k \in \;{\rm r}_1 \cup\; {\rm r}_2 / \{1\}}{\sum}{\bar{\pi}_k}= \gamma' \bar{\pi}_1$.
\end{theorem}
\begin{IEEEproof}
See Appendix \ref{app:MGF_PS_harvall}.
\end{IEEEproof}
\begin{cor}\label{cor:PS_harvall}
Given that $B=1$ and the transmitter can harvest energy anytime, the first and second moments of AoI for the LCFS-PS queueing discipline can be respectively expressed as 
\begin{align}\label{mom1_PS_harvall_B1}
\overset{(1)}{\Delta}_{\rm PS}= \dfrac{\beta^2\left(1+\rho\right)^3 + \beta \left(\rho^3+3\rho^2+\rho\right)+\rho^3+\rho^2}{\mu\rho\left(1+\rho\right)\beta\big[\beta\left(1+\rho\right)+\rho\big]},
\end{align}
\begin{align}\label{mom2_PS_harvall_B1}
\overset{(2)}{\Delta}_{\rm PS} =  \dfrac{2\sum_{n=0}^3{\beta^n\zeta_n}}{\mu^2\rho^2\left(1+\rho\right)^2 \beta^2 \big[\beta\left(1+\rho\right)+\rho\big]},
\end{align}
\begin{align*}
&\zeta_3 =  \left(1+\rho\right)^3\left(\rho^2+\rho+1\right), \zeta_2= \rho\left(\rho^4+4\rho^3+7\rho^2+4\rho+1\right), \\&\zeta_1 = \rho^2\left(1+\rho\right)\left(\rho^2+3\rho+1\right), \zeta_0 = \rho^3\left(1+\rho\right)^2.
\end{align*}

Further, the first and second moments of AoI for $B=2$ ($\bar{\pi}_2$ is given by (\ref{pi_2_B2})) can be respectively expressed as
\begin{align}\label{mom1_PS_harvall_B2}
\overset{(1)}{\Delta}_{\rm PS} = \dfrac{\bar{\pi}_2\sum_{n=0}^6{\beta^n \bar{\zeta}_n}}{\mu\rho^2\beta^2\left(1+\beta\right)\left(1+\rho+\beta\right)^2},
\end{align}
\begin{align*}
&\bar{\zeta}_6 = \big(1+\rho\big)^2,\;\bar{\zeta}_5= 3
\big(1+\rho\big)^3, \;\bar{\zeta}_4 = \left(1+\rho\right)^2\big(3\rho^2+7\rho+3\big),\\&\bar{\zeta}_3 = \big(1+\rho\big)\big(\rho^2+3\rho+1\big)^2,\;\bar{\zeta}_2 = \rho^5+6\rho^4+12\rho^3+6\rho^2+\rho,\\&\bar{\zeta}_1 = \rho^2\big(1+\rho\big)\big(\rho^2+5\rho+1\big),\;\bar{\zeta}_0 = \rho^3\left(1+\rho\right)^2,
\end{align*}
\begin{align}\label{mom2_PS_harvall_B2}
\overset{(2)}{\Delta}_{\rm PS} = \dfrac{2\bar{\pi}_2\Big[\sum_{n=0}^{8}{\beta^n\zeta'_n}\Big]}{\mu^2\rho^3\beta^3\left(1+\beta\right)\left(1+\rho+\beta\right)^3},
\end{align}
\begin{align*}
&\zeta'_8 = \big(1+\rho\big)\big(\rho^2+\rho+1\big),\; \zeta'_7 = 4\big(1+\rho\big)^2\big(\rho^2+\rho+1\big),\\& \zeta'_6 = \big(1+\rho\big)\big(\rho^2+\rho+1\big)\big(2\rho+3\big)\big(3\rho+2\big),\\
&\zeta'_5 = \big(\rho^2+\rho+1\big)\big(4\rho^4+19\rho^3+31\rho^2+19\rho+4\big),\\
& \zeta'_4 = \big(1+\rho\big)\big(\rho^2+3\rho+1\big)\big(\rho^4+5\rho^3+7\rho^2+5\rho+1\big),
\\&\zeta'_3=\rho\big(\rho^6+8\rho^5+25\rho^4+39\rho^3+25\rho^2+8\rho+1\big),\\
&\zeta'_2 = \rho^2\big(1+\rho\big)\big(\rho^4+7\rho^3+18\rho^2+7\rho+1\big),\;\zeta'_1= \rho^3 \big(1+\rho\big)^2\\
&\big(\rho^2+6\rho+1\big),\;\zeta'_0 = \rho^4\big(1+\rho\big)^3.
\end{align*}
\end{cor}
\begin{IEEEproof}
The result follows from (\ref{MGF_derav}) with noting that $\gamma' = \frac{\beta}{\rho}\left(1 + \rho\right)$ for $B = 1$, and $\gamma' = \frac{\beta}{\rho^2}\left(1 + \rho\right)\big[\rho + \beta \left(1 + \rho + \beta\right)\big]$ for $B = 2$.
\end{IEEEproof}
\begin{remark}\label{rem:4}
Similar to Remark \ref{rem:3}, note that as $\beta \rightarrow \infty$, $\overset{(1)}{\Delta}_{\rm PS}$ and $\overset{(2)}{\Delta}_{\rm PS}$ expressions derived for $B = 1$ and $B = 2$ (in (\ref{mom1_PS_harvall_B1})- (\ref{mom2_PS_harvall_B2})) reduce to the first and second moments of AoI for the LCFS-PS in \cite[Thereom~2(a)]{yates2018age}.
\end{remark}
\begin{remark}\label{rem:comp_WPandPS_harvall}
When the transmitter node can simultaneously harvest energy and serve status updates, it can be shown from Corollaries \ref{cor:WP_harvall} and \ref{cor:PS_harvall} for $B = 1$ that $\overset{(1)}{\Delta}_{\rm NP} - \overset{(1)}{\Delta}_{\rm PS}$ and $\overset{(2)}{\Delta}_{\rm NP} - \overset{(2)}{\Delta}_{\rm PS}$ are similar to the corresponding differences (given by (\ref{WPandPS_diff_mom1}) and (\ref{WPandPS_diff_mom2}), $\rho \neq \beta$) in the case where the transmitter is able to harvest energy only if its server is idle. This happens due to the fact that whether the transmitter is able or unable to harvest energy anytime, the continuous-time Markov chain modeling the system discrete state does not change when $B = 1$. Further, for $B = 2$, we can verify from Corollaries \ref{cor:WP_harvall} and \ref{cor:PS_harvall} that the LCFS-PS queueing discipline yields smaller values of the first and second moments of AoI at the destination node than the LCFS-NP. In particular, $\overset{(1)}{\Delta}_{\rm NP} - \overset{(1)}{\Delta}_{\rm PS}$ monotonically increases with $\rho$ from $\underset{\rho \rightarrow 0}{\rm lim} \overset{(1)}{\Delta}_{\rm NP} - \overset{(1)}{\Delta}_{\rm PS} = 0$ until it approaches $\underset{\rho \rightarrow \infty}{\rm lim} \overset{(1)}{\Delta}_{\rm NP} - \overset{(1)}{\Delta}_{\rm PS} = \dfrac{\beta^3 + 2\beta^2 + \beta + 1}{\mu\left(\beta^3 + 2\beta^2 + 2 \beta + 1\right)}$. On the other hand, $\overset{(2)}{\Delta}_{\rm NP} - \overset{(2)}{\Delta}_{\rm PS}$ monotonically increases as a function of $\rho$ from $\underset{\rho \rightarrow 0}{\rm lim} \overset{(2)}{\Delta}_{\rm NP} - \overset{(2)}{\Delta}_{\rm PS} = \frac{2}{\mu^2}$ until it approaches $\underset{\rho \rightarrow \infty}{\rm lim} \overset{(2)}{\Delta}_{\rm NP} - \overset{(2)}{\Delta}_{\rm PS} = \dfrac{4\beta^5 + 12 \beta^4 + 12 \beta^3 + 6 \beta^2 + 6 \beta + 2}{\mu^2\left(\beta^5 + 3 \beta^4 + 4 \beta^3 + 3 \beta^2 + \beta \right)}$.
\end{remark}
\begin{remark}\label{rem:asyminsight}
By inspecting the asymptotic results of the differences between the achievable first and second moments of AoI by the LCFS-NP and LCFS-PS queueing disciplines (in Remarks \ref{rem:comp_WPandPS} and \ref{rem:comp_WPandPS_harvall}), one can deduce that the ability of the transmitter node to simultaneously harvest energy and serve update packets reduces the gap between the achievable AoI performances by the two queueing disciplines.
\end{remark}
\begin{figure}[t!]
\centering
\includegraphics[width=0.85\columnwidth]{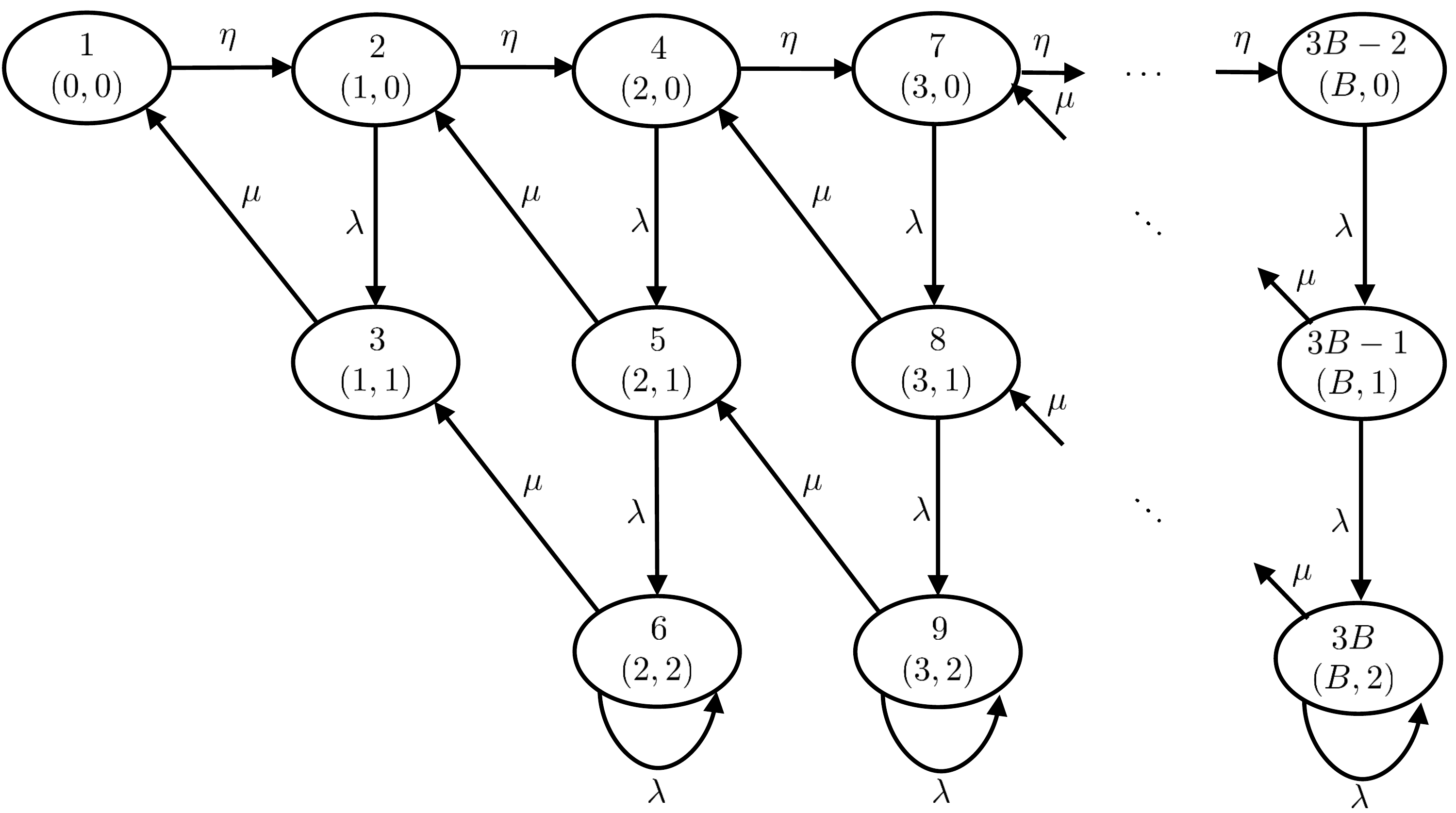}
\caption{The Markov chain modeling the discrete state in the LCFS-PW queueing discipline when the transmitter node can harvest energy only if the system is empty.}
\label{f:PW_MC}
\end{figure}
\section{The MGF Analysis of LCFS-PW Queueing Discipline}\label{sec:PW}
The analysis of the MGF of AoI for the LCFS-PW queueing discipline is presented in this section. As observed from Figs. \ref{f:PW_MC} and \ref{f:PW_harvall_MC}, the discrete state space $\ncalQ$ in the LCFS-PW queueing discipline is given by $\ncalQ = \{1,2,\cdots,3B\}$. Similar to the LCFS-NP and LCFS-PS queueing disciplines (presented in Sections \ref{sec:WP} and \ref{sec:PS}), the discrete state of the system $q(t)$ keeps tracking the numbers of status update and energy packets in the system. However, in the LCFS-PW queueing discipline, the system can contain at most two status updates; one is being served and the other is waiting in the queue. Therefore, the continuous process $\nbx(t)$ in this queueing discipline is given by $\nbx(t)=[x_0(t),x_1(t),x_2(t)]$, where $x_0(t)$ tracks the value of AoI at the destination node, and $x_1(t)$ $\left(x_2\left(t\right)\right)$ keeps the value that the AoI at the destination node will be reset to if the update packet in service (waiting in the queue for service) is delivered to the destination. Using the above SHS structure associated with the LCFS-PW queueing discipline, we proceed by driving the MGF of AoI at the destination for both the scenarios when the transmitter node is able/unable to simultaneously harvest energy and serve status update packets.
\begin{table*}\caption{Transitions of the LCFS-PW queueing discipline in Fig. \ref{f:PW_MC} $(2 \leq k \leq B-1).$} 
\label{table:PW}
\centering
{
\scalebox{.9}
{ \begin{tabular}{ |c |c|c|c|c|c|c|c|}
\hline
 $l$   & $q_l\rightarrow q'_l$  & $\lambda^{(l)}$ & $\nbx \nbA_l$ & $\nbA_l$ & $\hat{\nbA}_l$ & $\bar{\nbv}^s_{q_l} \nbA_l$ & $\bar{\pi}_{q_l} {\bf 1} \hat{\nbA}_l$\\ \hline
1& $1 \rightarrow 2$&$\eta$ &$[x_0,0,0]$ &$\begin{bmatrix}1&0&0\\0&0&0\\0&0&0\\ \end{bmatrix}$ &$\begin{bmatrix}0&0&0\\0&1&0\\0&0&1\\ \end{bmatrix}$ &$[\bar{v}^s_{10},0,0]$ & $[0,\bar{\pi}_1,\bar{\pi}_1]$\\ \hline
2& $2 \rightarrow 4$&$\eta$&$[x_0,0,0]$ &$\begin{bmatrix}1&0&0\\0&0&0\\0&0&0\\ \end{bmatrix}$ &$\begin{bmatrix}0&0&0\\0&1&0\\0&0&1\\ \end{bmatrix}$ &$[\bar{v}^s_{20},0,0]$ & $[0,\bar{\pi}_2,\bar{\pi}_2]$\\ \hline
3& $5 \rightarrow 2$&$\mu$ &$[x_1,0,0]$ &$\begin{bmatrix}0&0&0\\1&0&0\\0&0&0\\ \end{bmatrix}$ &$\begin{bmatrix}0&0&0\\0&1&0\\0&0&1\\ \end{bmatrix}$ &$[\bar{v}^s_{51},0,0]$ & $[0,\bar{\pi}_5,\bar{\pi}_5]$\\ \hline
4& $6 \rightarrow 3$&$\mu$ &$[x_1,x_2,0]$ &$\begin{bmatrix}0&0&0\\1&0&0\\0&1&0\\ \end{bmatrix}$ &$\begin{bmatrix}0&0&0\\0&0&0\\0&0&1\\ \end{bmatrix}$ & $[\bar{v}^s_{61},\bar{v}^s_{62},0]$ & $[0,0,\bar{\pi}_6]$\\ \hline
5& $2 \rightarrow 3$&$\lambda$&$[x_0,0,0]$ &$\begin{bmatrix}1&0&0\\0&0&0\\0&0&0\\ \end{bmatrix}$ &$\begin{bmatrix}0&0&0\\0&1&0\\0&0&1\\ \end{bmatrix}$ &$[\bar{v}^s_{20},0,0]$ & $[0,\bar{\pi}_2,\bar{\pi}_2]$\\ \hline
$6k-6$&$3k-2 \rightarrow 3k+1$&$\eta$ &$[x_0,0,0]$ & $\begin{bmatrix}1&0&0\\0&0&0\\0&0&0\\ \end{bmatrix}$&$\begin{bmatrix}0&0&0\\0&1&0\\0&0&1\\ \end{bmatrix}$ &$[\bar{v}^s_{3k-2,0},0,0]$ & $[0,\bar{\pi}_{3k-2},\bar{\pi}_{3k-2}]$ \\ \hline
$6k-5$&$3k+2 \rightarrow 3k-2$ & $\mu$&$[x_1,0,0]$ &$\begin{bmatrix}0&0&0\\1&0&0\\0&0&0\\ \end{bmatrix}$ &$\begin{bmatrix}0&0&0\\0&1&0\\0&0&1\\ \end{bmatrix}$&$[\bar{v}^s_{3k+2,1},0,0]$ &$[0,\bar{\pi}_{3k+2},\bar{\pi}_{3k+2}]$ \\ \hline
$6k-4$&$3k+3 \rightarrow 3k-1$ &$\mu$ &$[x_1,x_2,0]$ &$\begin{bmatrix}0&0&0\\1&0&0\\0&1&0\\ \end{bmatrix}$ &$\begin{bmatrix}0&0&0\\0&0&0\\0&0&1\\ \end{bmatrix}$ &$[\bar{v}^s_{3k+3,1},\bar{v}^s_{3k+3,2},0]$ &$[0,0,\bar{\pi}_{3k+3}]$ \\ \hline
$6k-3$&$3k-2 \rightarrow 3k-1$ &$\lambda$&$[x_0,0,0]$ &$\begin{bmatrix}1&0&0\\0&0&0\\0&0&0\\ \end{bmatrix}$ &$\begin{bmatrix}0&0&0\\0&1&0\\0&0&1\\ \end{bmatrix}$ &$[\bar{v}^s_{3k-2,0},0,0]$ &$[0,\bar{\pi}_{3k-2},\bar{\pi}_{3k-2}]$ \\ \hline
$6k-2$&$3k-1 \rightarrow 3k$ &$\lambda$ &$[x_0,x_1,0]$ &$\begin{bmatrix}1&0&0\\0&1&0\\0&0&0\\ \end{bmatrix}$ &$\begin{bmatrix}0&0&0\\0&0&0\\0&0&1\\ \end{bmatrix}$ &$[\bar{v}^s_{3k-1,0},\bar{v}^s_{3k-1,1},0]$ &$[0,0,\bar{\pi}_{3k-1}]$ \\ \hline
$6k-1$&$3k \rightarrow 3k$ &$\lambda$ &$[x_0,x_1,0]$ &$\begin{bmatrix}1&0&0\\0&1&0\\0&0&0\\ \end{bmatrix}$ &$\begin{bmatrix}0&0&0\\0&0&0\\0&0&1\\ \end{bmatrix}$ &$[\bar{v}^s_{3k,0},\bar{v}^s_{3k,1},0]$ &$[0,0,\bar{\pi}_{3k}]$ \\ \hline
$6B-6$&$3B-2 \rightarrow 3B-1$ &$\lambda$&$[x_0,0,0]$ &$\begin{bmatrix}1&0&0\\0&0&0\\0&0&0\\ \end{bmatrix}$ &$\begin{bmatrix}0&0&0\\0&1&0\\0&0&1\\ \end{bmatrix}$ &$[\bar{v}^s_{3B-2,0},0,0]$ &$[0,\bar{\pi}_{3B-2},\bar{\pi}_{3B-2}]$ \\ \hline
$6B-5$&$3B-1 \rightarrow 3B$ &$\lambda$ &$[x_0,x_1,0]$ &$\begin{bmatrix}1&0&0\\0&1&0\\0&0&0\\ \end{bmatrix}$ &$\begin{bmatrix}0&0&0\\0&0&0\\0&0&1\\ \end{bmatrix}$ &$[\bar{v}^s_{3B-1,0},\bar{v}^s_{3B-1,1},0]$ &$[0,0,\bar{\pi}_{3B-1}]$ \\ \hline
$6B-4$&$3B \rightarrow 3B$ &$\lambda$ &$[x_0,x_1,0]$ &$\begin{bmatrix}1&0&0\\0&1&0\\0&0&0\\ \end{bmatrix}$ &$\begin{bmatrix}0&0&0\\0&0&0\\0&0&1\\ \end{bmatrix}$ &$[\bar{v}^s_{3B,0},\bar{v}^s_{3B,1},0]$ &$[0,0,\bar{\pi}_{3B}]$ \\ \hline
\end{tabular}}} 
\end{table*} 
\subsection{EH is Only Allowed When the System is Empty}\label{sub:PW_a}
When the transmitter node is able to harvest energy only if its server is idle, the discrete state of the system $q(t)$ in the LCFS-PW queueing discipline is modeled by the Markov chain in Fig. \ref{f:PW_MC}. Now, we describe the set of transitions $\ncalL$ (listed in Table \ref{table:PW}) between the states of the Markov chain in Fig. \ref{f:PW_MC}.
\begin{itemize}
    \item $l=6k-6$: A transition from this subset of transitions is induced by the arrival of a new harvested energy packet when the server is idle (i.e., $q(t)$ belongs to one of the states in the first row ${\rm r}_1$ of the Markov chain in Fig. \ref{f:PW_MC}). Since the AoI value at the destination is not affected by this transition, and the components $x_1(t)$ and $x_2(t)$ are irrelevant for the states in ${\rm r}_1$, we observe that the updated age vector $\nbx \nbA_{6k-k} = [x_0,0,0]$.
    \item $l=6k-5$: This subset of transitions takes place when the system contains only one status update in service, and its service time is completed. Clearly, as a result of this transition, the AoI at the destination will reset to the age of the received packet $x_1$, the age $x_1(t)$ becomes irrelevant (since the system becomes empty), and the age $x_2(t)$ is irrelevant (since the queue storing the latest arriving status update while the server is busy was empty before the transition). Thus, the age vector is updated as $[x_1,0,0]$.
    \item $l=6k-4$: This subset of transitions is similar to the previous one labeled by $l=6k-5$ with the only difference that there is a status update in the queue waiting for service at the moment when the current packet in service is delivered to the destination. As a result of this transition, the status update in the queue will enter the service, and thus $x_1(t)$ (quantifying the age of the packet in service) is reset to its age $x_2$ just before the transition, as indicated by the updated age vector $\nbx \nbA_{6k-4}$.
    \item $l=6k-3$: A transition from this subset of transitions occurs when the system is empty and a new update packet arrives at the transmitter node. Clearly, the new arriving update packet will immediately start its service time, and thus $x_1(t)$ will be 0 (i.e., the age of the new arriving packet).
    \item $l=6k-2$: Similar to the previous subset of transitions, this subset of transitions is induced by the arrival of a new status update packet at the transmitter node. However, the system now contains one status update packet in service upon the arrival of the new update packet. Thus, the new arriving packet will enter the queue and $x_2(t)$ is reset to 0.
    \item $l=6k-1$: A transition from this subset of transitions occurs when the system has two status update packets (one in service and the other waiting in the queue) and a new update packet arrives at the transmitter node. According to the mechanism of the LCFS-PW queueing discipline, the new arriving update packet will replace the old one in the queue, and hence the age $x_2(t)$ is reset to 0 as indicated by $\nbx \nbA_{6k-1}$.
\end{itemize}

Next, we characterize the steady state probabilities of the Markov chain modeling $q(t)$ in the following proposition.
\begin{prop}\label{prop2}
The steady state probabilities $\{\bar{\pi}_i\}$ can be expressed as
\begin{align}\label{prop2_1}
\nonumber&\bar{\pi}_2 = \frac{\beta \left(1 + \rho\right)}{\rho \left(1 + \rho + \beta\right)}\bar{\pi}_1,\;\; \bar{\pi}_3 = \beta \bar{\pi}_1,\\& \bar{\pi}_{3B-2} = \frac{\beta}{\rho}\Big[\frac{\beta \left(1 + \rho\right)}{\rho \left(1 + \rho + \beta\right)} \Big]^{B-1} \bar{\pi}_1,
\end{align}
\begin{align}\label{prop2_2}
\bar{\pi}_{3k-2} = \Big[\frac{\beta \left(1 + \rho\right)}{\rho \left(1 + \rho + \beta\right)} \Big]^{k} \bar{\pi}_1,\; \; 2 \leq k \leq B - 1,
\end{align}
\begin{align}\label{prop2_3}
\bar{\pi}_{3k-1} = \frac{\rho\left(1 + \rho + \beta\right)}{\left(1 + \rho\right)^{2}}\Big[\frac{\beta \left(1 + \rho\right)}{\rho \left(1 + \rho + \beta\right)} \Big]^{k}\bar{\pi}_1,\;\;  2 \leq k \leq B,
\end{align}
\begin{align}\label{prop2_4}
\bar{\pi}_{3k} = \frac{\rho^{2}\left(1 + \rho + \beta\right)}{\left(1 + \rho\right)^{2}}\Big[\frac{\beta \left(1 + \rho\right)}{\rho \left(1 + \rho + \beta\right)} \Big]^{k}\bar{\pi}_1,\;\;  2 \leq k \leq B,
\end{align}
where $\bar{\pi}_1$ is given by (\ref{prop2_pi1}) [at the top of the next page].
\end{prop}
\begin{figure*}
\begin{align}\label{prop2_pi1}
\bar{\pi}_1 = \begin{cases}
\dfrac{\rho}{\beta + \rho \left(1 + \beta\right) B},\; & {\rm if}\; \beta = \rho\left(1 + \rho\right),\\
\dfrac{\rho \Big[\beta - \rho \left(1 + \rho\right)\Big] \Big[\rho\left(1 + \rho + \beta\right)\Big]^{B-1}}{\beta^{B+1} \left(1 + \rho\right)^{B-1} \left(\rho^2+\rho+1\right) - \rho^{B+1} \left(1 + \rho + \beta\right)^B \left(1 + \beta\right)},\; & {\rm otherwise}.
\end{cases}
\end{align}
\end{figure*}
\begin{IEEEproof}
See Appendix \ref{app:prop2}.
\end{IEEEproof}

Using Proposition \ref{prop2} and the set of transitions $\ncalL$ in Table \ref{table:PW}, we are now ready to derive the MGF of AoI at the destination in the following theorem.
\begin{theorem}\label{MGF_PW}
When the transmitter node can only harvest energy if the system is empty, the MGF of AoI at the destination node for the LCFS-PW queueing discipline is given by
\begin{align}\label{theorem_MGF_PW_1}
\overset{\rm PW}{M}\left(\bar{s}\right) = \frac{\rho \bar{\pi}_1 \left(\overset{\rm PW}{M_1}\left(\bar{s}\right) + \overset{\rm PW}{M_2}\left(\bar{s}\right)\right)}{\left(1 + \rho + \beta\right)\left(1 - \bar{s}\right)^{3}  \left(1 + \rho - \bar{s}\right)^{2}\left(\rho - \bar{s}\right)\left(\beta - \bar{s}\right)},
\end{align}
\begin{align*}
&\overset{\rm PW}{M_1}\left(\bar{s}\right) = \left(1+\rho+\beta\right)\left(1-\bar{s}\right)\left(\beta-\bar{s}\right) \times\\&\Big[\bar{s}^{2}-2\bar{s}\left(1+\rho\right)+\rho^2+\rho+1\Big]\Big[\bar{s}\bar{\theta}_2+\theta_1-\beta-\bar{\theta}_2\left(1+\rho\right)\Big],
\end{align*}
\begin{align*}
\overset{\rm PW}{M_2}\left(\bar{s}\right) &= \beta \left(1 + \rho\right) \Big[-\bar{s}\left(3+\rho+\beta\right)+1+\rho+\beta\Big] \times\\& \Big[-\bar{s}+1+\rho+\beta\Big],
\end{align*}
where $\bar{\theta}_2 = 1 - \theta_2$ and we have
\begin{align}\label{theorem_MGF_PW_theta1}
\theta_1 = \begin{cases}
\beta B,\;\;\; {\rm if}\; \beta = \rho (1 + \rho),\\
\dfrac{\beta^{B+1}\left(1+\rho\right)^{B} - \beta \rho^{B} \left(1+\rho+\beta\right)^B}{\rho^{B-1} \left(1+\rho+\beta\right)^{B-1} \Big[\beta-\rho\left(1+\rho\right)\Big]},\; {\rm otherwise}.
\end{cases}
\end{align}
\begin{align}\label{theorem_MGF_PW_theta2}
\theta_2 = \begin{cases}
\dfrac{\beta}{\rho}+B,\;\;\;  {\rm if}\; \beta = \rho (1 + \rho),\\
\dfrac{\beta^{B+1}\left(1+\rho\right)^{B-1} - \rho^{B+1} \left(1+\rho+\beta\right)^B}{\rho^{B} \left(1+\rho+\beta\right)^{B-1} \Big[\beta-\rho\left(1+\rho\right)\Big]},{\rm otherwise}.
\end{cases}
\end{align}
\end{theorem}
\begin{IEEEproof}
See Appendix \ref{app:MGF_PW}.
\end{IEEEproof}
\begin{cor}
When the transmitter node can only harvest energy if the system is empty, using $\overset{\rm PW}{M}(\bar{s})$ derived in Theorem \ref{MGF_PW} and given that $\beta = \rho \left(1 + \rho\right)$, the first and second moments of the LCFS-PW queueing discipline can be respectively expressed as
\begin{align}
\overset{(1)}{\Delta}_{\rm PW} = \dfrac{\bar{\pi}_1\Big[ \sum_{n=1}^{3}{\beta^{n}\psi_{n}} + \rho^2\left(1+\rho\right)^4\Big]}{\mu \rho^2 \beta \left(1+\rho \right)^3 \left(1+\rho+\beta \right)},
\end{align}
\begin{align*}
&\psi_3 = \rho\left(2\rho^4+5\rho^3+4\rho^2+4\rho+1\right)B + 3\rho^4 + 10\rho^3 + 8\rho^2 + \\&4\rho + 1, \;\psi_2 = \left(1+\rho\right)\big[\rho\left(4\rho^4+9\rho^3+7\rho^2+7\rho+2\right)B - \\&\left(2\rho^5-\rho^4-12\rho^3-9\rho^2-4\rho-1\right)\big],\;\psi_1 = \rho\left(1+\rho\right)^2 \\&\big[ \left(2\rho^4+4\rho^3+3\rho^2+3\rho+1\right)B - \rho\left(2\rho^3+2\rho^2-3\rho-2\right)\big].
\end{align*}
\begin{align}
\overset{(2)}{\Delta}_{\rm PW} = \dfrac{2\bar{\pi}_1\Big[\sum_{n=1}^{4}{\beta^n\bar{\psi}_n} + \rho^3 \left(1+\rho\right)^5\Big]}{\mu^2\rho^3\beta^2\left(1+\rho\right)^4\left(1+\rho+\beta\right)},  
\end{align}
\begin{align*}
&\bar{\psi}_4 = \rho\left(3\rho^6+11\rho^5+15\rho^4+11\rho^3+11\rho^2+5\rho+1\right)B +\\& 4\rho^6+19\rho^5+35\rho^4+26\rho^3+13\rho^2+5\rho+1,\;\bar{\psi}_3 = \rho\left(1+\rho\right)\\&\left(6\rho^6+20\rho^5+25\rho^4+18\rho^3+18\rho^2+9\rho+2\right)B - 
\big(1+\rho\big)\\&\big(3\rho^7+2\rho^6-20\rho^5-43\rho^4-30\rho^3-14\rho^2-5\rho -1\big),\\& \bar{\psi}_2 = \rho \big(1 + \rho\big)^2\big(3\rho^6+9\rho^5+10\rho^4+7\rho^3+7\rho^2 +4\rho+1\big) B\\& - \rho^2\big(1+\rho\big)^2\left(3\rho^5+6\rho^4-3\rho^3-15\rho^2-10\rho-2\right),\\& \bar{\psi}_1 = \rho^2\left(1+\rho\right)^4\left(2\rho^2+4\rho+1\right).
\end{align*}

On the other hand, when $\beta \neq \rho\left(1+\rho\right)$, the first and second moments can be respectively expressed as
\begin{align}\label{mom1_PW}
\overset{(1)}{\Delta}_{\rm PW}= \frac{\bar{\pi}_1\Big[\sum_{n=0}^{3}{\beta^n\psi'_n} +  \theta_1 \psi'_4 - \bar{\theta}_2 \psi'_5\Big]}{\mu\rho\beta\left(1+\rho\right)^3\left(1+\rho+\beta\right)},
\end{align}
\begin{align*}
&\psi'_0= \rho\left(1+\rho\right)^4,\; \psi'_1 = \left(1 + \rho\right)^3\left(2\rho^2+4\rho+1\right),\\& \psi'_2 = - \left(1 + \rho\right) \left(2\rho^4+\rho^3-8\rho^2-6\rho-1\right),\\& \psi'_3 = - \rho\left(2\rho^3+3\rho^2-3\rho-2\right), \;\psi'_4 = \beta \left(1+\rho+\beta\right)\\&\left(2\rho^4+5\rho^3+4\rho^2+4\rho+1\right),\; \psi'_5 = \beta \left(1+\rho\right) \left(1+\rho+\beta\right)\\& \left(2\rho^4+4\rho^3+3\rho^2+3\rho+1\right),
\end{align*}

\begin{align}\label{mom2_PW}
\overset{(2)}{\Delta}_{\rm PW} = \frac{2\bar{\pi}_1\Big[\sum_{n=0}^{4}{\beta^n\psi_n^*} +  \theta_1 \psi_5^* - \bar{\theta}_2 \psi_6^*\Big]}{\mu^2\rho^2\beta^2\left(1+\rho\right)^4\left(1+\rho+\beta\right)},
\end{align}
\begin{align*}
&\psi^*_0 = \rho^2 \left(1 + \rho\right)^5,\;\psi^*_1 = \rho \left(1 + \rho\right)^4\left(2\rho^2+4\rho+1\right),\;\psi^*_2 = \\&\left(1+\rho\right)^3 \left(3\rho^4+10\rho^3+12\rho^2+5\rho+1\right),\;
\psi^*_3 = -\left(1+\rho\right)\\&\left(3\rho^6+5\rho^5-11\rho^4-33\rho^3-23\rho^2-7\rho-1\right),\;\psi^*_4 = -\rho\\
&\big(3\rho^5+8\rho^4-18\rho^2-12\rho-2\big),\;\psi^*_5= \beta^2 
\left(1+\rho+\beta\right)\\&\left(3\rho^6+11\rho^5+15\rho^4+11\rho^3+11\rho^2+5\rho+1\right), \;\psi^*_6= \beta^2\big(1 \\&+ \rho\big)\big(1 + \rho + \beta\big)\big(3\rho^6+9\rho^5+10\rho^4 + 7\rho^3+7\rho^2+4\rho+1\big).
\end{align*}
\end{cor}
\begin{cor}\label{cor_PW_betainf}
When $\beta \rightarrow \infty$, $\overset{(1)}{\Delta}_{\rm PW}$ and $\overset{(2)}{\Delta}_{\rm PW}$ (in (\ref{mom1_PW}) and (\ref{mom2_PW})) respectively reduce to (\ref{cor_PW_mom1_betainf}) and (\ref{cor_PW_mom2_betainf}) [at the top of this page],
\begin{figure*}
\begin{align}\label{cor_PW_mom1_betainf}
\underset{\beta \rightarrow \infty}{\rm lim}\overset{(1)}{\Delta}_{\rm PW} = \dfrac{\left(1+\rho\right)^{B-2}\left(2\rho^5+7\rho^4+8\rho^3+7\rho^2+4\rho+1\right) - \rho^{B+1}\left(2\rho^2+3\rho-1\right)}{\mu\rho\left(1+\rho\right)\Big[\left(1+\rho\right)^{B-1}\left(\rho^2+\rho+1\right) - \rho^{B+1}\Big]},    
\end{align}
\begin{align}\label{cor_PW_mom2_betainf}
\underset{\beta \rightarrow \infty}{\rm lim}\overset{(2)}{\Delta}_{\rm PW} =\dfrac{2\Big[\psi^{*}_{7} \big(1+\rho\big)^{B-2} - \psi^{*}_{8} \rho^{B+1}\Big]}{\mu^2\rho^2\big(1+\rho\big)^2\Big[\big(1+\rho\big)^{B-1}\big(\rho^2+\rho+1\big) - \rho^{B+1}\Big]},
\end{align}
\hrulefill
\end{figure*}
where $\psi^*_7 = 3\rho^7+14\rho^6+24\rho^5+ 21\rho^4+18\rho^3+12\rho^2+5\rho+1$, and $\psi^*_8 = 3\rho^4+8\rho^3+4\rho^2-5\rho-1$.
\end{cor}
\begin{remark}\label{rem:PW_limit}
When the arrival rate of harvested energy packets goes to infinity $(\beta \rightarrow \infty)$, we note from Corollary \ref{cor_PW_betainf} that unlike the other queueing disciplines studied in this paper, $\underset{\beta \rightarrow \infty}{\rm lim}\overset{(1)}{\Delta}_{\rm PW}$ and $\underset{\beta \rightarrow \infty}{\rm lim}\overset{(2)}{\Delta}_{\rm PW}$  (when the transmitter node can harvest energy only if the system is empty) are functions of the battery capacity $B$. This can be interpreted from Fig. \ref{f:PW_MC} by noting that whenever the discrete state of the system $q(t)$ is $q_3$, having $\beta \rightarrow \infty$ does not impact the fact that all the new arrivals of status updates will be discarded until the packet in service is delivered to the destination (i.e., $q(t)$ transitions from $q_3$ to $q_1$). 
\end{remark}
\subsection{EH is Allowed Anytime}\label{sub:PW_b}
\begin{figure}[t!]
\centering
\includegraphics[width=0.85\columnwidth]{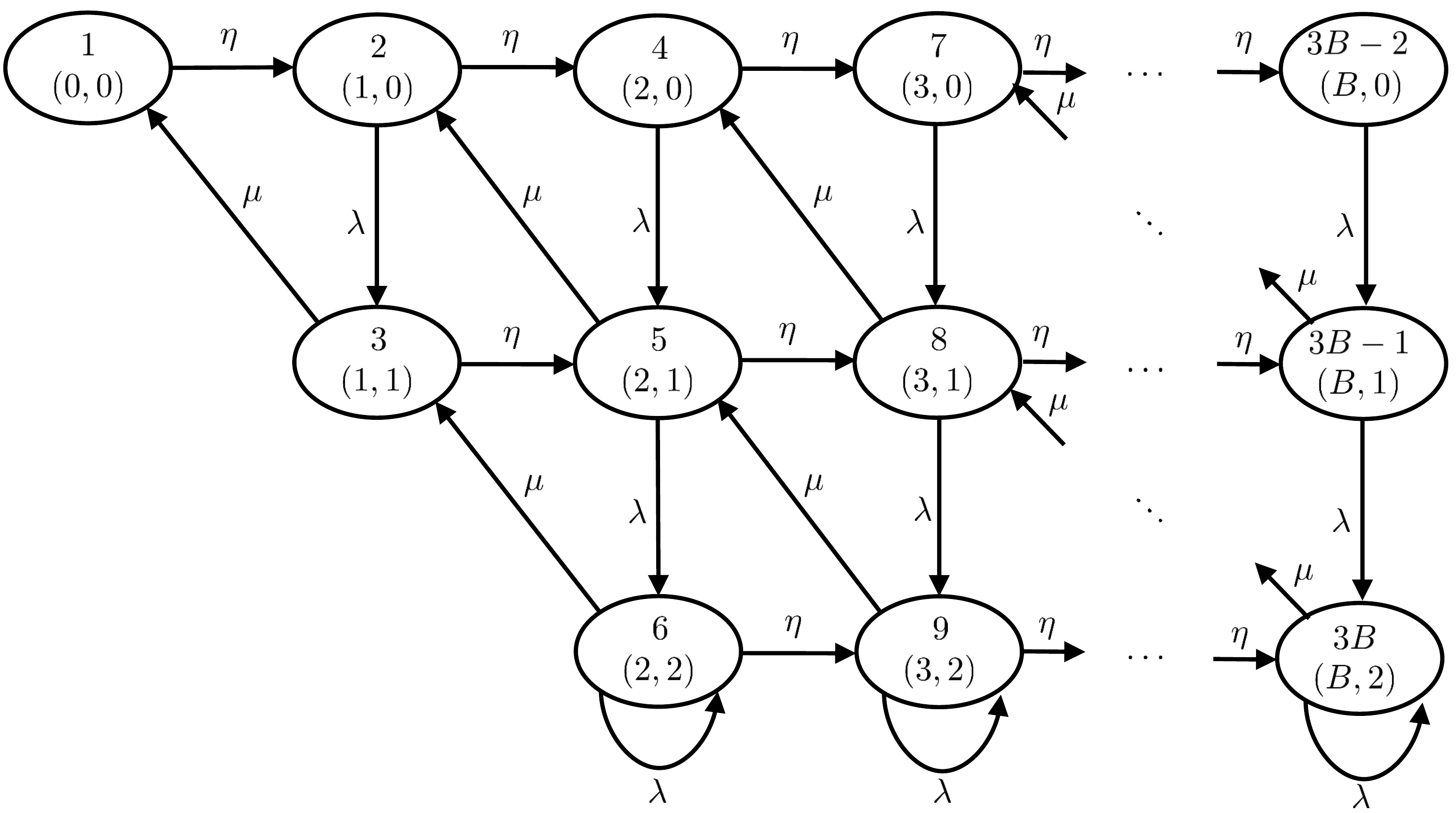}
\caption{The Markov chain modeling the discrete state in the LCFS-PW queueing discipline when the transmitter node can simultaneously harvest energy and serve status updates.}
\label{f:PW_harvall_MC}
\end{figure}

When the transmitter node can simultaneously harvest energy and send status update packets to the destination, the discrete state of the system $q(t)$ in the LCFS-PW queueing discipline is modeled by the Markov chain in Fig. \ref{f:PW_harvall_MC}. The set of transitions $\ncalL$ in this case include the transitions listed in Table \ref{table:PW} along with a new subset of transitions given by Table \ref{table:PW_harvall}. In particular, this new subset of transitions is associated with the arrival of a new energy harvested packet at the transmitter when the system is not empty, i.e., the system either contains one update packet in service (as indicated by $l=6B+2k-8$ in Table \ref{table:PW_harvall}) or two update packets ($l=6B+2k-7$). For this case, the MGF of AoI at the destination is given by the following theorem.
\begin{theorem}\label{MGF_PW_harvall}
When the transmitter node can harvest energy anytime, the MGF of AoI at the destination node for the LCFS-PW queueing discipline is given by (\ref{theorem_MGF_PW_harvall_1}) [at the top of the next page],
\begin{figure*}
\begin{align}\label{theorem_MGF_PW_harvall_1}
\overset{\rm PW}{M}\left(\bar{s}\right) = \frac{\rho \bar{\pi}_1 \left(\overset{\rm PW}{M_1}\left(\bar{s}\right) + \overset{\rm PW}{M_2}\left(\bar{s}\right)\right)}{\left(1 + \rho + 2\beta\right)\left(1 - \bar{s}\right)^{2} \left(1 + \rho - \bar{s}\right)^{2} \left(\rho - \bar{s}\right) \left(\beta - \bar{s}\right)\left(1+\beta-\bar{s}\right)\left(1+\rho+\beta-\bar{s}\right)},
\end{align}
\begin{align*}
\overset{\rm PW}{M_1}\left(\bar{s}\right) = \left(1+\rho+2\beta\right)\left(\beta-\bar{s}\right)\left(1+\beta-\bar{s}\right)\left(1+\rho+\beta-\bar{s}\right)\Big[\bar{s}^{2}-2\bar{s}\left(1+\rho\right)+\rho^2+\rho+1\Big]\Big[\bar{s}\bar{\gamma}_2+\gamma_1-\beta-\bar{\gamma}_2\left(1+\rho\right)\Big],  
\end{align*}
\begin{align*}
\overset{\rm PW}{M_2}\left(\bar{s}\right) = \beta \left(1 + \rho - \bar{s}\right) \Big[\big[\left(1+\beta\right)^2+\rho\big]\left(1+\rho+\beta-\bar{s}\right)+\beta\left(1+\rho+\beta\right)^2\Big]\Big[-\bar{s}\left(3+\rho+\beta\right)+1+\rho+\beta\Big].
\end{align*}
\end{figure*}
\begin{table*}[h!]
\centering
{\caption{Transitions of the LCFS-PW queueing discipline in Fig. \ref{f:PW_harvall_MC} $(3\leq k \leq B)$.} 
\label{table:PW_harvall}
\scalebox{.9}
{ \begin{tabular}{ |c |c|c|c|c|c|c|c|}
\hline
$l$   & $q_l\rightarrow q'_l$  & $\lambda^{(l)}$ & $\nbx \nbA_l$ & $\nbA_l$ & $\hat{\nbA}_l$ & $\bar{\nbv}^s_{q_l} \nbA_l$ & $\bar{\pi}_{q_l} {\bf 1} \hat{\nbA}_l$\\ \hline
 $6B-3$&$3 \rightarrow 5$&$\eta$ &$[x_0,x_1,0]$ & $\begin{bmatrix}1&0&0\\0&1&0\\0&0&0\\ \end{bmatrix}$&$\begin{bmatrix}0&0&0\\0&0&0\\0&0&1\\ \end{bmatrix}$ &$[\bar{v}^s_{30},\bar{v}^s_{31},0]$ & $[0,0,\bar{\pi}_{3}]$ \\ \hline
 $6B+2k-8$&$3k-4 \rightarrow 3k-1$&$\eta$ &$[x_0,x_1,0]$ & $\begin{bmatrix}1&0&0\\0&1&0\\0&0&0\\ \end{bmatrix}$&$\begin{bmatrix}0&0&0\\0&0&0\\0&0&1\\ \end{bmatrix}$ &$[\bar{v}^s_{3k-4,0},\bar{v}^s_{3k-4,1},0]$ & $[0,0,\bar{\pi}_{3k-4}]$ \\ \hline
 $6B+2k-7$&$3k-3 \rightarrow 3k$&$\eta$ &$[x_0,x_1,x_2]$ & $\begin{bmatrix}1&0&0\\0&1&0\\0&0&1\\ \end{bmatrix}$&$\begin{bmatrix}0&0&0\\0&0&0\\0&0&0\\ \end{bmatrix}$ &$[\bar{v}^s_{3k-3,0},\bar{v}^s_{3k-3,1},\bar{v}^s_{3k-3,2}]$ & $[0,0,0]$ \\ \hline
\end{tabular}}} 
\end{table*} 
where $\bar{\gamma}_2 = 1 - \gamma_2$, $\underset{k \in \;{\rm r}_2 \;\cup\;{\rm r}_3}{\sum}{\bar{\pi}_k} = \gamma_1 \bar{\pi}_1$, and $\underset{k \in \; {\rm r}_1}{\sum}{\bar{\pi}_k} = \gamma_2 \bar{\pi}_1$.
\end{theorem}
\begin{IEEEproof}
See Appendix \ref{app:MGF_PW_harvall}.
\end{IEEEproof}
\begin{cor}\label{cor:PW_harvall}
Given that $B=2$ and the transmitter can harvest energy anytime, the first and second moments of AoI for the LCFS-PW queueing discipline can be respectively expressed as 
\begin{align}\label{mom1_PW_harvall_B2}
\overset{(1)}{\Delta}_{\rm PW} \nonumber&= \dfrac{1}{\mu\rho\beta\left(1+\rho\right)^2\left(1+\beta\right)^2} \\& \times\dfrac{\sum_{n=1}^{5}{\beta^{n}\alpha^*_n} + \rho^3 \big(1+\rho\big)^2}{\Big[\beta^2\left(\rho^2+\rho+1\right) + \beta\rho\left(1+\rho\right)+\rho^2\big]},
\end{align}
\begin{align*}
&\alpha^*_5 = 2\rho^5+7\rho^4+8\rho^3+7\rho^2+4\rho+1,\; \alpha^*_4 = 6\rho^5+19\rho^4+\\&23\rho^3+18\rho^2+9\rho+2,\; \alpha^*_3 = \big(1+\rho\big)\big(7\rho^4+15\rho^3+11\rho^2+\\&5\rho+1\big),\;\alpha^*_2 = \rho \big(1+\rho\big)\big(6\rho^3+11\rho^2+5\rho+1\big),\\& \alpha^*_1 = \rho^2\big(4\rho^3+9\rho^2+6\rho+1\big).
\end{align*}
\begin{align}\label{mom2_PW_harvall_B2}
\overset{(2)}{\Delta}_{\rm PW} \nonumber&= \dfrac{1}{\mu^2\rho^2\beta^2\left(1+\rho\right)^3\left(1+\beta\right)^3} \\& \times \dfrac{2\sum_{n=0}^{7}{\beta^{n}\zeta^*_n}}{\Big[\beta^2\left(\rho^2+\rho+1\right) + \beta\rho\left(1+\rho\right)+\rho^2\big]},
\end{align}
\begin{align*}
&\zeta^*_7 =3\rho^7+14\rho^6+24\rho^5+21\rho^4+18\rho^3+12\rho^2+5\rho+1,\\& \zeta^*_6 = 12\rho^7+52\rho^6+87\rho^5+81\rho^4+66\rho^3+41\rho^2+16\rho+3, \\&
\zeta^*_5 = 19\rho^7+78\rho^6+132\rho^5+129\rho^4+95\rho^3+52\rho^2+18\rho+3,\\& \zeta^*_4 = \big(1+\rho\big)\big(17\rho^6+54\rho^5+69\rho^4+47\rho^3+23\rho^2+7\rho+1\big),
 \\&\zeta^*_3=\rho\big(1+\rho\big)\big(14\rho^5+43\rho^4+47\rho^3+23\rho^2+7\rho+1\big),\\&\zeta^*_2 = \rho^2\big(1+\rho\big)^2\big(11\rho^3+17\rho^2+6\rho+1\big),\\&\zeta^*_1= \rho^3 \big(1+\rho\big)^3\big(5\rho+1\big),\;\zeta^*_0 = \rho^4\big(1+\rho\big)^3.
\end{align*}
\end{cor}
\begin{IEEEproof}
The result follows from (\ref{MGF_derav}) with noting that $\gamma_1 = \frac{\beta}{\rho}\big[\beta + \rho \left(1 + \beta\right)\big]$ and $\gamma_2 = \frac{\beta^2+\beta\rho+\rho^2}{\rho^2}$.
\end{IEEEproof}
\begin{remark}\label{rem:6}
When $\beta \rightarrow \infty$, $\overset{(1)}{\Delta}_{\rm PW}$ and $\overset{(2)}{\Delta}_{\rm PW}$ (in (\ref{mom1_PW_harvall_B2}) and (\ref{mom2_PW_harvall_B2})) reduce to
\begin{align}
\nonumber&\underset{\beta \rightarrow \infty}{\rm lim} \overset{(1)}{\Delta}_{\rm PW} = \dfrac{2\rho^5+7\rho^4+8\rho^3+7\rho^2+4\rho+1}{\mu\rho\left(1+\rho\right)^2\left(\rho^2+\rho+1\right)},\\
&\underset{\beta \rightarrow \infty}{\rm lim} 
\overset{(2)}{\Delta}_{\rm PW} \nonumber= \dfrac{2}{\mu^2\rho^2\left(1+\rho\right)^3\left(\rho^2+\rho+1\right)} \times \\&\left(3\rho^7+14\rho^6+24\rho^5+21\rho^4+18\rho^3+12\rho^2+5\rho+1\right),
\end{align}
which match the first and second moments of AoI for the LCFC-PW queueing discipline in \cite[Theorem~2(b)]{yates2018age}, where the transmitter node does not have energy limitations. 
\end{remark}
\begin{remark}\label{rem:comp_PWandPS_harvall}
When the transmitter node can simultaneously harvest energy and serve status updates, it can be verified from Corollaries \ref{cor:PS_harvall} and \ref{cor:PW_harvall} for $B = 2$ that the LCFS-PS queueing discipline yields smaller values of the first and second moments of AoI at the destination node than the LCFS-PW one. In particular, $\overset{(1)}{\Delta}_{\rm PW} - \overset{(1)}{\Delta}_{\rm PS}$ monotonically increases with $\rho$ from $\underset{\rho \rightarrow 0}{\rm lim} \overset{(1)}{\Delta}_{\rm PW} - \overset{(1)}{\Delta}_{\rm PS} = 0$ until it approaches $\underset{\rho \rightarrow \infty}{\rm lim} \overset{(1)}{\Delta}_{\rm PW} - \overset{(1)}{\Delta}_{\rm PS} = \dfrac{\beta^3 + 2\beta^2 + \beta + 1}{\mu\left(\beta^3 + 2\beta^2 + 2 \beta + 1\right)}$. On the other hand, $\overset{(2)}{\Delta}_{\rm PW} - \overset{(2)}{\Delta}_{\rm PS}$ monotonically increases as a function of $\rho$ from $\underset{\rho \rightarrow 0}{\rm lim} \overset{(2)}{\Delta}_{\rm PW} - \overset{(2)}{\Delta}_{\rm PS} = 0$ until it approaches $\underset{\rho \rightarrow \infty}{\rm lim} \overset{(2)}{\Delta}_{\rm PW} - \overset{(2)}{\Delta}_{\rm PS} = \dfrac{4\beta^5 + 12 \beta^4 + 12 \beta^3 + 6 \beta^2 + 6 \beta + 2}{\mu^2\left(\beta^5 + 3 \beta^4 + 4 \beta^3 + 3 \beta^2 + \beta \right)}$.
\end{remark}
\begin{remark}\label{rem:comp_WPandPW_harvall}
From Remarks \ref{rem:comp_WPandPS_harvall} and \ref{rem:comp_PWandPS_harvall}, we obtain the following asymptotic results (when the transmitter node can harvest energy anytime and $B = 2$) regarding the comparison between the achievable AoI performances by the LCFS-NP and LCFS-PW queueing disciplines: $\underset{\rho \rightarrow 0}{\rm lim} \overset{(1)}{\Delta}_{\rm NP} - \overset{(1)}{\Delta}_{\rm PW} = \underset{\rho \rightarrow \infty}{\rm lim} \overset{(1)}{\Delta}_{\rm NP} - \overset{(1)}{\Delta}_{\rm PW}  = \underset{\rho \rightarrow \infty}{\rm lim} \overset{(2)}{\Delta}_{\rm NP} - \overset{(2)}{\Delta}_{\rm PW} = 0$, and $ \underset{\rho \rightarrow 0}{\rm lim} \overset{(2)}{\Delta}_{\rm NP} - \overset{(2)}{\Delta}_{\rm PW} = \frac{2}{\mu^2}$. We will also show in Section \ref{sec:numerical} that the insight obtained by \cite{costa2016age} indicating the superiority of the LCFS-PW queuing discipline over the LCFS-NP (when the transmitter does not have energy limitations), only holds for a certain range of values of $\beta$ when the transmitter is subject to energy constraints.
\end{remark}
The insights obtained in Remarks \ref{rem:comp_WPandPS}, \ref{rem:comp_WPandPS_harvall}, \ref{rem:comp_PWandPS_harvall} and \ref{rem:comp_WPandPW_harvall} are quite useful from an engineering perspective since they allow one to: i) quantify improvement of one queueing discipline over another in terms of the achievable AoI performance, ii) specify the ranges of system parameter values over which a certain queueing discipline achieves a better AoI performance than another queueing discipline, and iii) characterize the asymptotic behavior of the differences between the first/second AoI expressions associated with different queueing disciplines.
\begin{figure*}[t!]
\centerline{
\subfloat[]{\includegraphics[width=0.35\textwidth]{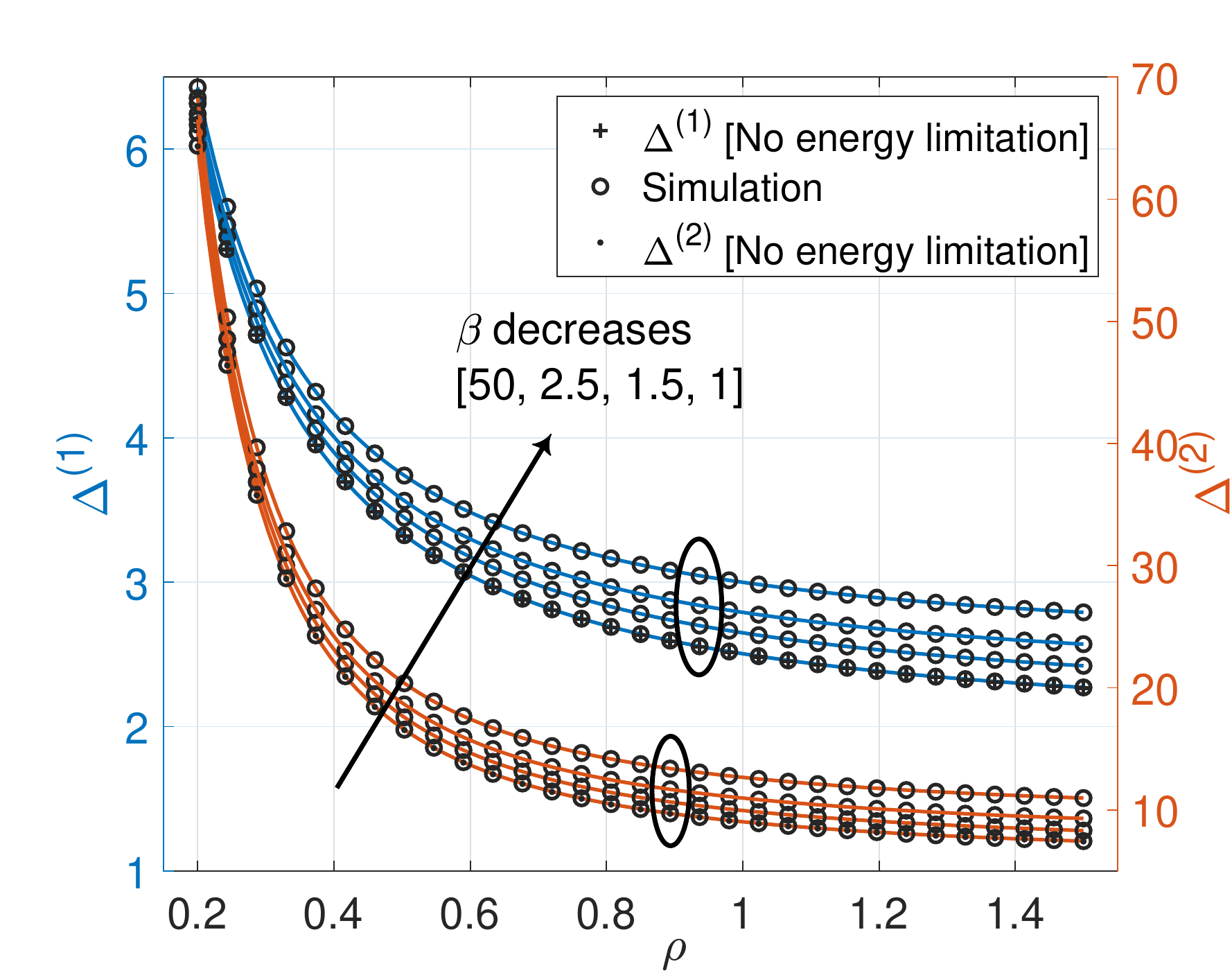}%
\label{f:B1_without_ver}} \hfil
\subfloat[]{\includegraphics[width=0.35\textwidth]{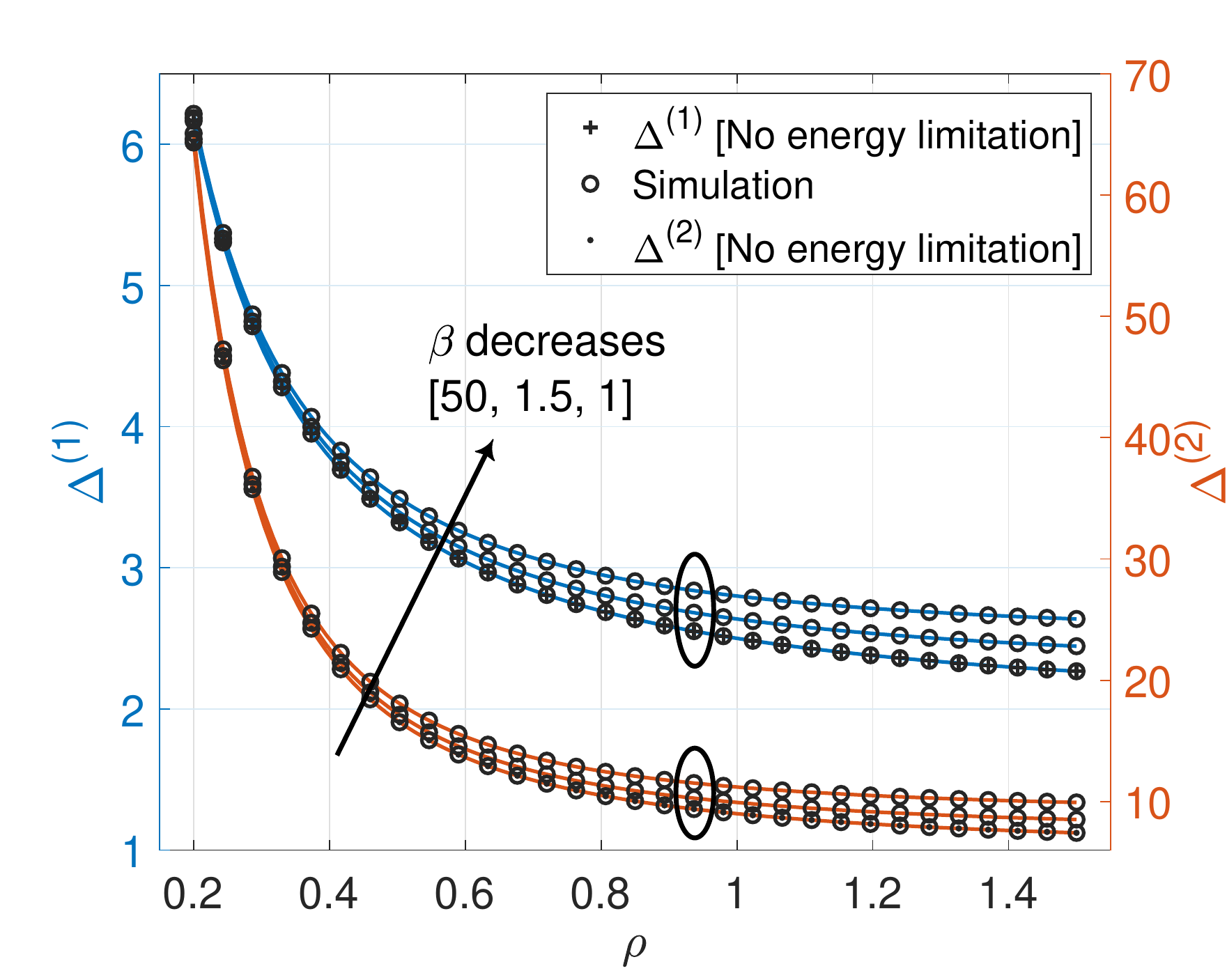}%
\label{f:B2_without_ver}} \hfil
\subfloat[]{\includegraphics[width=0.337\textwidth]{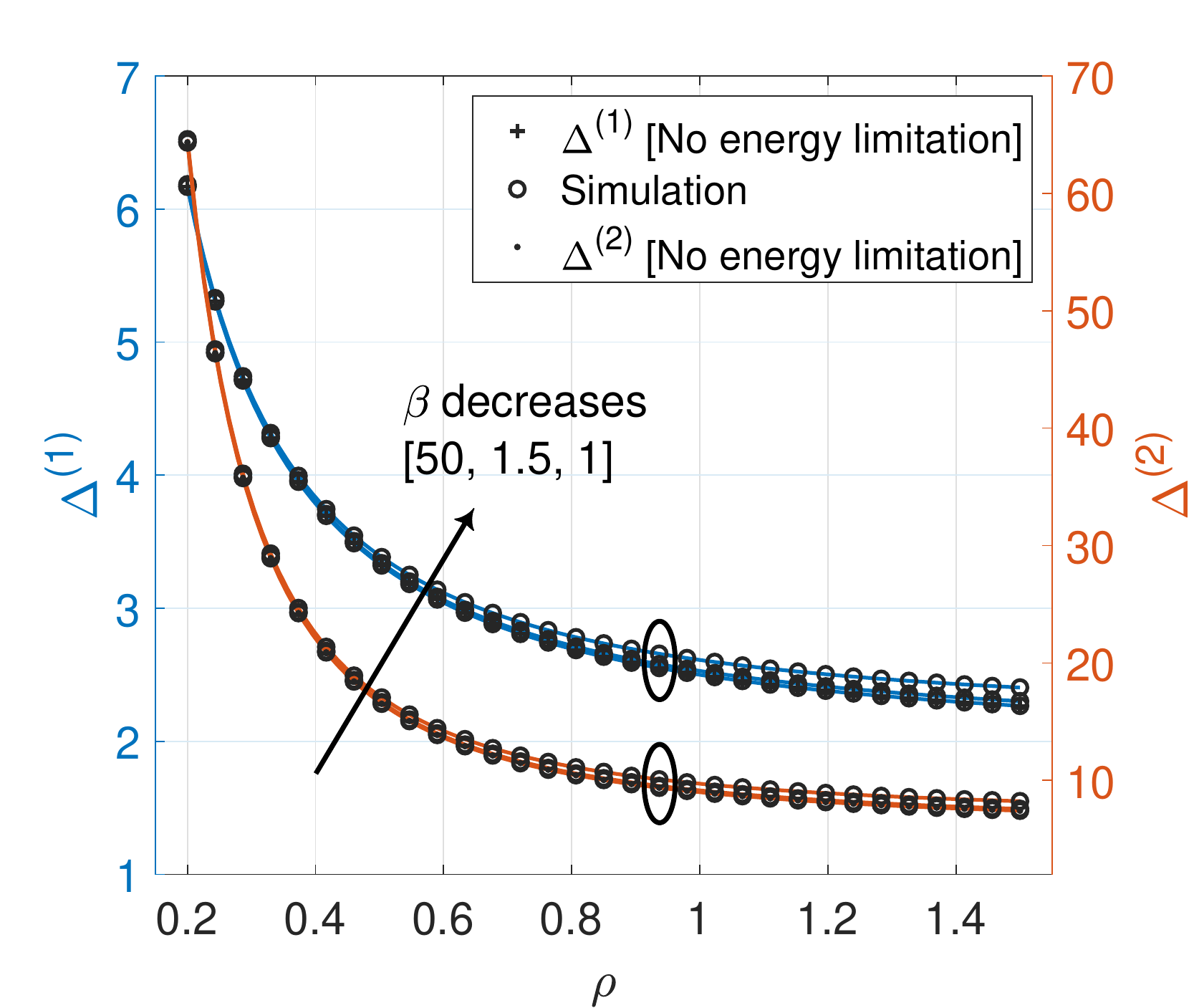}%
\label{f:B2_without_harvall_ver}}} \vfil
\centerline{
\subfloat[]{\includegraphics[width=0.35\textwidth]{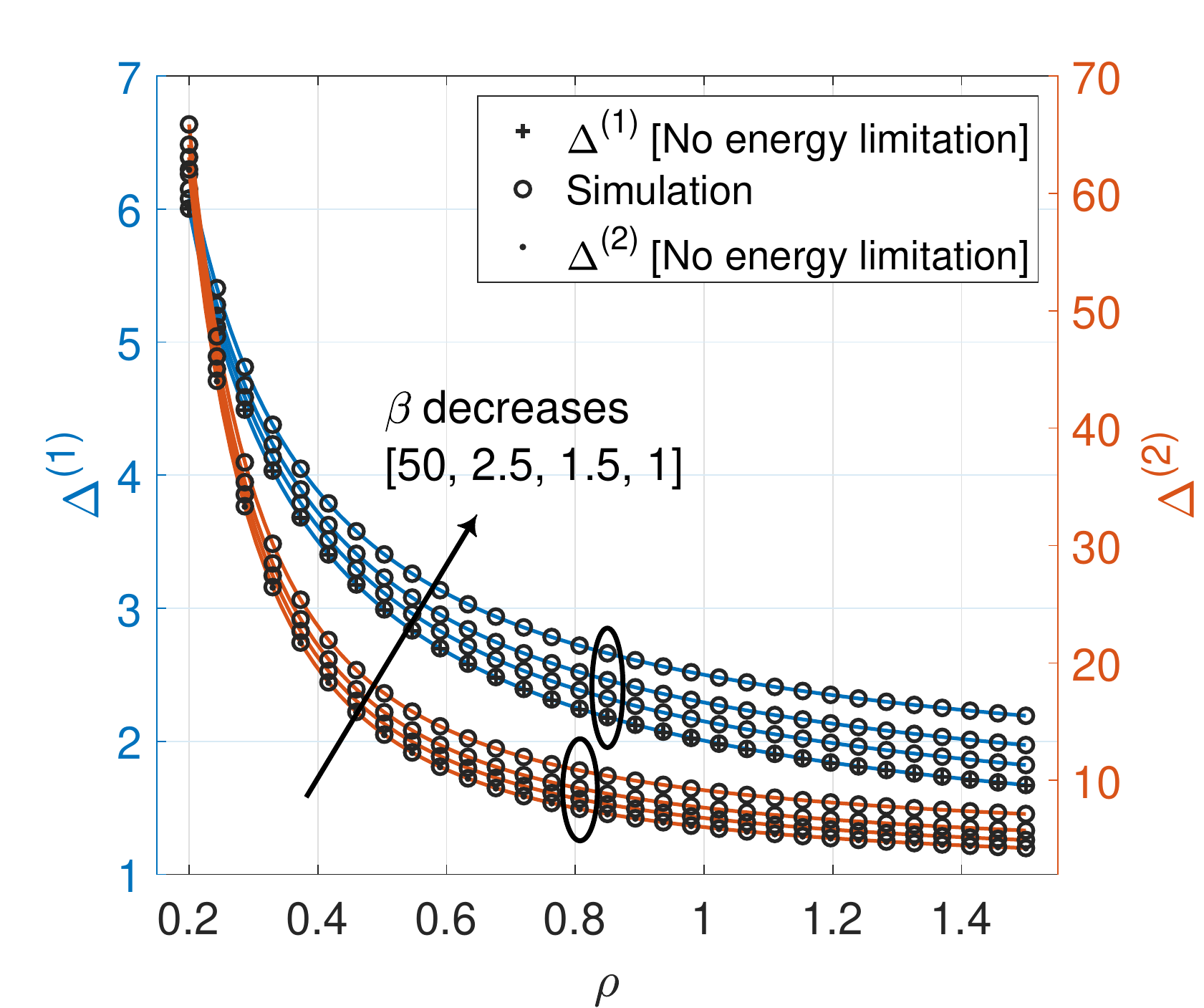}%
\label{f:B1_with_ver}} \hfil
\subfloat[]{\includegraphics[width=0.35\textwidth]{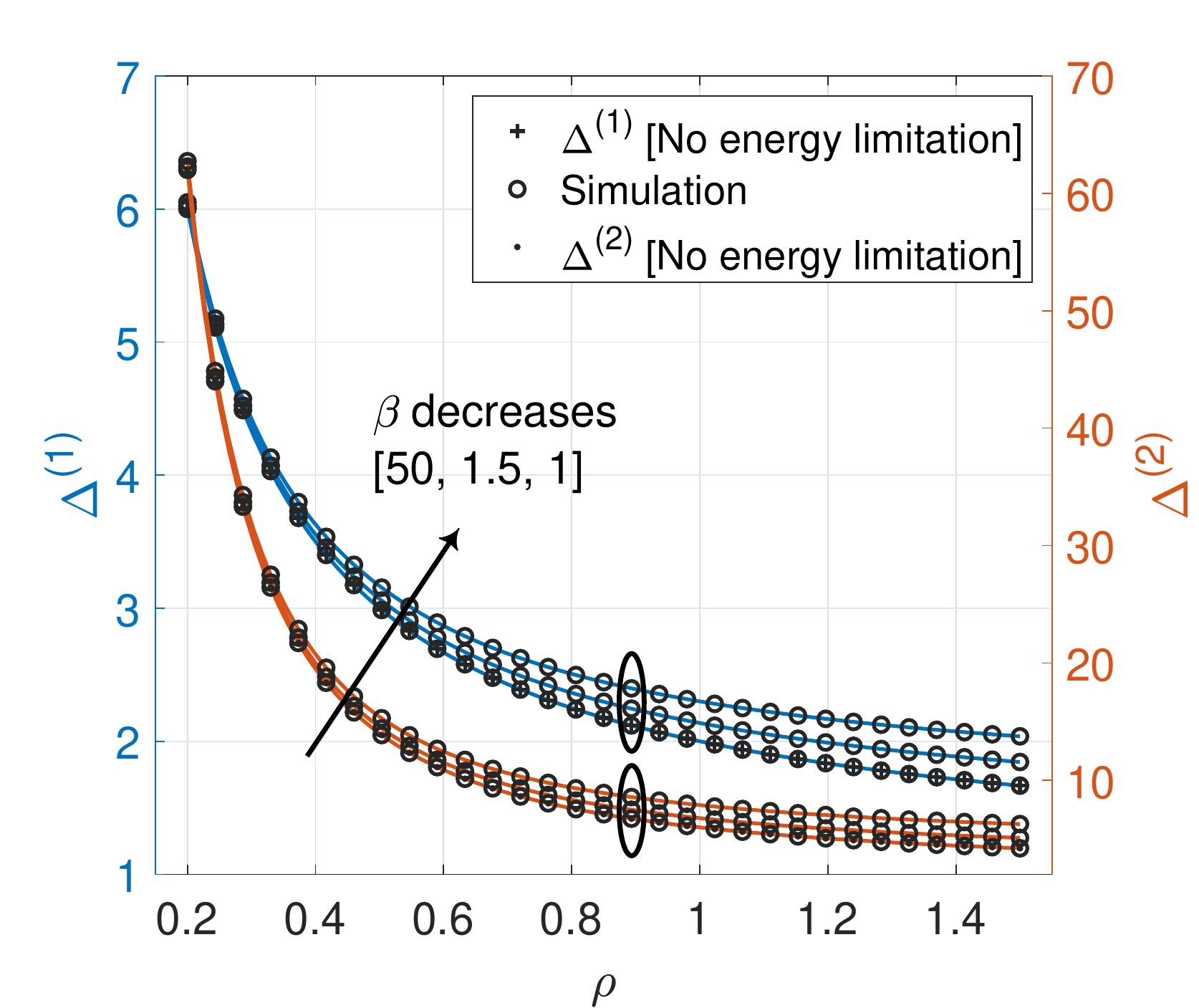}%
\label{f:B2_with_ver}} \hfil
\subfloat[]{\includegraphics[width=0.35\textwidth]{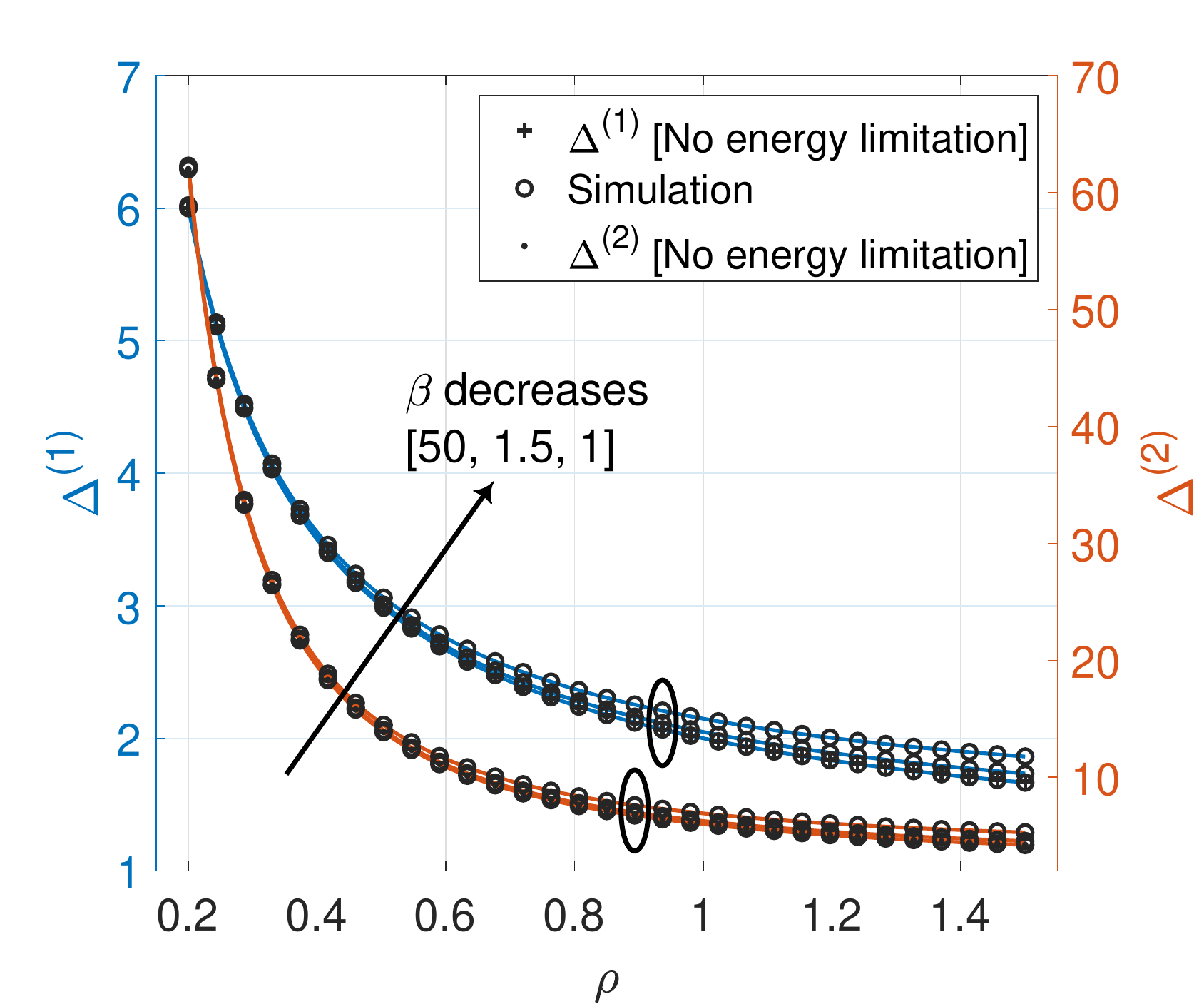}%
\label{f:B2_with_harvall_ver}}
} \caption{Verification of the analytical results derived for the LCFS-NP and LCFS-PS queueing disciplines. For the LCFS-NP [LCFS-PS] queueing discipline, we use: i) $B = 1$ in (a) [(d)], ii) $B = 2$ in (b) [(e)] and the transmitter can only harvest energy if the system is empty, and iii) $B = 2$ in (c) [(f)] and the transmitter can simultaneously harvest energy and serve status updates. }\label{f:without_with_verification}
\end{figure*} 

\begin{figure*}[t!]
\centerline{
\subfloat[]{\includegraphics[width=0.35\textwidth]{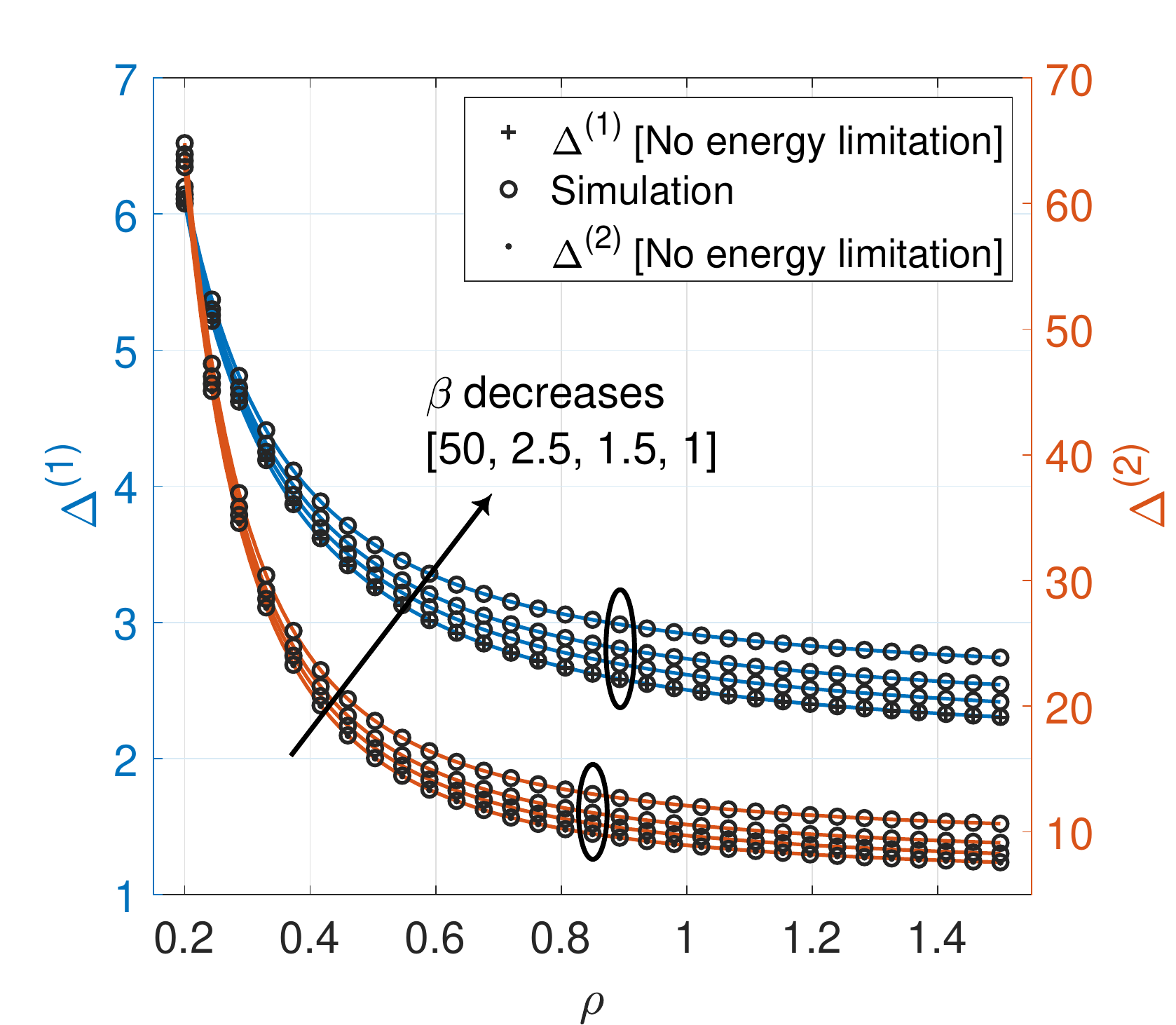}%
\label{f:B2_waiting_ver}} \hfil
\subfloat[]{\includegraphics[width=0.35\textwidth]{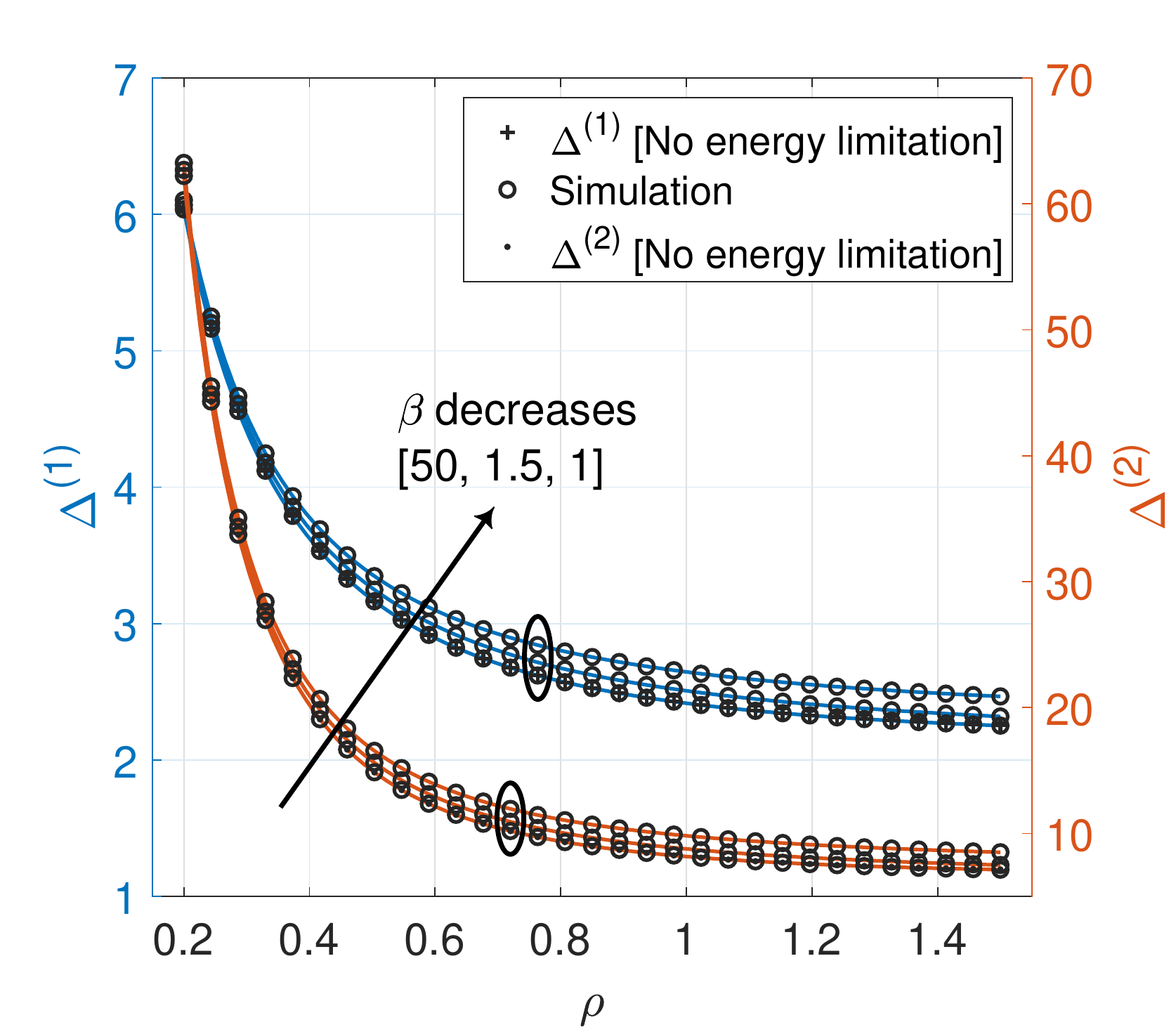}%
\label{f:B2_waiting_harvall_ver}} \hfil
\subfloat[]{\includegraphics[width=0.36\textwidth]{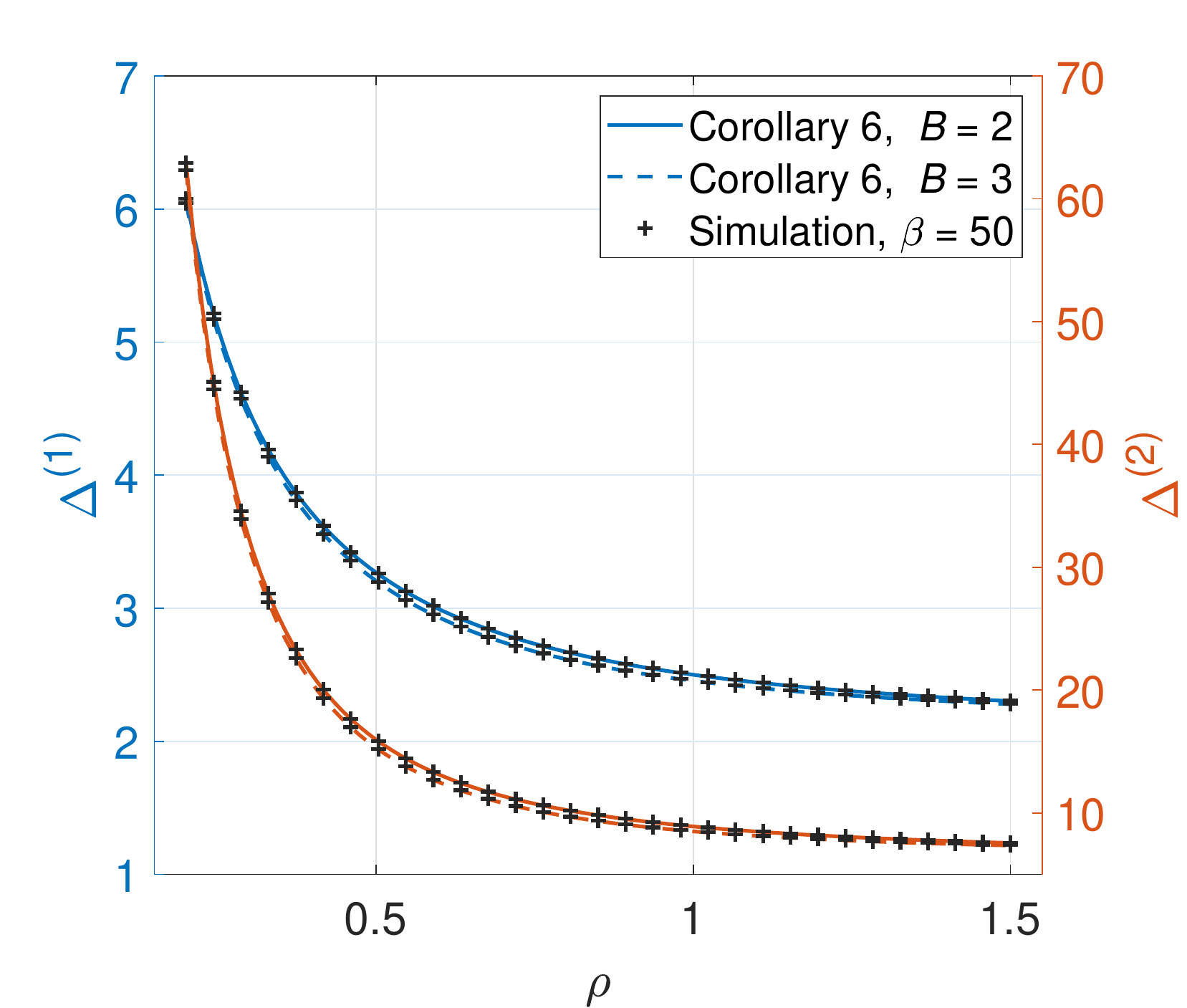}%
\label{f:limit_waiting_ver}}}  \caption{Verification of the analytical results derived for the LCFS-PW queueing discipline. For $B=2$, the transmitter node can only harvest energy if the system is empty in (a) whereas it can simultaneously harvest energy and serve status updates in (b).}\label{f:waiting_verification}
\end{figure*} 

\begin{figure*}[t!]
\centerline{
\subfloat[]{\includegraphics[width=0.35\textwidth]{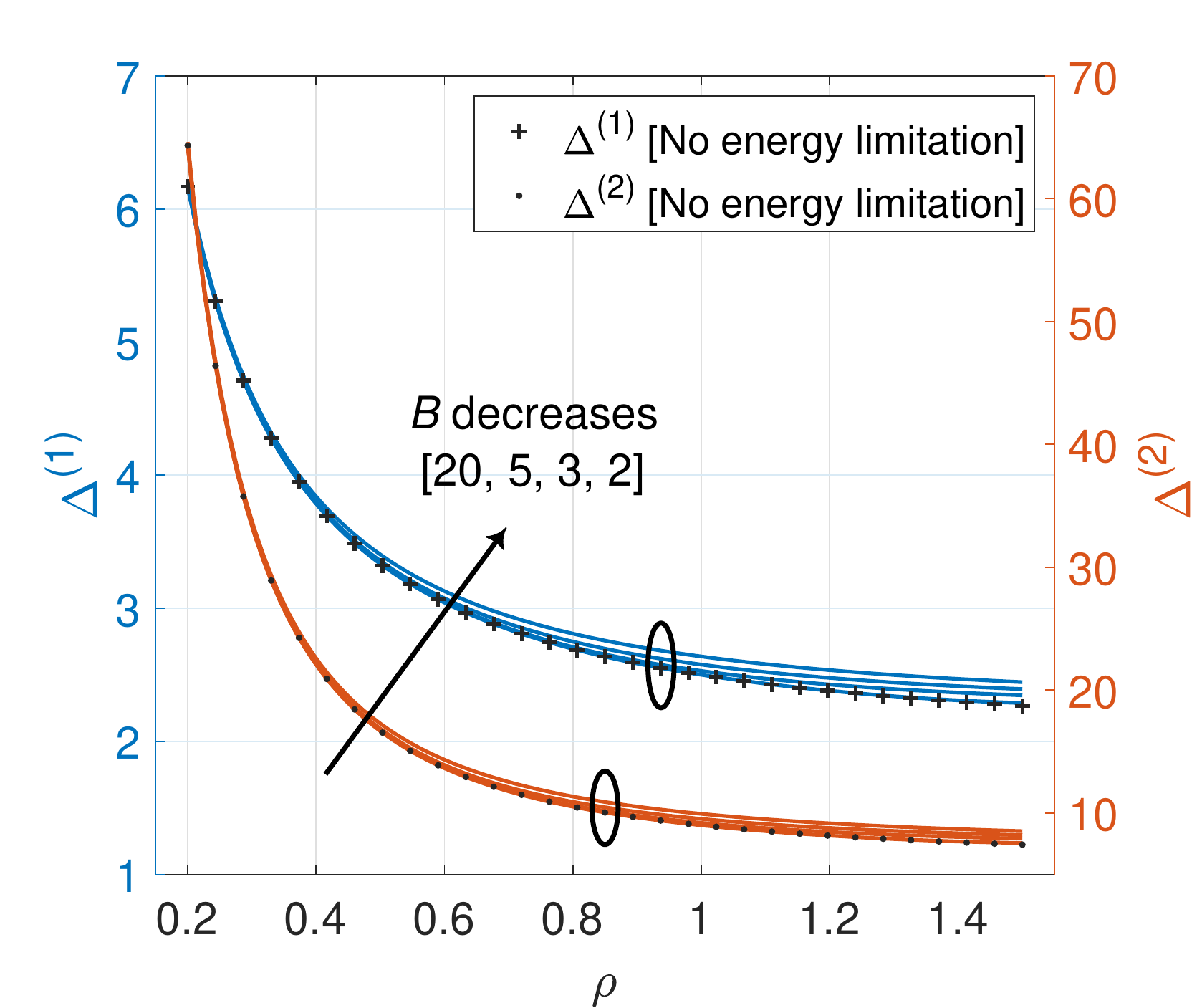}%
\label{f:Bimpact_without}} \hfil
\subfloat[]{\includegraphics[width=0.35\textwidth]{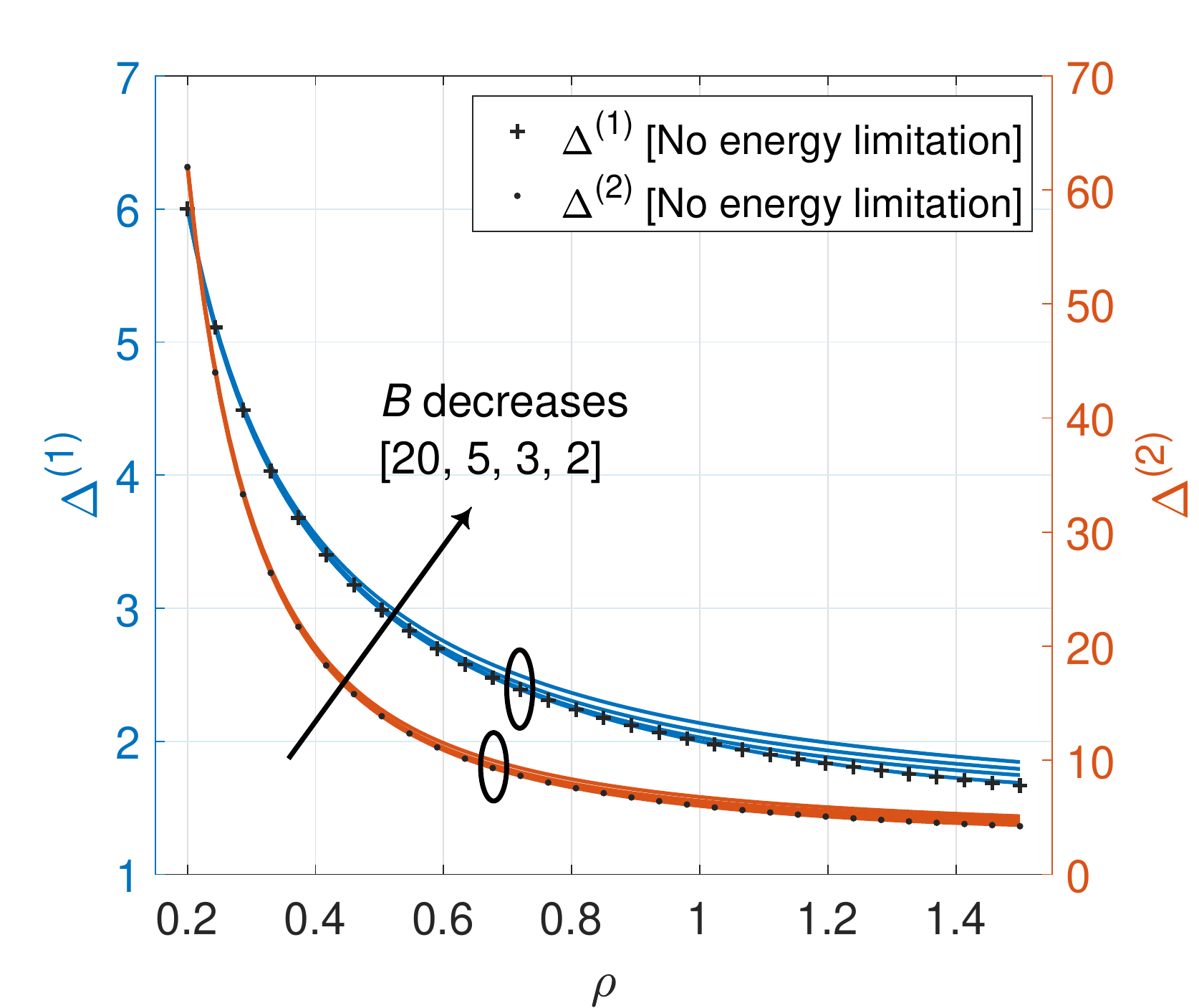}%
\label{f:Bimpact_with}} \hfil
\subfloat[]{\includegraphics[width=0.35\textwidth]{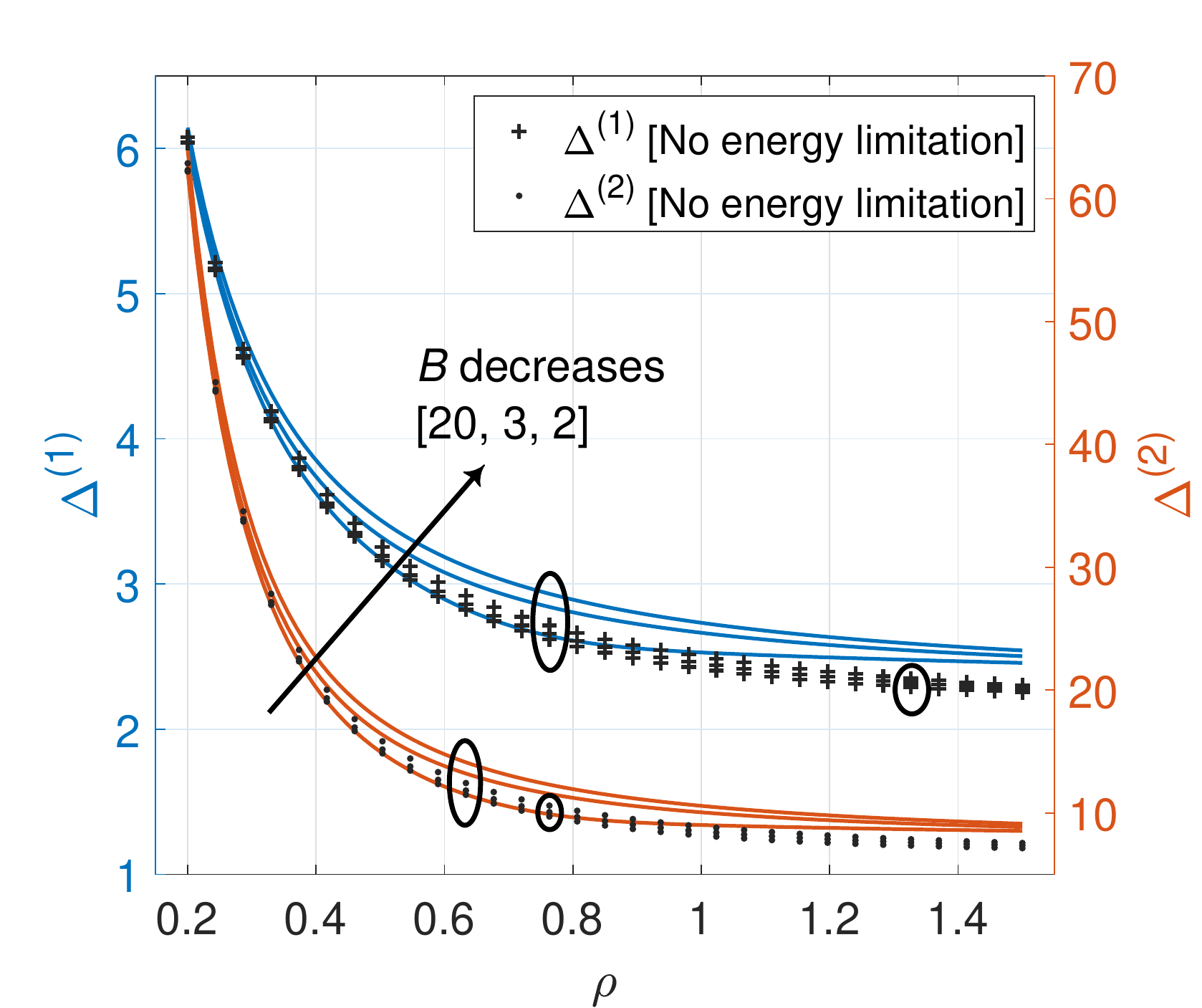}%
\label{f:Bimpact_waiting}}}  \caption{Impact of the size of battery $B$ on $\Delta^{(1)}$ and $\Delta^{(2)}$ when the transmitter node can only harvest energy if the system is empty: (a) the LCFS-NP queueing discipline, (b)  the LCFS-PS queueing discipline, and (c) the LCFS-PW queueing discipline. We use $\beta = 1.5$.}\label{f:B_impact}
\end{figure*} 

\begin{figure*}[t!]
\centerline{
\subfloat[]{\includegraphics[width=0.35\textwidth]{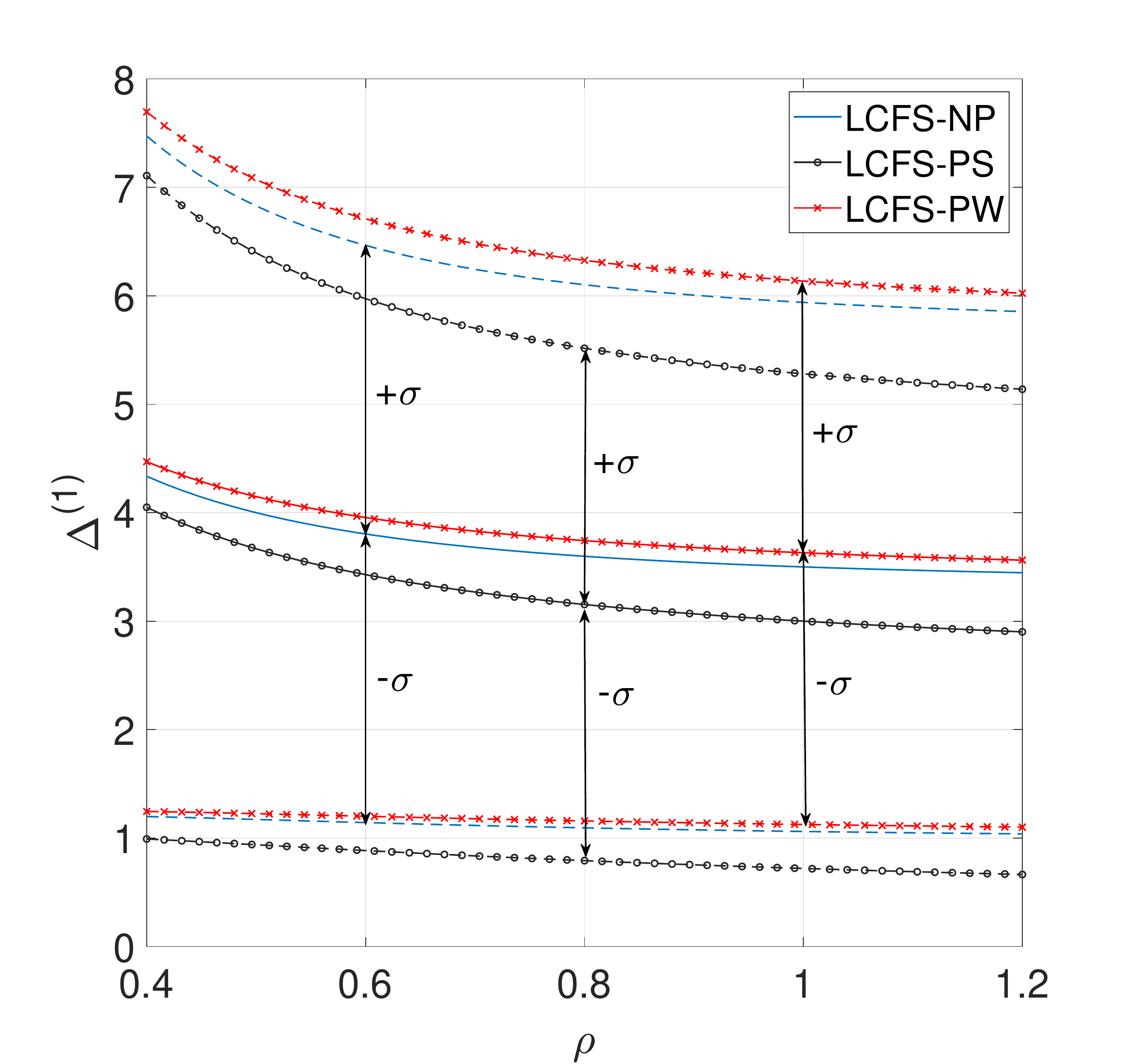}%
\label{f:comp_EH_empty_5}} \hfil
\subfloat[]{\includegraphics[width=0.35\textwidth]{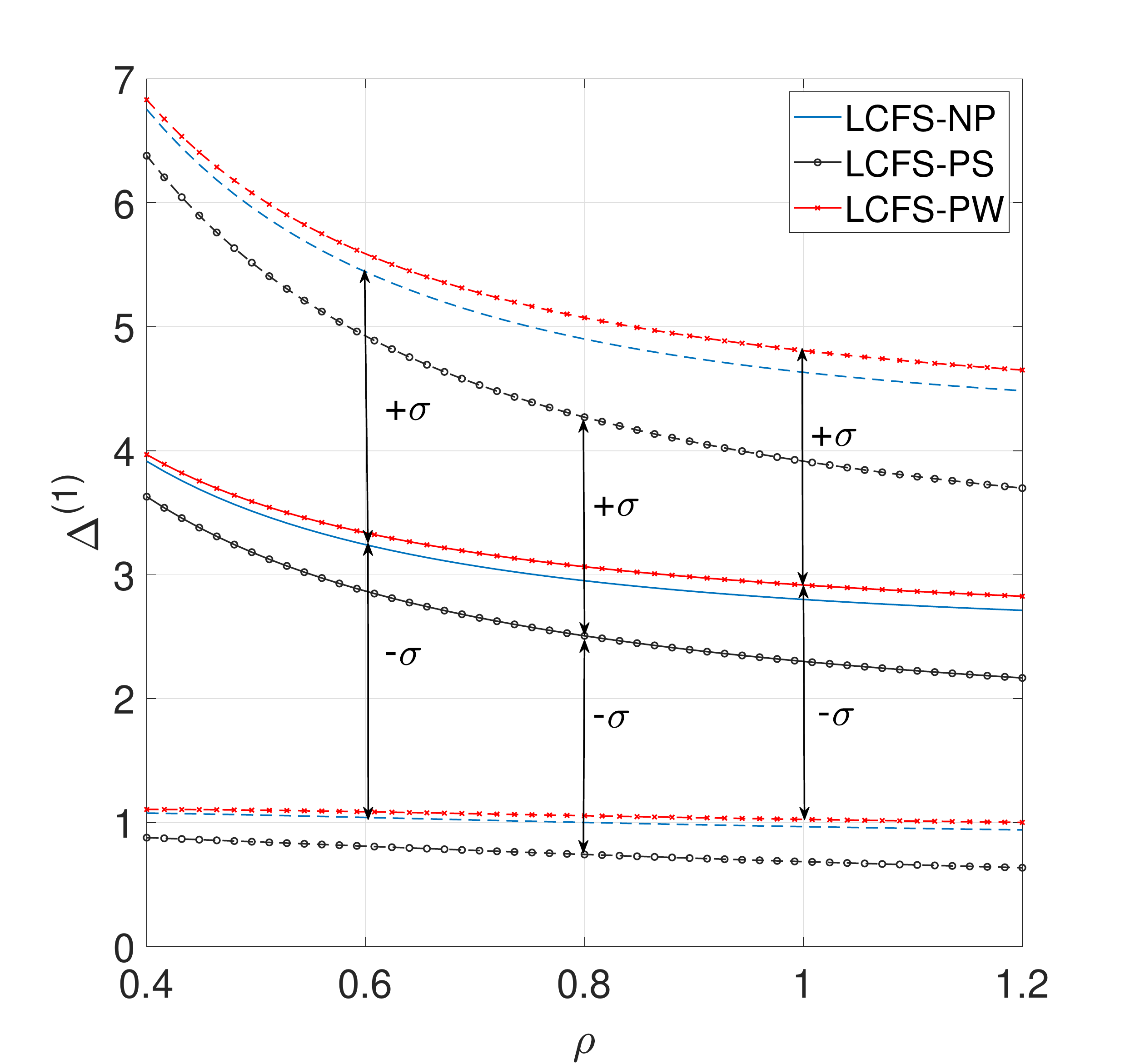}%
\label{f:comp_EH_empty_1}} \hfil
\subfloat[]{\includegraphics[width=0.35\textwidth]{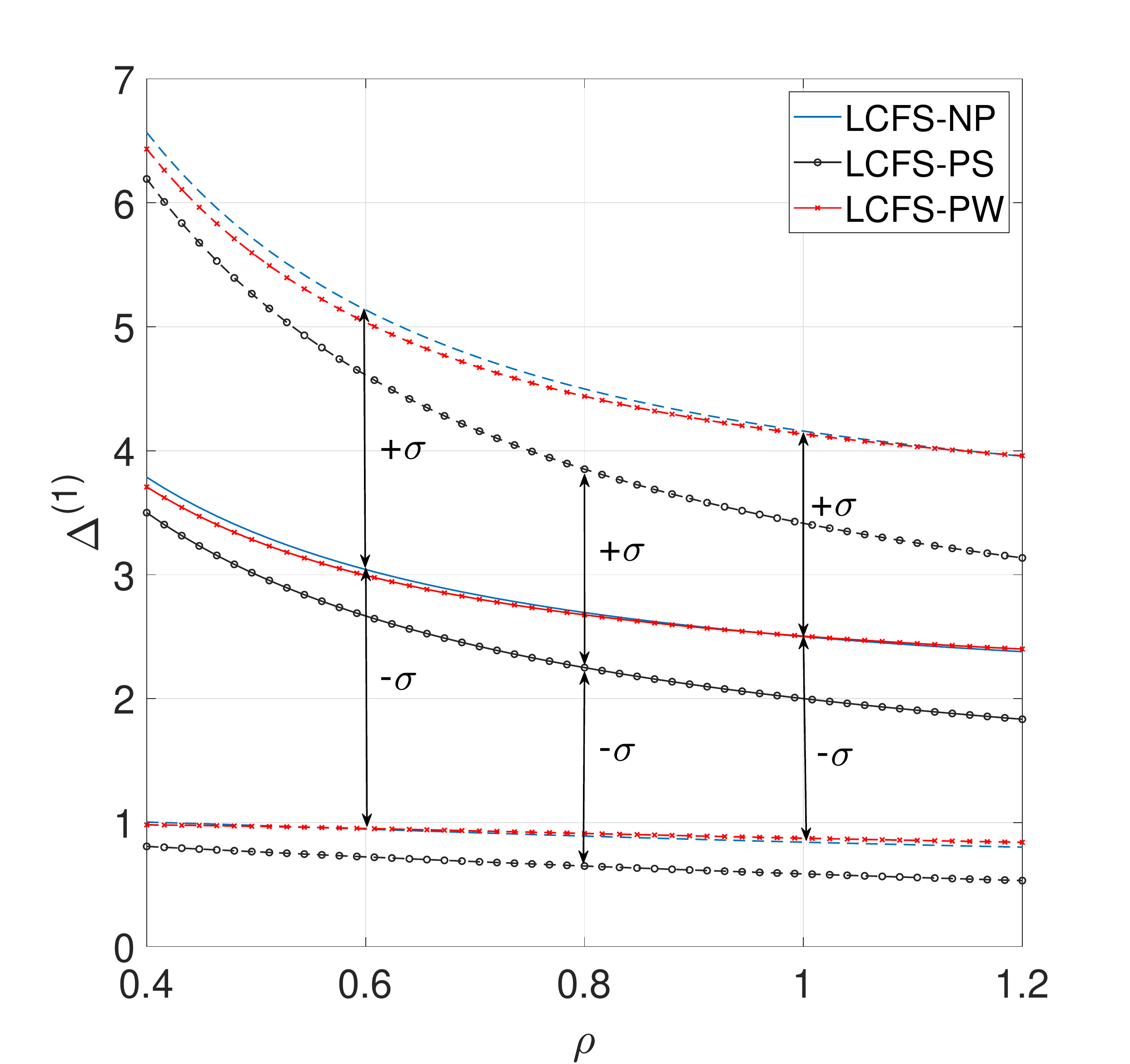}%
\label{f:comp_EH_empty_50}}} \vfil
\centerline{
\subfloat[]{\includegraphics[width=0.35\textwidth]{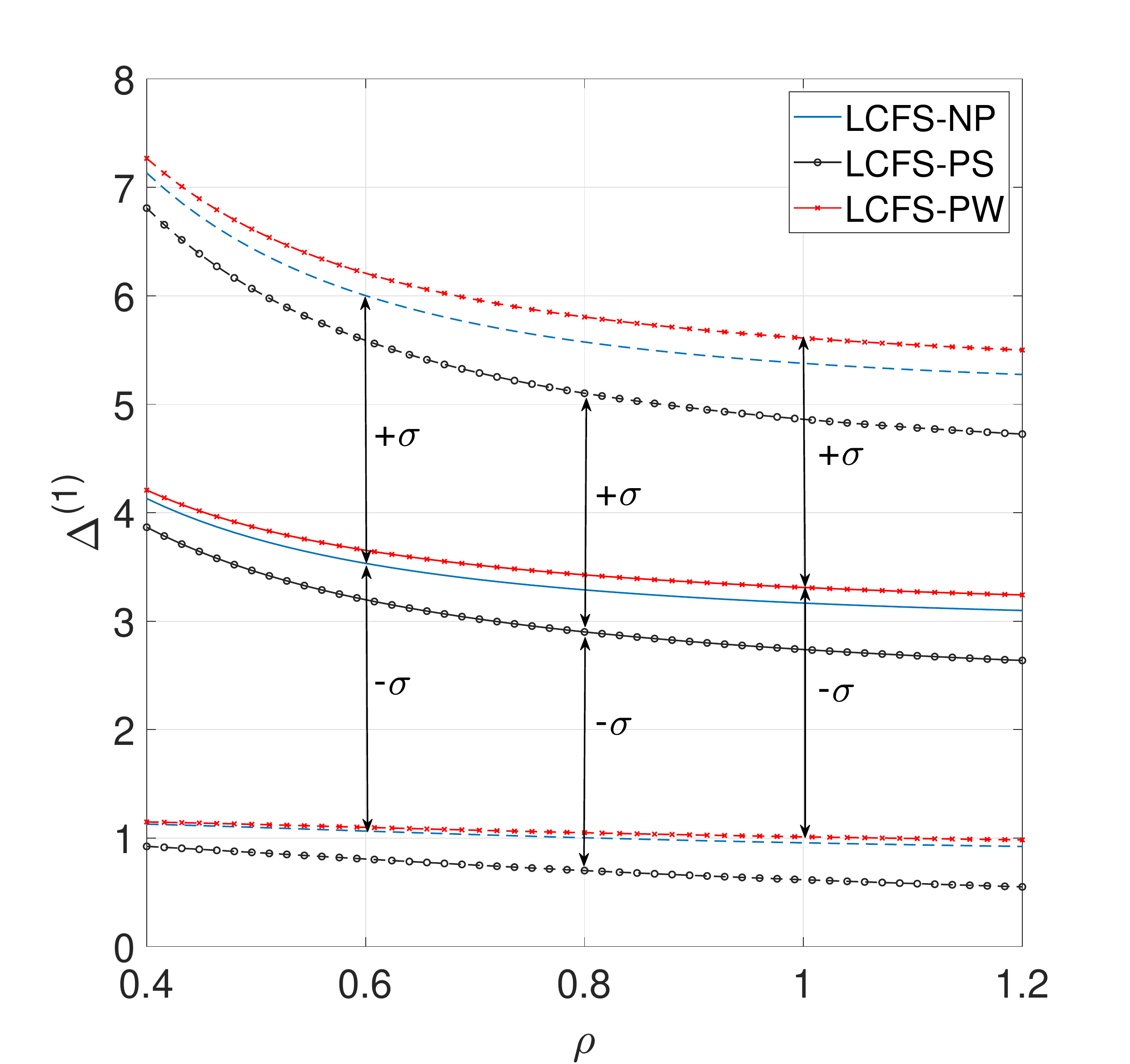}%
\label{f:comp_EH_any_5}} \hfil
\subfloat[]{\includegraphics[width=0.35\textwidth]{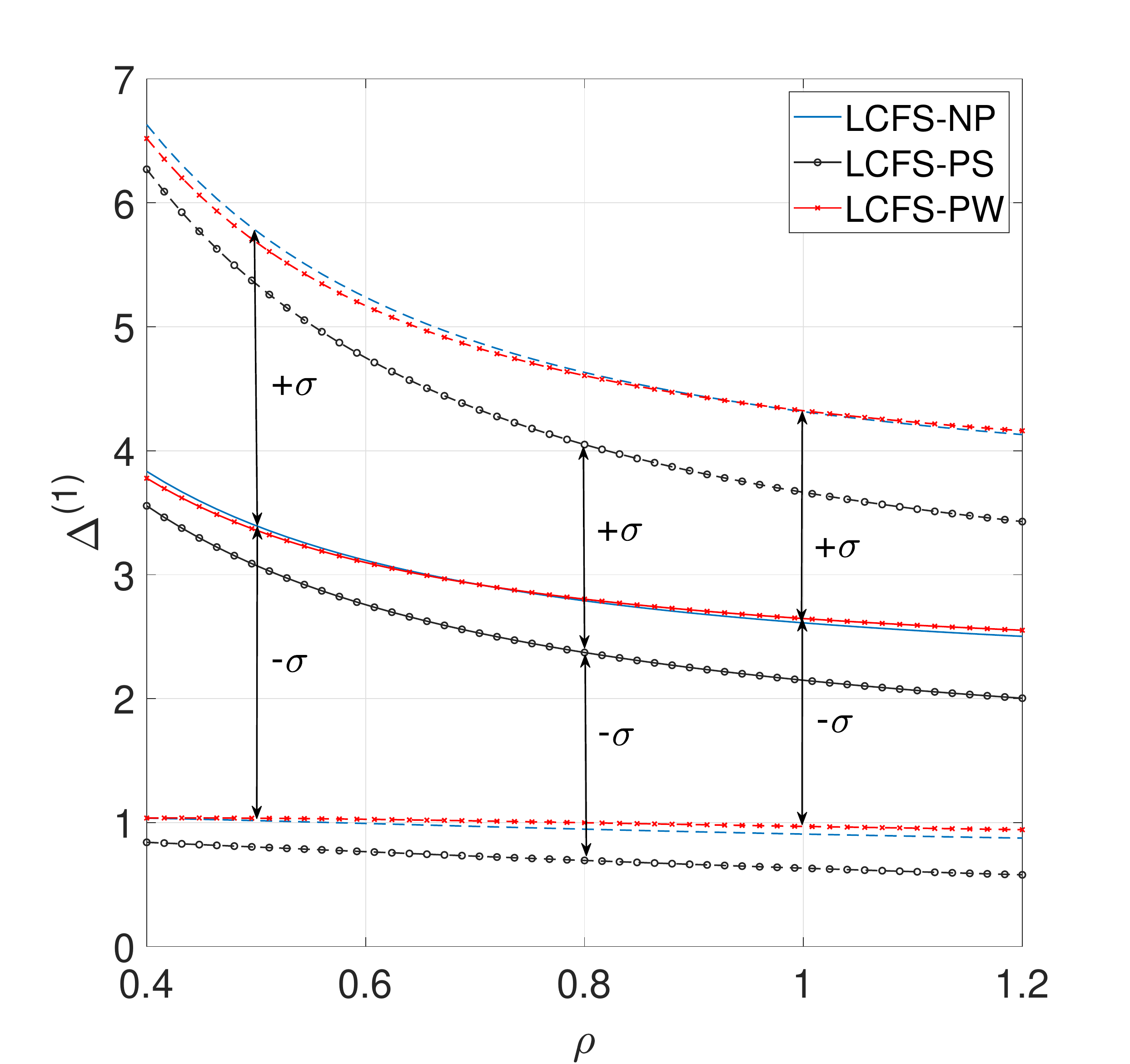}%
\label{f:comp_EH_any_1}} \hfil
\subfloat[]{\includegraphics[width=0.35\textwidth]{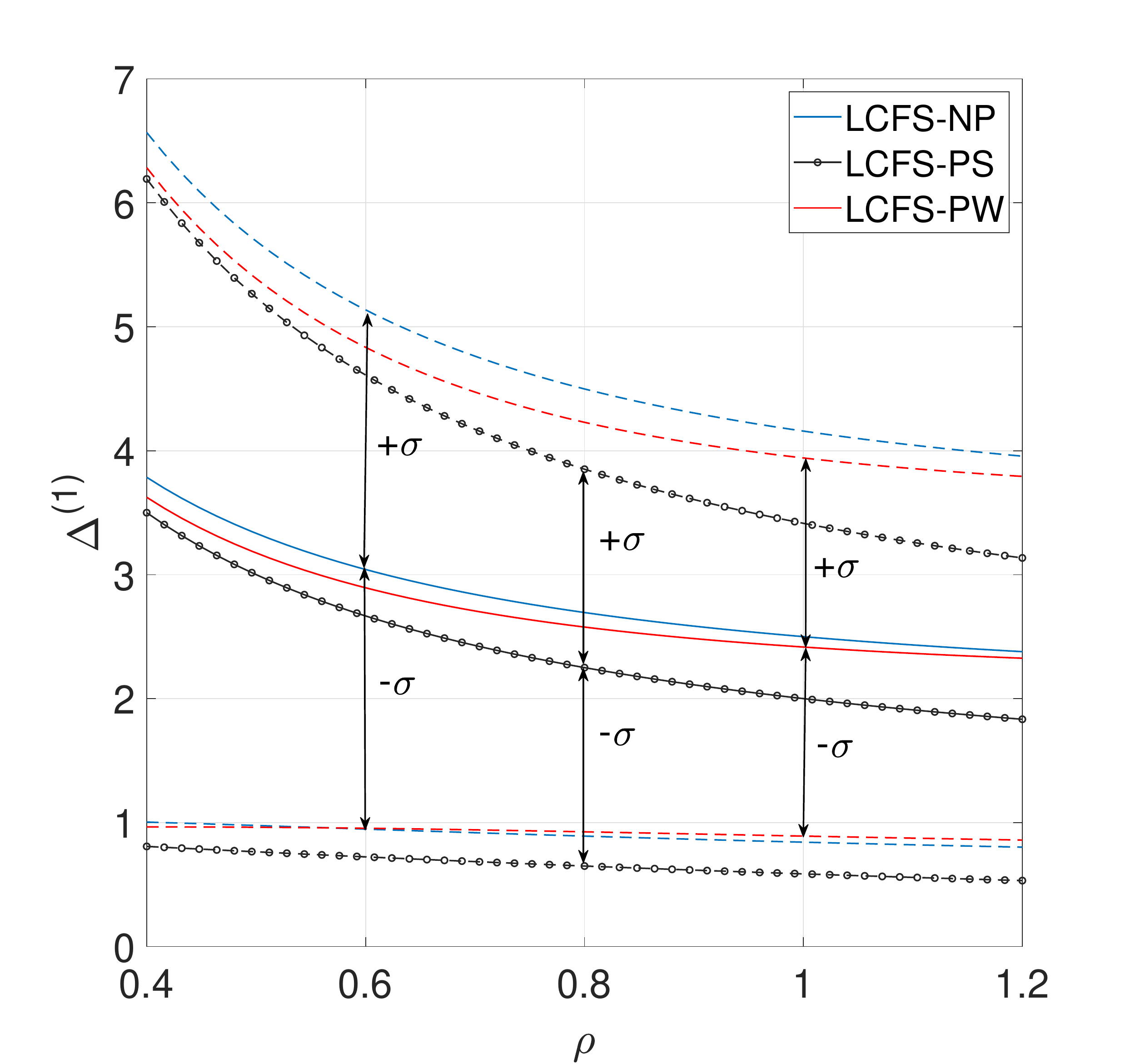}%
\label{f:comp_EH_any_50}}
} \caption{Comparison between the considered queueing disciplines for $B = 2$. When the transmitter node can only harvest energy if the system is empty, we use: i) $\beta = 0.5$ in (a), ii) $\beta = 1$ in (b), and iii) $\beta = 50$ in (c). When the transmitter node can simultaneously harvest energy and serve status updates, we use: i) $\beta = 0.5$ in (d), ii) $\beta = 1$ in (e), and iii) $\beta = 50$ in (f). }\label{f:comparison}
\end{figure*} 
\section{Numerical Results}\label{sec:numerical}
In this section, we verify our analytical results derived in Sections \ref{sec:WP}, \ref{sec:PS} and \ref{sec:PW}, and show the impact of the system parameters on the achievable AoI performance in the three queueing disciplines considered in this paper. We use $\mu = 1$ in all the figures. In Figs. \ref{f:without_with_verification} and \ref{f:waiting_verification}, we verify the accuracy of the analytical expressions of the first and second moments of AoI for all the queueing disciplines (obtained using the MGFs derived in Theorems (\ref{MGF_WP})-(\ref{MGF_PW_harvall})) by comparing them to their simulated counterparts (obtained numerically using \cite[Theorem~1]{yates2020age}). As expected, we observe a perfect match between the analytical and simulated curves. In Fig. \ref{f:limit_waiting_ver}, we also verify the asymptotic results of Corollary \ref{cor_PW_betainf} derived for the LCFS-PW queueing discipline when the transmitter node can harvest energy only if the system is empty. As stated in Remark \ref{rem:PW_limit}, we notice from Fig \ref{f:limit_waiting_ver} that $\underset{\beta \rightarrow \infty}{\rm lim}\overset{(1)}{\Delta}_{\rm PW}$ and $\underset{\beta \rightarrow \infty}{\rm lim}\overset{(2)}{\Delta}_{\rm PW}$ are functions of the battery capacity $B$. As expected, we further observe from Figs. \ref{f:without_with_verification} and \ref{f:waiting_verification} that as $\beta$ increases, the achievable AoI performance by each queueing discipline improves until it converges to its counterpart when the transmitter does not have energy limitations (represented by $\beta = 50$). In particular, we notice from: i) Figs. (\ref{f:B1_without_ver})-(\ref{f:B2_without_harvall_ver}) that $\overset{(1)}{\Delta}_{\rm NP}$ and $\overset{(2)}{\Delta}_{\rm NP}$ converge to their counterparts in the M/M/1/1 case of \cite{costa2016age} (as stated in Remarks \ref{rem:1} and \ref{rem:3a}), ii) Figs. (\ref{f:B1_with_ver})-(\ref{f:B2_with_harvall_ver}) that $\overset{(1)}{\Delta}_{\rm PS}$ and $\overset{(2)}{\Delta}_{\rm PS}$ converge to their counterparts in the single-source case of \cite[Theorem~2(a)]{yates2018age} (as stated in Remarks \ref{rem:3} and \ref{rem:4}), and iii) Fig. \ref{f:B2_waiting_harvall_ver} that $\overset{(1)}{\Delta}_{\rm PW}$ and $\overset{(2)}{\Delta}_{\rm PW}$ converge to their counterparts in the single-source case of \cite[Theorem~2(b)]{yates2018age} (as stated in Remark \ref{rem:6}).

In Fig. \ref{f:B_impact}, we show the impact of battery capacity $B$ on the achievable AoI performance for the considered queueing disciplines. Clearly, increasing $B$ increases the likelihood that the battery queue will have sufficient energy required for serving update packets upon their arrivals (when the server is idle), and hence the achievable AoI performance by each queueing discipline improves with increasing $B$. It is also worth noting that as $B \rightarrow \infty$, the achievable AoI performance by each queueing discipline approaches the AoI performance of its counterpart when the transmitter node does not have energy limitations. But, there may still be a slight gap between the two due to the finite rate of harvesting energy at the transmitter.

 Fig. \ref{f:comparison} compares the considered queueing disciplines in terms of the achievable AoI performance. First, we observe the superiority of the LCFS-PS queueing discipline over the LCFS-NP one in terms of the achievable first and second moments of AoI at the destination, which supports our arguments in Remarks \ref{rem:comp_WPandPS} and \ref{rem:comp_WPandPS_harvall}. Second, as it could be verified from \cite[Theorem~2]{yates2018age} for the single-source case (where there are no energy constraints at the transmitter), the achievable AoI performance by the LCFS-PS queueing discipline outperforms that of the LCFS-PW (for all different combinations of $\rho$ and $\beta$) when the transmitter is subject to energy constraints. This also agrees with our arguments in Remark \ref{rem:comp_PWandPS_harvall}. Third, different from the insight obtained by \cite{costa2016age} regarding the superiority of the LCFS-PW queuing discipline over the LCFS-NP (when the transmitter does not have energy limitations), we here note from Fig. \ref{f:comparison} that this ordering only holds for a certain range of values of $\beta$. In particular, the AoI performance of the LCFS-NP queueing discipline is better than that of LCFS-PW for small values of $\beta$ (for instance, $\beta =0.5$ and $\beta = 1$, as observed from Figs. \ref{f:comp_EH_empty_5}, \ref{f:comp_EH_empty_1} and \ref{f:comp_EH_any_5}). Finally, we observe that the standard deviation of AoI $\sigma$ associated with each queueing discipline is relatively large (with respect to the average value $\overset{(1)}{\Delta}$). This is a key insight indicating the necessity of incorporating the higher moments of AoI in the implementation/optimization of real-time status update systems rather than just relying on the average value (as has been mostly done in the existing literature on AoI). This insight demonstrates the importance of the analytical expressions of the second moment of AoI derived in this paper for different queueing disciplines.
\section{Conclusion}\label{sec:con}
This paper presented a queueing theory-based analysis of AoI to understand its distributional properties in EH status update systems, where an EH-powered transmitter node frequently sends status update packets (about its observed physical process) to a destination node. The status update and harvested energy packets were assumed to arrive at the transmitter according to independent Poisson processes, and the service time of each status update was assumed to be exponentially distributed. For this setup, the SHS approach was used to derive closed-form expressions of the MGF of AoI under several queueing disciplines at the transmitter, including non-preemptive (LCFS-NP) and preemptive in service/waiting (LCFS-PS/LCFS-PW) strategies. Under the non-preemptive queueing discipline, an arriving status update at the transmitter is discarded if another status update is being served, whereas the arriving status update preempts the current one in service (i.e., preemption in service) or waiting in the queue for service (i.e., preemption in waiting) under the preemptive queueing disciplines. Using the MGF results, closed-form expressions of the first and second moments of AoI were also obtained for each queueing discipline.

Our analytical results allowed us to obtain several interesting (and some counterintuitive) insights regarding the achievable AoI under the considered queueing disciplines. For instance, our results demonstrated that the second moment of AoI under the LCFS-NP queueing discipline is invariant to exchanging the arrival rates of status update and harvested energy packets. Our results also quantified the superiority of the LCFS-PS queueing discipline over the LCFS-NP and LCFS-PW queueing disciplines in terms of the achievable AoI performance. Further, our results revealed that the gap between the achievable AoI performances by the LCFS-PS and LCFS-NP queueing disciplines is reduced when the transmitter node can simultaneously harvest energy and serve update packets, compared to the other case where the transmitter can harvest energy only if the system is empty. Finally, our asymptotic results demonstrated the generality of the analysis presented in this paper by indicating that our derived expressions reduce to several existing results when the harvested energy arrival rate at the transmitter node becomes large (i.e., the transmitter node does not have energy limitations).

Several key system design insights were also drawn from our numerical results. For instance, our results quantified the improvement in the achievable AoI performance by each queueing discipline associated with the increase in either the battery capacity or the harvested energy arrival rate at the transmitter node. They also revealed that the implementation/optimization of real-time status update systems based on just the average value of AoI (as has been mostly done in the existing literature) does not ensure reliability, and thus it is crucial to incorporate the higher moments of AoI in the design of such systems. This insight demonstrated the importance of the AoI second moment expressions derived in this paper for different queueing disciplines. 

While our focus in this paper was on deriving the AoI distribution for a variety of queuing disciplines, it is also important to devise age-optimal policies for this setup. One approach is to use stochastic ordering arguments by extending the analysis of \cite{bedewy2016optimizing} to the scenario where the transmitter node is powered by EH. Given a completely different nature of this analysis compared to the SHS-based analysis presented in this paper, we have left this problem as a promising direction for future work.
\appendix

\subsection{Proof of Proposition~\ref{prop1}} \label{app:prop1}
The steady state probabilities are the unique solution of the set of equations in (\ref{gen_steady}), which can be expressed as
\begin{align}\label{prop1_proof_1}
\eta \bar{\pi}_1 = \mu \bar{\pi}_3,\; \left(\eta+\lambda\right)\bar{\pi}_2 = \eta \bar{\pi}_1 + \mu \bar{\pi}_5,\; \mu \bar{\pi}_3 = \lambda\bar{\pi}_2,
\end{align}
\begin{align} \label{prop1_proof_2}
\left(\eta+\lambda\right)\bar{\pi}_{2i} = \eta\bar{\pi}_{2i-2} + \mu \bar{\pi}_{2i+3}, 2 \leq i \leq B-1,
\end{align}
\begin{align}\label{prop1_proof_3}
\mu \bar{\pi}_{2i+1} = \lambda \bar{\pi}_{2i}, 2 \leq i \leq B,
\end{align}
\begin{align}\label{prop1_proof_4}
\sum_{i=1}^{2B+1}{\bar{\pi}_i} = 1.
\end{align}

From (\ref{prop1_proof_1}), we can express $\bar{\pi}_2$ and $\bar{\pi}_3$ as a function of $\bar{\pi}_1$ as
\begin{align}\label{prop1_proof_5}
\bar{\pi}_2 = \dfrac{\beta}{\rho}\bar{\pi}_1,\; \bar{\pi}_3 = \beta \bar{\pi}_1.
\end{align}

Next, using (\ref{prop1_proof_2}) and (\ref{prop1_proof_3}), $\bar{\pi}_{2i}$ and $\bar{\pi}_{2i+1}$, $1 \leq i \leq B$, can be expressed as in (\ref{prop1_1}) and (\ref{prop1_2}). Finally, by substituting (\ref{prop1_1}) and (\ref{prop1_2}) into (\ref{prop1_proof_4}), $\bar{\pi}_1$ can be expressed as
\begin{align}
\bar{\pi}_1 = \dfrac{1}{1 + \theta\left(1 + \rho\right)},
\end{align}
where $\theta = \sum_{i=1}^{B}{\left(\frac{\beta}{\rho}\right)^i}$. The final expression of $\bar{\pi}_1$ in (\ref{prop1_3}) can be obtained by noting that $\theta = B$ if $\beta = \rho$, and $\theta = \frac{\left(\frac{\beta}{\rho}\right)^{B+1}-\frac{\beta}{\rho}}{\frac{\beta}{\rho}-1}$ otherwise. \hfill 
\IEEEQED

\subsection{Proof of Theorem~\ref{MGF_WP}} \label{app:MGF_WP}
The set of equations in (\ref{gen_vMGF}) can be expressed as
\begin{align}\label{theorem_MGF_WP_proof_1}
q_1: \;\; \left(\eta - s\right)[\bar{v}^{s}_{10},\bar{v}^{s}_{11}] = \mu[\bar{v}^{s}_{31},\bar{\pi}_3],
\end{align}
\begin{align}\label{theorem_MGF_WP_proof_2}
q_2: \;\; \left(\eta + \lambda - s\right)[\bar{v}^{s}_{20},\bar{v}^{s}_{21}] = \mu[\bar{v}^{s}_{51},\bar{\pi}_5] + \eta[\bar{v}^{s}_{10},\bar{\pi}_1],
\end{align}
\begin{align}\label{theorem_MGF_WP_proof_3}
q_{2k}, 2 \leq k \leq B-1\nonumber&: \;\; \left(\eta + \lambda - s\right)[\bar{v}^{s}_{2k,0},\bar{v}^{s}_{2k,1}] =\\& \mu[\bar{v}^{s}_{2k+3,1},\bar{\pi}_{2k+3}] + \eta[\bar{v}^{s}_{2k-2,0},\bar{\pi}_{2k-2}],
\end{align}
\begin{align}\label{theorem_MGF_WP_proof_4}
q_{2B}: \;\; \left(\lambda - s\right)[\bar{v}^{s}_{2B,0},\bar{v}^{s}_{2B,1}] = \eta[\bar{v}^{s}_{2B-2,0},\bar{\pi}_{2B-2}].
\end{align}
\begin{align}\label{theorem_MGF_WP_proof_5}
q_{2k+1}, 1 \leq k \leq B: \;\; \left(\mu - s\right)[\bar{v}^{s}_{2k+1,0},\bar{v}^{s}_{2k+1,1}] = \lambda[\bar{v}^{s}_{2k,0},\bar{\pi}_{2k}],
\end{align}

Summing the set of equations in (\ref{theorem_MGF_WP_proof_5}) gives
\begin{align}\label{WP_v0_r2}
\left(\mu - s\right)\sum_{k \in \;{\rm r}_2} {\bar{v}^s_{k0}}= \lambda \sum_{k \in \;{\rm r}_1}{\bar{v}^s_{k0}} - \lambda  \bar{v}^s_{10}.
\end{align}
\begin{align}\label{WP_v1_r2}
\left(\mu - s\right)\sum_{k \in \;{\rm r}_2} {\bar{v}^s_{k1}}= \lambda \sum_{k \in \;{\rm r}_1/\{1\}}{\bar{\pi}_{k}}.
\end{align}

Further, by summing (\ref{theorem_MGF_WP_proof_1})-(\ref{theorem_MGF_WP_proof_4}), we have
\begin{align}\label{WP_v0_r1}
\left(\lambda - s\right)\sum_{k \in \;{\rm r}_1} {\bar{v}^s_{k0}}= \mu \sum_{k \in\;{\rm r}_2}{\bar{v}^s_{k1}} + \lambda \bar{v}^s_{10}. 
\end{align}

Now, from (\ref{gen_vMGF}), the MGF of AoI at the destination can be evaluated as
\begin{align}\label{MGF_WP_v0}
\nonumber & \overset{\rm NP}{M}(\bar{s}) = \sum_{k \in\; {\rm r}_1 \cup \;{\rm r}_2}{\bar{v}^s_{k0}} \a \frac{\lambda + \mu - s}{\mu - s} \sum_{k \in \;{\rm r}_1}{\bar{v}^s_{k0}} - \frac{\lambda}{\mu - s} \bar{v}^s_{10} ,\\
&\nonumber\b \frac{\rho \left(1 + \rho - \bar{s}\right)}{\left(1 - \bar{s}\right)^2 \left(\rho - \bar{s}\right)} \sum_{k \in \;{\rm r}_1 / \{1\}}{\bar{\pi}_k} + \frac{\rho}{\left(1 - \bar{s}\right)\left(\rho - \bar{s}\right) } \bar{v}^s_{10}, \\&\c \frac{\rho \left(1 + \rho - \bar{s}\right)\bar{\pi}_1 \theta}{\left(1 - \bar{s}\right)^2 \left(\rho - \bar{s}\right)} + \frac{\rho}{\left(1 - \bar{s}\right)\left(\rho - \bar{s}\right)} \bar{v}^s_{10},
\end{align}
where step (a) follows from substituting (\ref{WP_v0_r2}) into (\ref{MGF_WP_v0}), step (b) follows from substituting (\ref{WP_v1_r2}) and (\ref{WP_v0_r1})  into (\ref{MGF_WP_v0}), and step (c) follows from (\ref{prop1_1}) in Proposition \ref{prop1}. Note that $\bar{v}^s_{10}$ can be obtained from (\ref{theorem_MGF_WP_proof_1}) and (\ref{theorem_MGF_WP_proof_5}) as $\frac{\beta \bar{\pi}_1}{\left(1 - \bar{s}\right) \left(\beta - \bar{s}\right)}$. The final expression of $\overset{\rm NP}{M}(\bar{s})$ in (\ref{theorem_MGF_WP_1}) can be derived by substituting $\bar{v}^s_{10}$ into (\ref{MGF_WP_v0}), followed by some algebraic simplifications.
\hfill 
\IEEEQED

\subsection{Proof of Theorem~\ref{MGF_WP_harvall}} \label{app:MGF_WP_harvall}
We start the proof by noticing that the equations of (\ref{gen_vMGF}) corresponding to the states in ${\rm r}_1$ are still given by (\ref{theorem_MGF_WP_proof_1})-(\ref{theorem_MGF_WP_proof_4}) and so (\ref{WP_v0_r1}) also holds. Clearly, this happens due to the fact that the new subset of transitions in Table \ref{table:WP_harvall} (indicating the server's capability of simultaneously harvesting energy and sending status update packets) only occur between the states in ${\rm r}_2$. Regarding the states in ${\rm r}_2$, we have
\begin{align}\label{theorem_MGF_WP_harvall_proof_1}
q_3:\;\; \left(\eta + \mu - s\right)[\bar{v}^s_{30},\bar{v}^s_{31}] = \lambda [\bar{v}^s_{20},\bar{\pi}_2],
\end{align}
\begin{align}\label{theorem_MGF_WP_harvall_proof_2}
q_{2k+1}, 2 \leq k \leq B-1\nonumber&:\;\; \left(\eta + \mu - s\right)[\bar{v}^s_{2k+1,0},\bar{v}^s_{2k+1,1}] = \\&\lambda [\bar{v}^s_{2k,0},\bar{\pi}_{2k}] + \eta [\bar{v}^s_{2k-1,0},\bar{v}^s_{2k-1,1}],
\end{align}
\begin{align}\label{theorem_MGF_WP_harvall_proof_3}
q_{2B+1}:\;\; \left(\mu - s\right)[\bar{v}^s_{2B+1,0},\bar{v}^s_{2B+1,1}] \nonumber&= \lambda [\bar{v}^s_{2B,0},\bar{\pi}_{2B}] \\&+ \eta [\bar{v}^s_{2B-1,0},\bar{v}^s_{2B-1,1}].
\end{align}

Further, the sum of  (\ref{theorem_MGF_WP_harvall_proof_1})-(\ref{theorem_MGF_WP_harvall_proof_3}) also gives (\ref{WP_v0_r2}) and (\ref{WP_v1_r2}). Thus, the MGF of AoI at the destination can be expressed as
\begin{align}\label{MGF_WP_harvall_v0}
\nonumber &\overset{\rm NP}{M}(\bar{s})
= \frac{\rho \left(1 + \rho - \bar{s}\right)}{\left(1 - \bar{s}\right)^2 \left(\rho - \bar{s}\right)} \sum_{k \in \;{\rm r}_1/\{1\}}{\bar{\pi}_k}+ \frac{\rho}{\left(1 - \bar{s}\right)\left(\rho - \bar{s}\right)} \bar{v}^s_{10},\\
& \a \frac{ \rho \left(1 + \rho - \bar{s}\right)\gamma\bar{\pi}_1}{\left(1 - \bar{s}\right)^2 \left(\rho - \bar{s}\right)} + \frac{\rho^2 \bar{\pi}_2}{\left(1 - \bar{s}\right)\left(\rho - \bar{s}\right)\left(\beta - \bar{s}\right)\left(1 + \beta - \bar{s}\right)},
\end{align}
where step (a) follows from substituting $\bar{v}^s_{10} = \frac{\rho \bar{\pi}_2}{\left(\beta - \bar{s}\right)\left(1 + \beta - \bar{s}\right)}$ (obtained from (\ref{theorem_MGF_WP_proof_1}) and (\ref{theorem_MGF_WP_harvall_proof_1})) into (\ref{MGF_WP_harvall_v0}), and replacing $\sum_{k \in\;{\rm r}_1 /\{1\}}{\bar{\pi}_k}$ by $\gamma \bar{\pi}_1$. The final expression of $\overset{\rm NP}{M}(\bar{s})$ in (\ref{theorem_MGF_WP_harvall_1}) can be derived after applying some algebraic simplifications to (\ref{MGF_WP_harvall_v0}) along with noting that $\bar{\pi}_2 = \frac{\beta\left(1+\beta\right)}{\rho}\bar{\pi}_1$.
\hfill 
\IEEEQED

\subsection{Proof of Theorem~\ref{MGF_PS}} \label{app:MGF_PS}
We note from Fig. \ref{f:PS_MC} that the set of equations in (\ref{gen_vMGF}) corresponding to the states in ${\rm r}_1$ are given by (\ref{theorem_MGF_WP_proof_1})-(\ref{theorem_MGF_WP_proof_4}). Therefore, $\sum_{k \in\;{\rm r}_1}{\bar{v}^s_{k0}}$ can be expressed as in (\ref{WP_v0_r1}). On the other hand, for the states in ${\rm r}_2$, we have 
\begin{align}\label{theorem_MGF_PS_proof_1}
q_{2k+1}, 1 \leq k \leq B\nonumber&:\;\; 
\left(\lambda + \mu - s\right)[\bar{v}^s_{2k+1,0},\bar{v}^s_{2k+1,1}] = \\& \lambda [\bar{v}^s_{2k,0},\bar{\pi}_{2k}] + \lambda [\bar{v}^s_{2k+1,0},\bar{\pi}_{2k+1}].
\end{align}

From (\ref{theorem_MGF_PS_proof_1}), $\sum_{k \in\;{\rm r}_2}{\bar{v}^s_{k0}}$ is given by (\ref{WP_v0_r2}), and $\sum_{k \in\;{\rm r}_2}{\bar{v}^s_{k1}}$ can be expressed as
\begin{align}\label{PS_v1_r2}
\sum_{k \in\;{\rm r}_2}{\bar{v}^s_{k1}} = \dfrac{\lambda\sum_{k \in\;{\rm r}_1 \cup\;{\rm r}_2/\{1\}}{\bar{\pi}_k}}{\lambda + \mu - s} \a \dfrac{\rho\left(1 + \rho\right)\theta \bar{\pi}_1}{1 + \rho - \bar{s}},
\end{align}
where step (a) follows from Proposition \ref{prop1}. Hence, the MGF of AoI at the destination can be evaluated as
\begin{align}\label{MGF_PS_v0}
\nonumber  \overset{\rm PS}{M}(\bar{s}) &= \sum_{k \in\; {\rm r}_1 \cup \;{\rm r}_2}{\bar{v}^s_{k0}} \a \frac{\lambda + \mu - s}{\mu - s} \sum_{k \in \;{\rm r}_1}{\bar{v}^s_{k0}} - \frac{\lambda}{\mu - s} \bar{v}^s_{10}, \\&\nonumber \b \frac{\rho \left(1+\rho\right)\theta \bar{\pi}_1 + \rho \bar{v}^s_{10}}{\left(1 - \bar{s}\right) \left(\rho - \bar{s}\right)},\\
& \c \dfrac{\rho \left(1 + \rho\right)\bar{\pi}_1\Big[\theta\left(\beta - \bar{s}\right)\left(1 + \rho - \bar{s}\right) + \beta\Big]}{\left(1 - \bar{s}\right) \left(\rho - \bar{s}\right) \left(1 + \rho - \bar{s}\right)\left(\beta - \bar{s}\right)},
\end{align}
where step (a) follows from substituting (\ref{WP_v0_r2}) into (\ref{MGF_PS_v0}), and step (b) follows from substituting (\ref{WP_v0_r1}) and (\ref{PS_v1_r2}) into (\ref{MGF_PS_v0}). Step (c) follows from substituting $\bar{v}^s_{10}$, which is obtained from (\ref{theorem_MGF_WP_proof_1}), (\ref{theorem_MGF_PS_proof_1}) and Proposition \ref{prop1} as $\frac{\beta \left(1 + \rho\right)\bar{\pi}_1}{\left(\beta-\bar{s}\right)\left(1+\rho-\bar{s}\right)}$. This completes the proof. 
\hfill 
\IEEEQED

\subsection{Proof of Theorem~\ref{MGF_PS_harvall}} \label{app:MGF_PS_harvall}
First, it is worth noting that (\ref{theorem_MGF_WP_proof_1})-(\ref{theorem_MGF_WP_proof_4}) still express the set of equations in (\ref{gen_vMGF}) corresponding to the states in ${\rm r}_1$. Further, the set of equations in (\ref{gen_vMGF}) corresponding to the states in ${\rm r}_2$ can be expressed as
\begin{align}\label{theorem_MGF_PS_harvall_proof_1}
q_3:\;\; \left(\eta + \lambda + \mu - s\right)[\bar{v}^s_{30},\bar{v}^s_{31}] = \lambda [\bar{v}^s_{20},\bar{\pi}_2] + \lambda [\bar{v}^s_{30},\bar{\pi}_3],
\end{align}
\begin{align}\label{theorem_MGF_PS_harvall_proof_2}
q_{2k+1}: &\left(\eta + \lambda + \mu - s\right)\nonumber[\bar{v}^s_{2k+1,0},\bar{v}^s_{2k+1,1}] = \lambda [\bar{v}^s_{2k,0},\bar{\pi}_{2k}] + \\&\lambda [\bar{v}^s_{2k+1,0},\bar{\pi}_{2k+1}] + \eta [\bar{v}^s_{2k-1,0},\bar{v}^s_{2k-1,1}],
\end{align}
\begin{align}\label{theorem_MGF_PS_harvall_proof_3}
q_{2B+1}: \nonumber&\left(\mu + \lambda - s\right)[\bar{v}^s_{2B+1,0},\bar{v}^s_{2B+1,1}] = \lambda [\bar{v}^s_{2B,0},\bar{\pi}_{2B}] + \\&\lambda [\bar{v}^s_{2B+1,0},\bar{\pi}_{2B+1}] + \eta [\bar{v}^s_{2B-1,0},\bar{v}^s_{2B-1,1}],
\end{align}
where $2 \leq k \leq B-1$. Clearly, $\sum_{k \in \;{\rm r}_1}{\bar{v}^s_{k0}}$ is given by (\ref{WP_v0_r1}), and $\sum_{k \in \;{\rm r}_2}{\bar{v}^s_{k0}}$ can be obtained from the sum of (\ref{theorem_MGF_PS_harvall_proof_1})-(\ref{theorem_MGF_PS_harvall_proof_3}) as in (\ref{WP_v0_r2}). Further, from (\ref{theorem_MGF_PS_harvall_proof_1})-(\ref{theorem_MGF_PS_harvall_proof_3}), $\sum_{k \in\;{\rm r}_2}{\bar{v}^s_{k1}}$ is given by
\begin{align}\label{PS_harvall_v1_r2}
\sum_{k \in\;{\rm r}_2}{\bar{v}^s_{k1}} = \dfrac{\lambda\sum_{k \in\;{\rm r}_1 \cup\;{\rm r}_2/\{1\}}{\bar{\pi}_k}}{\lambda + \mu - s} \a \dfrac{\rho \gamma' \bar{\pi}_1}{1 + \rho - \bar{s}},
\end{align}
where step (a) follows from expressing  $\underset{k \in \;{\rm r}_1 \cup\; {\rm r}_2 / \{1\}}{\sum}{\bar{\pi}_k}$ as $\gamma' \bar{\pi}_1$. Hence, the MGF of AoI at the destination can be evaluated as
\begin{align}\label{MGF_PS_harvall_v0}
  \overset{\rm PS}{M}(\bar{s}) \nonumber&= \sum_{k \in\; {\rm r}_1 \cup \;{\rm r}_2}{\bar{v}^s_{k0}} \a \frac{\rho \gamma' \bar{\pi}_1 + \rho \bar{v}^s_{10}}{\left(1 - \bar{s}\right) \left(\rho - \bar{s}\right)}, \\&\b \dfrac{\rho \bar{\pi}_1\Big[\gamma'\left(\beta - \bar{s}\right)\left(1 + \rho + \beta - \bar{s}\right) + \beta \left(1 + \rho + \beta\right)\Big]}{\left(1 - \bar{s}\right) \left(\rho - \bar{s}\right)\left(\beta - \bar{s}\right) \left(1 + \rho + \beta - \bar{s}\right)},
\end{align}
where step (a) follows from substituting (\ref{WP_v0_r2}), (\ref{WP_v0_r1}) and (\ref{PS_harvall_v1_r2}) into (\ref{MGF_PS_harvall_v0}). Step (b) follows from substituting $\bar{v}^s_{10}$ into (\ref{MGF_PS_harvall_v0}), which is obtained from (\ref{theorem_MGF_WP_proof_1}) and (\ref{theorem_MGF_PS_harvall_proof_1}) as $\frac{\rho\left(\bar{\pi}_2 + \bar{\pi}_3\right)}{\left(\beta-\bar{s}\right)\left(1+\rho+\beta-\bar{s}\right)}$, where $\bar{\pi}_3 = \beta \bar{\pi}_1$ and $\bar{\pi}_2 = \frac{\beta\left(1+\beta\right)\bar{\pi}_1}{\rho}$. This completes the proof. 
\hfill 
\IEEEQED

\subsection{Proof of Proposition~\ref{prop2}} \label{app:prop2}
We start the proof by expressing the set of equations in (\ref{gen_steady}) as
\begin{align}\label{prop2_proof_1}
\nonumber&\eta \bar{\pi}_1 = \mu \bar{\pi}_3,\; \left(\eta+\lambda\right)\bar{\pi}_2 = \eta \bar{\pi}_1 + \mu \bar{\pi}_5,\; \mu \bar{\pi}_3 = \lambda\bar{\pi}_2 + \mu \bar{\pi}_6,\\& \left(\eta + \lambda\right)\bar{\pi}_4 = \eta \bar{\pi}_2 + \mu \bar{\pi}_8,
\end{align}
\begin{align} \label{prop2_proof_2}
\left(\eta+\lambda\right)\bar{\pi}_{3i-2} = \eta\bar{\pi}_{3i-5} + \mu \bar{\pi}_{3i+2},\;\; 3 \leq i \leq B-1,
\end{align}
\begin{align}\label{prop2_proof_3}
\left(\lambda + \mu\right) \bar{\pi}_{3i-1} = \lambda \bar{\pi}_{3i-2} + \mu \bar{\pi}_{3i+3},\;\; 2 \leq i \leq B-1,
\end{align}
\begin{align}\label{prop2_proof_4}
\mu \bar{\pi}_{3i} = \lambda \bar{\pi}_{3i-1},\;\; 2 \leq i \leq B-1,
\end{align}
\begin{align}\label{prop2_proof_5}
\lambda\bar{\pi}_{3B-2} = \eta \bar{\pi}_{3B-5},\; \left(\lambda + \mu\right)\bar{\pi}_{3B-1} = \lambda \bar{\pi}_{3B-2},
\end{align}
\begin{align}\label{prop2_proof_6}
\sum_{i=1}^{3B}{\bar{\pi}_i} = 1.
\end{align}

From (\ref{prop2_proof_1})-(\ref{prop2_proof_5}), the steady state probabilities $\{\bar{\pi}\}$ can first be expressed in terms of $\bar{\pi}_1$ as in (\ref{prop2_1})-(\ref{prop2_4}). Afterwards, the expression of $\bar{\pi}_1$ in (\ref{prop2_pi1}) can be derived by substituting (\ref{prop2_1})-(\ref{prop2_4}) into the normalization equation in (\ref{prop2_proof_6}).
\hfill 
\IEEEQED

\subsection{Proof of Theorem~\ref{MGF_PW}} \label{app:MGF_PW}
We start the proof by expressing the set of equations in (\ref{gen_vMGF}) for different states $q \in \ncalQ$. For the states in ${\rm r}_1$, we have
\begin{align}\label{theorem_MGF_PW_proof_1}
q_1: \left(\eta - s\right)[\bar{v}^s_{10},\bar{v}^s_{11},\bar{v}^s_{12}] = \mu[\bar{v}^s_{31},\bar{\pi}_3,\bar{\pi}_3],
\end{align}
\begin{align}\label{theorem_MGF_PW_proof_2}
q_2: \left(\eta + \lambda - s\right)[\bar{v}^s_{20},\bar{v}^s_{21},\bar{v}^s_{22}] = \mu[\bar{v}^s_{51},\bar{\pi}_5,\bar{\pi}_5] + \eta [\bar{v}^s_{10},\bar{\pi}_1,\bar{\pi}_1],
\end{align}
\begin{align}\label{theorem_MGF_PW_proof_3}
q_4: \left(\eta + \lambda - s\right)[\bar{v}^s_{40},\bar{v}^s_{41},\bar{v}^s_{42}] = \mu[\bar{v}^s_{81},\bar{\pi}_8,\bar{\pi}_8] + \eta [\bar{v}^s_{20},\bar{\pi}_2,\bar{\pi}_2],
\end{align}
\begin{align}\label{theorem_MGF_PW_proof_4}
\nonumber&q_{3k-2}: \left(\eta + \lambda - s\right)[\bar{v}^s_{3k-2,0},\bar{v}^s_{3k-2,1},\bar{v}^s_{3k-2,2}] = \\&\mu[\bar{v}^s_{3k+2,1},\bar{\pi}_{3k+2},\bar{\pi}_{3k+2}] + \eta [\bar{v}^s_{3k-5,0},\bar{\pi}_{3k-5},\bar{\pi}_{3k-5}],
\end{align}
\begin{align}\label{theorem_MGF_PW_proof_5}
q_{3B-2}: \left(\lambda - s\right)\nonumber&[\bar{v}^s_{3B-2,0},\bar{v}^s_{3B-2,1},\bar{v}^s_{3B-2,2}] = \\&\eta [\bar{v}^s_{3B-5,0},\bar{\pi}_{3B-5},\bar{\pi}_{3B-5}],
\end{align}
where $3 \leq k \leq B-1$. Further, for the states in ${\rm r}_2$, we have
\begin{align}\label{theorem_MGF_PW_proof_6}
q_3: \left(\mu - s\right)[\bar{v}^s_{30},\bar{v}^s_{31},\bar{v}^s_{32}] = \mu[\bar{v}^s_{61},\bar{v}^s_{62},\bar{\pi}_6] + \lambda[\bar{v}^s_{20},\bar{\pi}_2,\bar{\pi}_2],
\end{align}
\begin{align}\label{theorem_MGF_PW_proof_7}
\nonumber&q_{3k-1}: \left(\lambda + \mu - s\right)[\bar{v}^s_{3k-1,0},\bar{v}^s_{3k-1,1},\bar{v}^s_{3k-1,2}] = \\&\mu[\bar{v}^s_{3k+3,1},\bar{v}^s_{3k+3,2},\bar{\pi}_{3k+3}] + \lambda[\bar{v}^s_{3k-2,0},\bar{\pi}_{3k-2},\bar{\pi}_{3k-2}],
\end{align}
\begin{align}\label{theorem_MGF_PW_proof_8}
\nonumber q_{3B-1}: \left(\lambda + \mu - s\right)&[\bar{v}^s_{3B-1,0},\bar{v}^s_{3B-1,1},\bar{v}^s_{3B-1,2}] =  \\&\lambda[\bar{v}^s_{3B-2,0},\bar{\pi}_{3B-2},\bar{\pi}_{3B-2}],
\end{align}
where $2 \leq k \leq B-1$. Finally, the set of equations corresponding to the states in ${\rm r}_3$ is given by
\begin{align}\label{theorem_MGF_PW_proof_9}
    q_{3k}: \left(\lambda + \mu - s\right)\nonumber&[\bar{v}^s_{3k,0},\bar{v}^s_{3k,1},\bar{v}^s_{3k,2}] = \lambda[\bar{v}^s_{3k,0},\bar{v}^s_{3k,1},\bar{\pi}_{3k}] + \\&\lambda[\bar{v}^s_{3k-1,0},\bar{v}^s_{3k-1,1},\bar{\pi}_{3k-1}],
\end{align}
where $1 \leq k \leq B$. Summing (\ref{theorem_MGF_PW_proof_1})-(\ref{theorem_MGF_PW_proof_5}) gives
\begin{align}\label{PW_v0_r1}
\left(\lambda - s\right)\sum_{k \in\;{\rm r}_1}{\bar{v}^s_{k0}} = \mu \sum_{k \in\;{\rm r}_2}{\bar{v}^s_{k1}} + \lambda \bar{v}^s_{10}.
\end{align}

Further, from (\ref{theorem_MGF_PW_proof_6})-(\ref{theorem_MGF_PW_proof_8}), we have
\begin{align}\label{PW_v0_r2}
\left(\lambda + \mu - s \right)\sum_{k \in {\rm r}_2}{\bar{v}^s_{k0}} = \lambda \sum_{k \in {\rm r}_1}{\bar{v}^s_{k0}} + \mu \sum_{k \in {\rm r}_3}{\bar{v}^s_{k1}} + \lambda \left(\bar{v}^s_{30} - \bar{v}^s_{10}\right),
\end{align}
\begin{align}\label{PW_v1_r2}
\nonumber&\bar{v}^s_{31} = \dfrac{\mu\bar{v}^s_{62} + \lambda \bar{\pi}_2}{\mu - s},\;\; \bar{v}^s_{3k-1,1} = \dfrac{\mu \bar{v}^s_{3k+3,2} + \lambda \bar{\pi}_{3k-2}}{\lambda + \mu - s}, \\& \bar{v}^s_{3B-1,1} = \dfrac{\lambda \bar{\pi}_{3B-2}}{\lambda + \mu - s},
\end{align}
where $2 \leq k \leq B-1$. In addition, we can obtain from (\ref{theorem_MGF_PW_proof_9}) the following
\begin{align}\label{PW_v0_r3}
\left(\mu - s\right)\sum_{k \in {\rm r}_3}{\bar{v}^s_{k0}} = \lambda \sum_{k \in {\rm r}_2}{\bar{v}^s_{k0}} - \lambda \bar{v}^{s}_{30}, 
\end{align}
\begin{align}\label{PW_v1_r3}
\left(\mu - s\right)\sum_{k \in {\rm r}_3}{\bar{v}^s_{k1}} = \lambda \sum_{k \in {\rm r}_2}{\bar{v}^s_{k1}} - \lambda \bar{v}^s_{31},
\end{align}
\begin{align}\label{PW_v2_r3}
 \bar{v}^s_{3k,2} = \dfrac{\lambda\left(\bar{\pi}_{3k-1}+\bar{\pi}_{3k}\right)}{\lambda + \mu - s},\;\; 1 \leq k \leq B. 
\end{align}

Now, the MGF of AoI at the destination can be evaluated as
\begin{align}\label{MGF_PW_v0}
\nonumber & \overset{\rm PW}{M}(\bar{s}) = \sum_{k \in\; {\rm r}_1}{\bar{v}^s_{k0}} + \sum_{k \in\; {\rm r}_2}{\bar{v}^s_{k0}} + \sum_{k \in\; {\rm r}_3}{\bar{v}^s_{k0}}, \\\nonumber&\a \sum_{k \in\; {\rm r}_1}{\bar{v}^s_{k0}} + \dfrac{\lambda + \mu - s}{\mu - s}\sum_{k \in\; {\rm r}_2}{\bar{v}^s_{k0}} -\dfrac{\lambda \bar{v}^s_{30}}{\mu - s},\\
&\nonumber \b \dfrac{\left(\lambda + \mu - s\right)\sum_{k \in\; {\rm r}_1}{\bar{v}^s_{k0}} + \mu \sum_{k \in\; {\rm r}_3}{\bar{v}^s_{k1}} - \lambda \bar{v}^s_{10}}{\mu - s},\\ 
&\nonumber \c \dfrac{\mu\big[\left(\lambda + \mu - s\right) \sum_{k \in\; {\rm r}_2}{\bar{v}^s_{k1}} + \left(\lambda - s\right) \sum_{k \in\; {\rm r}_3}{\bar{v}^s_{k1}} + \lambda \bar{v}^s_{10}\big]}{\left(\mu - s\right)\left(\lambda - s\right)},\\
&\nonumber \d \dfrac{\mu\left(\eta - s\right)\big[\left(\mu - s\right)\left(\lambda + \mu - s\right) - \lambda \left(\lambda - s\right)\big] \sum_{k \in\; {\rm r}_2}{\bar{v}^s_{k1}}}{\left(\mu - s\right)^2 \left(\lambda - s\right) \left(\eta - s\right)} \\& + \dfrac{\mu\lambda \bar{v}^s_{31}\big[\mu\left(\mu - s\right) - \left(\lambda - s\right)\left(\eta - s\right)\big]}{\left(\mu - s\right)^2 \left(\lambda - s\right) \left(\eta - s\right)},
\end{align}
where steps (a) follows from substituting $\sum_{k \in\; {\rm r}_3}{\bar{v}^s_{k0}}$ from (\ref{PW_v0_r3}) and step (b) follows from substituting $\sum_{k \in\; {\rm r}_2}{\bar{v}^s_{k0}}$ from (\ref{PW_v0_r2}). Step (c) follows from substituting $\sum_{k \in\; {\rm r}_1}{\bar{v}^s_{k0}}$ from (\ref{PW_v0_r1}), and step (d) follows from expressing $\sum_{k \in\; {\rm r}_3}{\bar{v}^s_{k1}}$ as a function of $\sum_{k \in\; {\rm r}_2}{\bar{v}^s_{k1}}$ using (\ref{PW_v1_r3}). Next, $\sum_{k \in\; {\rm r}_2}{\bar{v}^s_{k1}}$ and $\bar{v}^s_{31}$ can be respectively obtained from (\ref{PW_v1_r2}) and (\ref{PW_v2_r3}) as
\begin{align}\label{PW_v1_r2_sum}
\nonumber& \sum_{k \in\; {\rm r}_2}{\bar{v}^s_{k1}} = \dfrac{\lambda \bar{v}^s_{31}}{\lambda + \mu - s} + \\&\dfrac{\mu\lambda \sum_{k \in\;{\rm r}_2\;\cup\;{\rm r}_3\;/\{3\}}{\bar{\pi}_k}+\lambda \left(\lambda + \mu - s\right)\sum_{k \in\;{\rm r}_1\;/\{1\}}{\bar{\pi}_k}}{\left(\lambda + \mu - s\right)^2},
\end{align}
\begin{align}\label{PW_v31_r2}
\bar{v}^s_{31} = \dfrac{\mu \lambda\left(\bar{\pi}_5+\bar{\pi}_6\right) + \lambda \bar{\pi}_2 \left(\lambda + \mu - s\right)}{\left(\mu - s\right)\left(\lambda + \mu - s\right)},
\end{align}
where the steady state probabilities $\{\bar{\pi}_i\}$ can be expressed in terms of $\bar{\pi}_1$ using Proposition \ref{prop2}, and $\bar{\pi}_1$ is given by (\ref{prop2_pi1}). Also, $\theta_1$ and $\theta_2$ given by (\ref{theorem_MGF_PW_theta1}) and (\ref{theorem_MGF_PW_theta2}) satisfy $\sum_{k \in \; {\rm r}_2\;\cup\;{\rm r}_3}{\bar{\pi}_k} = \theta_1 \bar{\pi}_1$ and $\sum_{k \in \;{\rm r}_1}{\bar{\pi}_k} = \theta_2 \bar{\pi}_1$. The final expression of $\overset{\rm PW}{M}(\bar{s})$ in (\ref{theorem_MGF_PW_1}) can be derived by substituting (\ref{PW_v1_r2_sum}) and (\ref{PW_v31_r2}) into (\ref{MGF_PW_v0}), followed by some algebraic simplifications.
\hfill 
\IEEEQED

\subsection{Proof of Theorem~\ref{MGF_PW_harvall}} \label{app:MGF_PW_harvall}
First, we observe from Figs. \ref{f:PW_MC} and \ref{f:PW_harvall_MC} that the sets of incoming and outgoing transitions ($\ncalL'_q$ and $\ncalL_q$) associated with the states in ${\rm r}_{1}$ are not influenced by whether the server is able/unable to simultaneously harvest energy and serve status update packets. Therefore, the set of equations in (\ref{gen_vMGF}) corresponding to the states in ${\rm r}_1$ are still given by (\ref{theorem_MGF_PW_proof_1})-(\ref{theorem_MGF_PW_proof_5}), and hence $\sum_{k \in\;{\rm r}_1}{\bar{v}^s_{k0}}$ can be expressed as in (\ref{PW_v0_r1}). Now, coming to the states in ${\rm r}_2$, we have
\begin{align}\label{theorem_MGF_PW_harvall_proof_1}
q_3: \left(\eta + \mu  - s\right)[\bar{v}^s_{30},\bar{v}^s_{31},\bar{v}^s_{32}] = \mu[\bar{v}^s_{61},\bar{v}^s_{62},\bar{\pi}_6] + \lambda[\bar{v}^s_{20},\bar{\pi}_2,\bar{\pi}_2],
\end{align}
\begin{align}\label{theorem_MGF_PW_harvall_proof_2}
q_5: \nonumber&\left(\eta + \lambda + \mu - s\right)[\bar{v}^s_{50},\bar{v}^s_{51},\bar{v}^s_{52}] = \mu[\bar{v}^s_{91},\bar{v}^s_{92},\bar{\pi}_9] +\\& \lambda[\bar{v}^s_{40},\bar{\pi}_4,\bar{\pi}_4] + \eta [\bar{v}^s_{30},\bar{v}^s_{31},\bar{\pi}_3],
\end{align}
\begin{align}\label{theorem_MGF_PW_harvall_proof_3}
&q_{3k-1}: \left(\eta + \lambda + \mu - s\right)\nonumber[\bar{v}^s_{3k-1,0},\bar{v}^s_{3k-1,1},\bar{v}^s_{3k-1,2}] =\\\nonumber& \mu[\bar{v}^s_{3k+3,1},\bar{v}^s_{3k+3,2},\bar{\pi}_{3k+3}] + \lambda[\bar{v}^s_{3k-2,0},\bar{\pi}_{3k-2},\bar{\pi}_{3k-2}] + \\&\eta[\bar{v}^s_{3k-4,0},\bar{v}^s_{3k-4,1},\bar{\pi}_{3k-4}],
\end{align}
\begin{align}\label{theorem_MGF_PW_harvall_proof_4}
\nonumber&q_{3B-1}: \left(\lambda + \mu - s\right)[\bar{v}^s_{3B-1,0},\bar{v}^s_{3B-1,1},\bar{v}^s_{3B-1,2}] =  \\&\lambda[\bar{v}^s_{3
B-2,0},\bar{\pi}_{3B-2},\bar{\pi}_{3B-2}] + \eta[\bar{v}^s_{3B-4,0},\bar{v}^s_{3B-4,1},\bar{\pi}_{3B-4}],
\end{align}
where $3 \leq k \leq B-1$. Finally, regarding the states in ${\rm r}_3$, we have
\begin{align}\label{theorem_MGF_PW_harvall_proof_5}
    q_{6}: \left(\eta + \lambda + \mu - s\right)\nonumber&[\bar{v}^s_{6,0},\bar{v}^s_{6,1},\bar{v}^s_{6,2}] = \\&\lambda[\bar{v}^s_{6,0},\bar{v}^s_{6,1},\bar{\pi}_{6}] + \lambda[\bar{v}^s_{5,0},\bar{v}^s_{5,1},\bar{\pi}_{5}],
\end{align}
\begin{align}\label{theorem_MGF_PW_harvall_proof_6}
   \nonumber& q_{3k}: \left(\eta + \lambda + \mu - s\right)[\bar{v}^s_{3k,0},\bar{v}^s_{3k,1},\bar{v}^s_{3k,2}] = \lambda[\bar{v}^s_{3k,0},\bar{v}^s_{3k,1},\bar{\pi}_{3k}] \\
    &+ \lambda[\bar{v}^s_{3k-1,0},\bar{v}^s_{3k-1,1},\bar{\pi}_{3k-1}] + \eta[\bar{v}^s_{3k-3,0},\bar{v}^s_{3k-3,1},\bar{v}^s_{3k-3,2}],
\end{align}
\begin{align}\label{theorem_MGF_PW_harvall_proof_7}
   \nonumber& q_{3B}: \left(\lambda + \mu - s\right)[\bar{v}^s_{3B,0},\bar{v}^s_{3B,1},\bar{v}^s_{3B,2}] = \lambda[\bar{v}^s_{3B,0},\bar{v}^s_{3B,1},\bar{\pi}_{3B}] \\&+ \lambda[\bar{v}^s_{3B-1,0},\bar{v}^s_{3B-1,1},\bar{\pi}_{3B-1}] + \eta[\bar{v}^s_{3B-3,0},\bar{v}^s_{3B-3,1},\bar{v}^s_{3B-3,2}],
\end{align}
where $3 \leq k \leq B-1$. Note that $\sum_{k \in\;{\rm r}_2}{\bar{v}^s_{k0}}$ and $\sum_{k \in\;{\rm r}_2}{\bar{v}^s_{k1}}$ can be expressed from the sum of (\ref{theorem_MGF_PW_harvall_proof_1})-(\ref{theorem_MGF_PW_harvall_proof_4}) as in (\ref{PW_v0_r2}) and (\ref{PW_v1_r2_sum}), respectively. Further, $\sum_{k \in\;{\rm r}_3}{\bar{v}^s_{k0}}$ and $\sum_{k \in\;{\rm r}_3}{\bar{v}^s_{k1}}$ can be expressed from the sum of (\ref{theorem_MGF_PW_harvall_proof_5})-(\ref{theorem_MGF_PW_harvall_proof_7}) as in (\ref{PW_v0_r3}) and (\ref{PW_v1_r3}), respectively. Therefore, the MGF of AoI at the destination can be expressed as
\begin{align}\label{MGF_PW_harvall_v0}
\nonumber & \overset{\rm PW}{M}(\bar{s}) = \sum_{k \in\; {\rm r}_1}{\bar{v}^s_{k0}} + \sum_{k \in\; {\rm r}_2}{\bar{v}^s_{k0}} + \sum_{k \in\; {\rm r}_3}{\bar{v}^s_{k0}},\\
\nonumber&\a \dfrac{\mu\left(\eta - s\right)\big[\left(\mu - s\right)\left(\lambda + \mu - s\right) - \lambda \left(\lambda - s\right)\big] \sum_{k \in\; {\rm r}_2}{\bar{v}^s_{k1}} }{\left(\mu - s\right)^2 \left(\lambda - s\right) \left(\eta - s\right)}\\&+ \dfrac{\mu\lambda \bar{v}^s_{31}\big[\mu\left(\mu - s\right) - \left(\lambda - s\right)\left(\eta - s\right)\big]}{\left(\mu - s\right)^2 \left(\lambda - s\right) \left(\eta - s\right)},
\end{align}
where step (a) follows from step (d) in (\ref{MGF_PW_v0}) and $\sum_{k \in\;{\rm r}_2}{\bar{v}^s_{k1}}$ is given by (\ref{PW_v1_r2_sum}). In addition, $\bar{v}^s_{31}$ can be obtained from (\ref{theorem_MGF_PW_harvall_proof_1}) and (\ref{theorem_MGF_PW_harvall_proof_5}) as
\begin{align}\label{PW_harvall_v31_r2}
\bar{v}^s_{31} = \dfrac{\mu \lambda\left(\bar{\pi}_5+\bar{\pi}_6\right) + \lambda \bar{\pi}_2\left(\eta + \lambda + \mu - s\right)}{\left(\eta + \mu - s\right)\left(\eta + \lambda + \mu - s\right)},
\end{align}
where $\bar{\pi}_{2} = \frac{\beta\left(1+\beta\right)^2 + \rho\beta}{\rho\left(1 + \rho + 2\beta\right)} \bar{\pi}_1$, $\bar{\pi}_{5} = \frac{\beta^2\left(1+\beta\right) \left(1+\rho+\beta\right)}{\rho\left(1 + \rho + 2\beta\right)} \bar{\pi}_1$, and $\bar{\pi}_{6} = \frac{\beta^2 \left(1+\rho+\beta\right)}{\left(1 + \rho + 2\beta\right)} \bar{\pi}_1$. The final expression of $\overset{\rm PW}{M}(\bar{s})$ in (\ref{theorem_MGF_PW_harvall_1}) can be obtained by substituting (\ref{PW_v1_r2_sum}) and (\ref{PW_harvall_v31_r2}) into (\ref{MGF_PW_harvall_v0}) while expressing $\sum_{k\in\;{\rm r}_2\;\cup\;{\rm r}_3}{\bar{\pi}_k}$ and $\sum_{k \in\;{\rm r}_1}{\bar{\pi}_k}$ as $\underset{k \in \;{\rm r}_2 \;\cup\;{\rm r}_3}{\sum}{\bar{\pi}_k} = \gamma_1 \bar{\pi}_1$ and $\underset{k \in \; {\rm r}_1}{\sum}{\bar{\pi}_k} = \gamma_2 \bar{\pi}_1$, respectively.
\hfill 
\IEEEQED

\bibliographystyle{IEEEtran}
\bibliography{AoI_QT_submitted.bbl}
\end{document}